\documentclass[longauth]{aa}

\usepackage{xparse}
\usepackage{graphicx}
\usepackage{txfonts}
\usepackage{tikz}
\usepackage{siunitx}
\usepackage[switch]{lineno}
\usepackage{overpic}
\usepackage{xspace}
\usepackage{float}
\usepackage{booktabs}
\usepackage{amsmath}
\usepackage{nccmath}
\usepackage{subcaption}
\usepackage{etoolbox}
\apptocmd{\thebibliography}{\setlength{\itemsep}{0pt}\setlength{\parskip}{0pt}}{}{}
\usepackage{placeins}
\usepackage{multirow}
\usepackage{xcolor}
\usepackage{float}
\usepackage[allcolors=blue]{hyperref}
\hypersetup{
    colorlinks=true,
    citecolor=blue,
    filecolor=magenta,      
    urlcolor=black,
    linkcolor = blue,
    }
\usepackage{orcidlink}
\usepackage{lipsum}
\usepackage{stfloats}

\newcommand{\nhp}{N$_2$H$^+$\xspace}
\newcommand{\nthp}{N$_2$H$^+$\xspace}
\newcommand{\hco}{H$_2$CO\xspace}
\newcommand{\hdos}{H$_2$\xspace}

\newcommand{\msun}{\hbox{$\hbox{\rm M}_{\odot}$}\xspace}

\newcommand{\kms}{\rm km\,s^{-1}}

\newcommand{\minflow}{$\dot{\textup{M}}_{\textup{tot,in}}$\xspace}
\newcommand{\HII}{H\,{\sc ii}\ }
\newcommand{\novel}{\multicolumn{1}{c}{\textemdash}}
\newcommand{\ntot}{N$_{\rm tot}$\xspace}
\newcommand{\mtot}{M$_{\rm tot}$\xspace}

\newcommand{\msunyr}{\hbox{$\hbox{\rm M}_{\odot}\,\rm{yr}^{-1}$}\xspace}

\begin{document} 

   \title{ALMA-IMF XVIII: The assembly of a star cluster: Dense \nhp~(1-0)
     kinematics in the massive G351.77 protocluster}
   \titlerunning{\nhp~(1-0) in G351.77}
   \author{N.\ A.\ Sandoval-Garrido   \inst{1}          \orcidlink{0000-0001-9600-2796},
		A.\ M.\ Stutz              \inst{1,2}        \orcidlink{0000-0003-2300-8200}, 
		R.\ H.\ Álvarez-Gutiérrez  \inst{1}          \orcidlink{0000-0002-9386-8612},
		R.\ Galván-Madrid          \inst{3}          \orcidlink{0000-0003-1480-4643},
		F.\ Motte                  \inst{4}          \orcidlink{0000-0003-1649-8002},
		A.\ Ginsburg               \inst{5}          \orcidlink{0000-0001-6431-9633},
		N.\ Cunningham             \inst{6}          \orcidlink{0000-0003-3152-8564},
		S.\ Reyes-Reyes            \inst{1,7}        \orcidlink{0000-0003-0276-5368},
		E.\ Redaelli               \inst{8,9}        \orcidlink{0000-0002-0528-8125},
		M.\ Bonfand                \inst{10}         \orcidlink{0000-0001-6551-6444},
		J.\ Salinas                \inst{1}          \orcidlink{0009-0009-4976-4320}, 
		A.\ Koley                  \inst{1}          \orcidlink{0000-0003-2713-0211},
		G.\ Bernal-Mesina          \inst{1}          \orcidlink{0009-0004-1442-3060},
		J.\ Braine                 \inst{11}         \orcidlink{0000-0003-1740-1284},
		L.\ Bronfman               \inst{12}         \orcidlink{0000-0002-9574-8454},
		G.\ Busquet                \inst{13,14,15}   \orcidlink{0000-0002-2189-6278},
		T.\ Csengeri               \inst{16}         \orcidlink{0000-0002-6018-1371},
		J.\ Di Francesco           \inst{17}         \orcidlink{0000-0002-9289-2450},
		M.\ Fern\'andez-L\'opez    \inst{18}         \orcidlink{0000-0001-5811-0454},
		P.\ Garcia                 \inst{19,20}      \orcidlink{0000-0002-8586-6721},
		A.\ Gusdorf                \inst{21,22}      \orcidlink{0000-0002-0354-1684},
		H.-L.\ Liu                 \inst{23}         \orcidlink{0000-0003-3343-9645},
		P.\ Sanhueza               \inst{24,25}      \orcidlink{0000-0002-7125-7685}
          }
   \institute{Departamento de Astronom\'{i}a, Universidad de Concepci\'{o}n,
         	 Casilla 160-C, Concepci\'{o}n, Chile 
         	 \email{nsandovalgarrido@gmail.com}
         	 \and
         	 Franco-Chilean Laboratory for Astronomy, IRL 3386, CNRS and Universidad de Chile, Santiago, Chile
         	 \and 
         	 Instituto de Radioastronomía y Astrofísica, Universidad Nacional Autónoma de México, Morelia, Michoacán 58089, México
         	 \and 
         	 Univ. Grenoble Alpes, CNRS, IPAG, 38000 Grenoble, France
         	 \and
         	 Department of Astronomy, University of Florida, P.O. Box 112055, Gainesville, FL 32611
         	 \and
         	 SKA Observatory, Jodrell Bank, Lower Withington, Macclesfield, SK11 9FT, United Kingdom
         	 \and
         	 Max-Planck-Institut für Astronomie, Königstuhl 17, D-69117 Heidelberg, Germany
         	 \and
         	 European Southern Observatory, Karl-Schwarzschild-Strasse 2, 85748 Garching, Germany
         	 \and
         	 Max-Planck-Institut für Extraterrestrische Physik, Giessenbachstrasse 1, 85748 Garching, Germany 
         	 \and
         	 Departments of Astronomy and Chemistry, University of Virginia, Charlottesville, VA 22904, USA
         	 \and
         	 Laboratoire d'Astrophysique de Bordeaux, Univ. Bordeaux, CNRS, B18N, all\'ee Geoffroy Saint-Hilaire, 33615 Pessac, France 
         	 \and
         	 Astronomy Department, Universidad de Chile,  Casilla 36-D, Santiago, Chile
         	 \and
         	 Departament de Física Quàntica i Astrofísica (FQA), Universitat de Barcelona (UB ), Martí i Franquès 1, 08028 Barcelona, Catalonia, Spain
         	 \and
         	 Institut de Ciències del Cosmos (ICCUB ), Universitat de Barcelona, Martí i Franquès 1, 08028 Barcelona, Catalonia, Spain
         	 \and
         	 Institut d'Estudis Espacials (IEEC), Esteve Terradas 1, Edifici RDIT, Ofic. 212
         	 Parc Mediterrani de la Tecnologia (PMT) Campus del Baix Llobregat – UPC
             08860 Castelldefels (Barcelona), Catalonia, Spain
         	 \and
         	 Laboratoire d'Astrophysique de Bordeaux, Univ. Bordeaux, CNRS, UMR 5804, F-33615 Pessac, France
         	 \and
         	 NRC Herzberg Astronomy and Astrophysics Research Centre, 5071 West Saanich Road, Victoria, BC V9E 2E7
         	 \and
         	 Instituto Argentino de Radioastronom\'ia (CCT-La Plata, CONICET; UNLP; CICPBA), C.C. No. 5, 1894, Villa Elisa, Buenos Aires, Argentina
         	 \and
         	 Chinese Academy of Sciences South America Center for Astronomy, National Astronomical Observatories, CAS, Beijing 100101, China
         	 \and
         	 Instituto de Astronom\'ia, Universidad Cat\'olica del Norte, Av. Angamos 0610, Antofagasta, Chile
         	 \and
         	 Laboratoire de Physique de l'\'Ecole Normale Sup\'erieure, ENS, Universit\'e PSL, CNRS, Sorbonne Universit\'e, Universit\'e Paris Cit\'e, F-75005, Paris, France
         	 \and
         	 Observatoire de Paris, PSL University, Sorbonne Universit\'e, LERMA, 75014, Paris, France
         	 \and
         	 School of Physics and Astronomy, Yunnan University, Kunming, 650091, People’s Republic of China
         	 \and
         	 Department of Earth and Planetary Sciences, Institute of Science Tokyo, Meguro, Tokyo, 152-8551, Japan
         	 \and
         	 National Astronomical Observatory of Japan, National Institutes of Natural Sciences, 2-21-1 Osawa, Mitaka, Tokyo 181-8588, Japan
             }

   \authorrunning{N.\ A.\ Sandoval-Garrido et al.}
   

   \date{Received 15 October 2024; accepted XXX}

   \abstract{ ALMA-IMF observed 15 massive protoclusters capturing multiple
 spectral lines and the continuum emission. Here we focus on the
 massive protocluster G351.77 ($\sim$ 2500~\msun, estimated from
 single-dish continuum observations) located at 2~kpc. We trace the
 dense gas emission and kinematics with \nhp~(1-0) at $\sim$\,4~kau
 resolution.  We estimate an \nhp relative abundance $\sim\,(1.66 \pm
 0.46) \times 10^{-10}$. We decompose the \nhp emission into up to
 two velocity components, highlighting the kinematic complexity in the
 dense gas.  By examining the position-velocity (PV) and PPV diagrams
 on small scales, we observe clear inflow signatures (V-shapes)
 associated with 1.3~mm cores. The most prominent V-shape has a mass
 inflow rate of $\sim\,13.45 \times 10^{-4}$~\msunyr and a short
 timescale of $\sim$\,11.42~kyr. We also observe V-shapes without
 associated cores. This suggests both that cores or centers of
 accretion exist below the 1.3~mm detection limit, and that the
 V-shapes may be viable tracers of very early accretion and star
 formation on $\sim\,4$~kau scales. The large-scale PV diagram shows
 that the protocluster is separated into 2 principal velocity
 structures separate by $\sim\,$2\,$\kms$. Combined with smaller scale
 DCN and H$_2$CO emission in the center, we propose a scenario of
 larger scale slow contraction with rotation in the center based on
 simple toy models. This scenario is consistent with previous lines
 of evidence, and leads to the new suggestion of outside-in evolution
 of the protocluster as it collapses. The gas depletion times implied
 by the V-shapes are short ($\sim\,$0.3~Myr), requiring either very fast
 cluster formation, and/or continuous mass feeding of the protocluster.
 The latter is possible via the Mother Filament G351.77 is forming out of.
 The remarkable similarities in the properties of G351.77 and the recently 
 published work in G353.41 indicate that many of the physical conditions 
 inferred via the ALMA-IMF \nhp observations may be generic to protoclusters.}


   \maketitle
   

   \section{Introduction}\label{sec:introduction}

The fragmentation of molecular clouds leads to the formation of
clumps. Regions where gas and dust are densely concentrated give 
rise to prestellar cores, where stars will later form \citep{lada}. 
These regions, where the gas is actively turning into several young stars, 
are termed protoclusters, representing the gas-dominated bassinets
of stellar clusters. These structures offer valuable
insights into the initial stages of star cluster formation,
enabling us to characterize their early evolution \citep{peretto, yueh, anualmotte, amy2018, rodrigo24}.

\begin{figure}[h]
\centering
\includegraphics[width = \columnwidth]{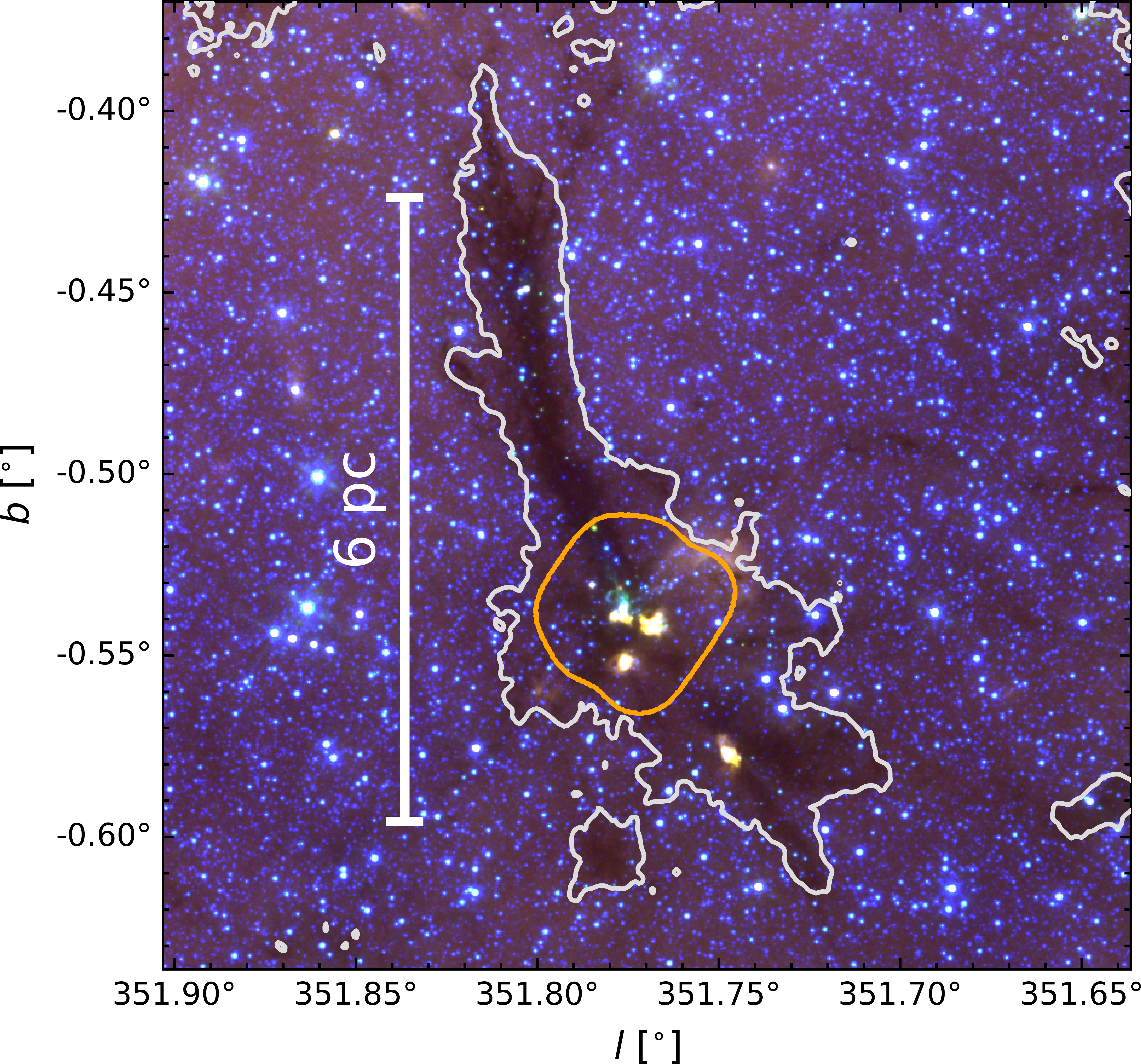}
\caption{G351.77-0.53 filamentary region IRAC color composite,
  where the 8.0~$\mu$m, 5.6~$\mu$m, and 3.6~$\mu$m are shown in red, green, and
  blue, respectively. The orange contour shows the coverage of the 3~mm data
  ALMA-IMF Large Program observations. The white contour
  represents the ATLASGAL emission (850~$\mu$m) at 0.62~Jy~beam$^{-1}$.
  The dominant central closed contour indicates
  the ``Mother Filament'', discussed below \citep[also referred to as
  the ``G351 filament'' in][]{Simon}. This filament is also observed
  in NIR extinction and submm emission. In its densest part the star formation
  activity is revealed by strong NIR emission.}
\vspace{-0.3cm}
\label{fig:rgbimage}
\end{figure} 

At the same time, filamentary structures are present in many star 
forming regions throughout the Galaxy. These have been shown to 
hold significant importance in both low and high mass star formation 
processes, presenting diverse and intricate morphologies 
\citep{john, Lee2014, Amy2015, amy2016, anualmotte, amy2018, hacar2023, hongli2023}. Their 
formation can be triggered by different processes, such as shock fronts, 
cloud collisions,
feedback, magnetic fields, gravitational instabilities, or global 
environmental effects \citep{amy2016, Louvet2016, montillaud, bonne, issac, shuo, hacar2023, hongli2023}. More specifically, kinematics studies have revealed a powerful means to probe the physical processes at work within filaments coveting a range of line-masses. These processes include fragmentation, rotation, infall, magnetic instabilities, 
to name a few \citep{Henshaw, Manuel2014, liu2015, Amy2015, amy2016, amy2018.2, amy2018,  hongli2019, Valentina19, Rodrigo21, Patricio2021, zhou2022, hacar2023, 
rodrigo24}. 
Meanwhile, a strong link has been established between filamentary 
structures, dense cores, and protoclusters 
\citep{Schneider, Roberto2010, amy2016,  amy2018, Plunkett, rodrigo24}, 
leading to the discovery of connections between the kinematics and 
chemistry of dense cores and the gas surrounding the filamentary 
structure \citep[]{ hacar2011, tafalla2014, amy2016, andre2019, hongli2020, 
kim2022, anirudh2023}. Therefore, understanding the kinematic processes
is crucial for characterizing the physical 
mechanisms associated with star formation over time. This 
understanding allows us to link mass, density, velocity
gradients, and other properties associated with the formation of 
stars, with the characteristics of the cores found in these regions 
\citep{Nichol}. These cores, which represent the sites where dust 
and gas will turn into individual or small numbers of stars through gravitational collapse, show 
strong  relations with denser regions in molecular clouds. They are 
preferentially embedded within dense filamentary structures, which are 
influenced by gravitational effects and processes such as infall and 
magnetic fields \citep[][]{ amy2016, anualmotte, li23, hongli2023, pirogov23, kirk24, rodrigo24}. 
Together, the above studies highlight the imperative need to connect the structures, namely filaments, within which cores and, ultimately, stars are born, to the cores themselves, including their kinematic properties. 

In this context, the ALMA-IMF Large Program\footnote{\url{https://www.almaimf.com/}}
observed 15 massive $(2.5 - 33~\times10^3$~\msun) and relatively nearby
($2 - 5.5$~kpc) protoclusters down to $\sim$~2~kau resolution, with the
main goal of understanding the origin of stellar masses. The ALMA-IMF
Large Program utilized the 12m-array, 7m-array and Total Power (TP)
antennas of the Atacama Large Millimeter/Submillimeter Array (ALMA),
employing the 1.3~mm and 3~mm bands, providing observations of the
continuum and spectral lines \citep{motte22, adam22, poetau1, brouillet, nony, poetau2, Nichol, danidiaz, towner, Armante, Melisse, Pierre, rodrigo24, Roberto, Fabien}.  
A key aspect of the ALMA-IMF data-set is that the protoclusters were 
observed at approximately matched resolution and sensitivity. The 15 
protoclusters were classified into different evolutionary stages, Young,
Intermediate, and Evolved, based on their observed 1.3~mm and 3~mm fluxes, as 
well as the free-free emission at the H41$\alpha$ frequency
\citep[][]{motte22, Roberto}. This classification takes into account 
the extent of dense gas impacted by local \HII regions \citep{motte22}. 

The 1.3~mm and 3~mm continuum images provide the possibility of
detecting and analyzing cores and their key parameters, such as
temperature, molecular composition, and masses, to name a
few \citep[][Motte et al., submmited]{poetau1, brouillet,
poetau2, Pierre, Fabien}. These studies demonstrated that cores populations
are influenced 
and characterized by the cloud formation process, their evolutionary
stage, and the history of star formation. Additionally, they show that
cores increase their masses during the protostellar phase through
inflowing material, with massive cores exhibiting greater mass
growth than their lower mass counterparts \citep{nony}. 

Simultaneously,
spectral lines enable us to analyze core kinematics via e.g., DCN
\citep{Nichol}, demonstrating that this molecule can trace structures 
with different morphologies and complex velocity. These structures are
more extended and filamentary in Evolved regions compared to
Intermediate and Young regions, where the DCN emission appears more
compact (see also \citealt{Nichol}). Additionally, the emission from SiO and CO
analyses have facilitated the detection of outflows in the protoclusters
\citep[][Valeille et al. submmited; Nony et al. in prep.]{towner, nony}, 
revealing that outflow properties are correlated with the total core 
mass and connections between outflow mass and the total mass of the 
protocluster.

Given the emergent nature of protoclusters, which are gas-dominated but
also are main factories for the stellar content we can so readily 
observe in both our Galaxy and external ones, a precise gas tracer is 
crucial to delineate and comprehend the intricate gas dynamics during 
these early dense and cold phases. Moreover, a critical yet 
under-exploited aspect in the above studies is pinpointing the 
properties of the cold, dense extended gas outside the relatively 
compact cores. This gas is accessed here via the \nhp~(1-0) line at 
$\sim$~93.173~GHz. 

More thorough research in laboratories and nearby star-forming regions 
has revealed \nhp~(1-0) emissions occurring at seven different frequencies 
\citep{green74, turner74, tadeus75, Caselli95, Ivanov}, resulting in 
hyperfine structure in the spectrum. \nhp~(1-0) has critical densities 
between $\sim~6.1~\times~10^4$~cm$^{-3}$ and $2.0~\times~10^4$~cm$^{-3}$ 
at kinetic temperatures from 10~K to 100~K \citep{shirley2015}. \nhp is 
formed during the gas-phase reactions from 
H$_3^{+}$~+~N$_2$~$\rightarrow$~\nhp~+~\hdos at temperatures of 20~K \citep{bergin2001, jorgensen2004,  Bergin, hoff2017, Naiping2018} 
where it is mostly resistant to freezing onto dust grains. However, at 
temperatures above 20~K (but see \citealt{Dale} for the theoretical side, and \citealt{Tatematsu}; \citealt{amy2016}; \citealt{Valentina19}; and \citealt{Hacar2024} for the observational side), it can be destroyed by CO molecules, which are 
desorbed from the dust grains, leading to reactions such as 
\nhp~+~CO~$\rightarrow$~HCO$^{+}$~+~N$_2$. Alternatively, it can be destroyed by 
free electrons in \HII regions, resulting in reactions like \nhp~+~$e^{-}$~$\rightarrow$~N$_2$~+~H or HN~+~N
\citep{jorgensen2004, Patricio2012, tobin2013, Lippok2013, hoff2017}. These 
characteristics make \nhp~(1-0) a reliable tracer for dense and cold gas. 
It allows us to investigate the initial stages in star-forming regions 
and understand the chemistry, kinematics, and dynamics of filamentary 
structures, protoclusters, and cores \citep{Daniel2005, fontani2006, Gemma2011,  Storm, Pety, Chen2019, Schwarz2019,
Valentina19, Rodrigo21, Elena22}.

An important analysis that can be applied to the velocity structure of 
the dense gas derived from the \nhp~(1-0) line emission is the study of 
intensity-weighted Position-Velocity 
(PV) diagrams, utilizing integrated intensity, velocities, and positions 
\citep[][Salinas et al.\ in prep.; Stutz et al.\ in prep.]
{Valentina19, Rodrigo21, rodrigo24}. These diagrams enable the 
characterization of important kinematic patterns that can provide insight into 
rotation, infall, cloud-cloud collisions, or transversal velocity gradients 
within filaments \citep{Henshaw, Manuel2014, amy2016, montillaud, Rodrigo21, Elena22, rodrigo24}. Additionally, PV diagrams provide a valuable tool for characterizing velocity 
gradients, measuring timescales, and estimating mass inflow rates at small scales, enabling us to link these characterizations to core scales. Notably, recent 
research of the G353.41 protocluster, one of the ALMA-IMF targets, has revealed 
correlations between cores and velocity structures observed in PV space at 
small scales. Meanwhile, large-scale analysis of the velocity structures on 
PV space suggest an ongoing infalling process \citep{rodrigo24}. This approach allow us to associate these characterizations with core evolution, and infer the behavior and timescales of future processes. 

The high-mass star-forming region G351.77-0.53 (IRAS 17233-3606) is 
a filamentary infrared dark cloud (IRDC) (see Fig.~\ref{fig:rgbimage}) 
located at $\sim\,2\,\pm\,0.14$~kpc and whose estimated mass is $\sim\,10200$~\msun 
\xspace for the entire filament within an area of 11~pc$^2$ \citep[][]{Simon}. Recent observations
of several molecular tracers 
such as CO, HCO$^+$, CH$_3$OH, CH$_3$CN, SiO, $^{13}$CO, 
C$^{17}$O, C$^{18}$O, H$_2$O and \hdos have shown a variety of physical 
processes along the filament. These processes encompass fragmentation 
into different clumps and the generation of turbulence, attributed to 
ongoing star formation activity within the region, magnetic fields, 
and gravitational effects produced by the ongoing high-mass star formation 
\citep{Leurini2011_1, Naiping2018, Leurini2019, Sabatini2019}. Intriguingly, 
it has been estimated that the star formation efficiency and the star 
formation rate along the filament is extremely low compared to the mass reservoir (see above) and low compared to local clouds, and compared specifically to 
Orion A \citep[even when accounting for incompleteness,][]{Simon}. 

\begin{figure}[h]
\centering
\includegraphics[width = \columnwidth]{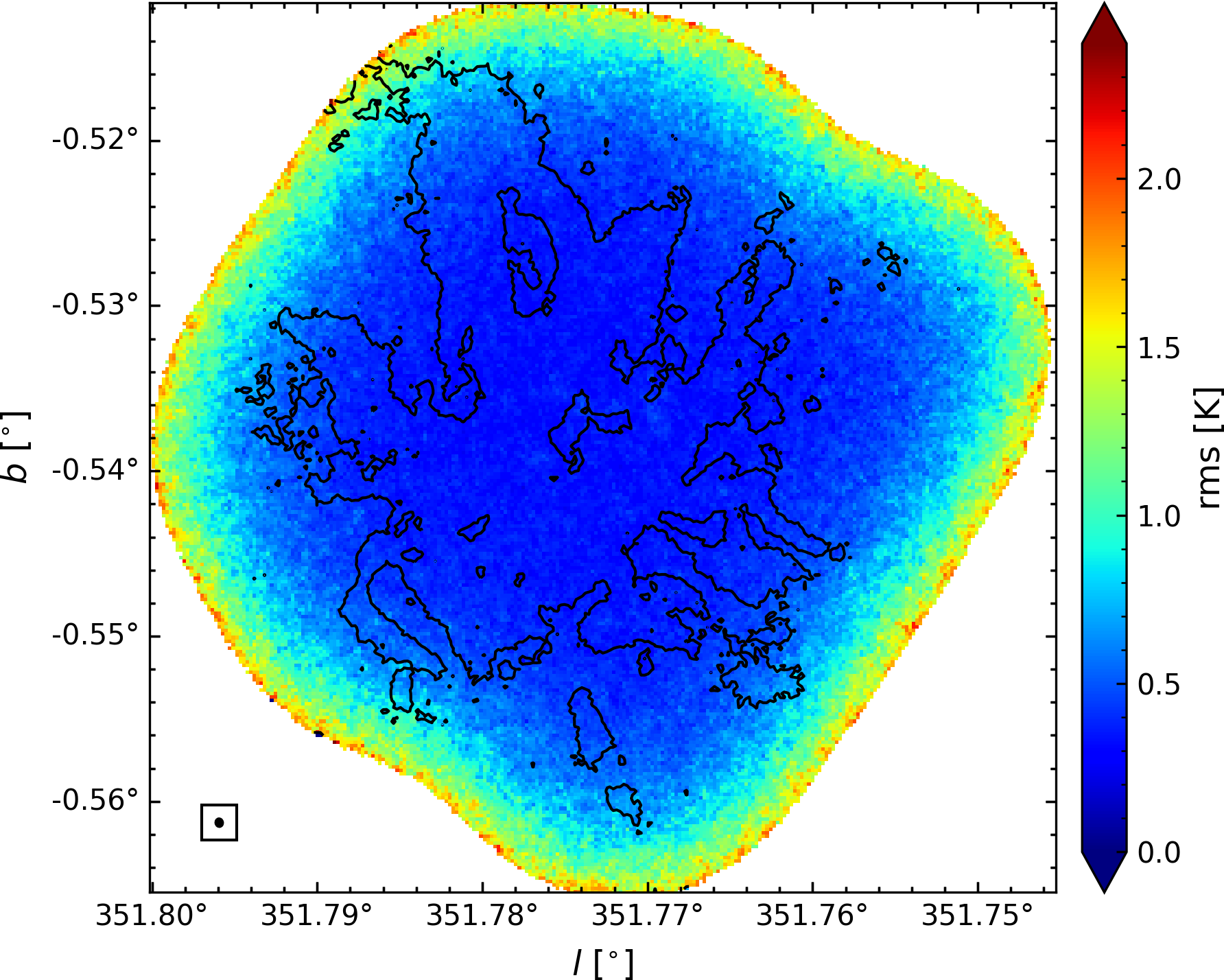}
\includegraphics[width = \columnwidth]{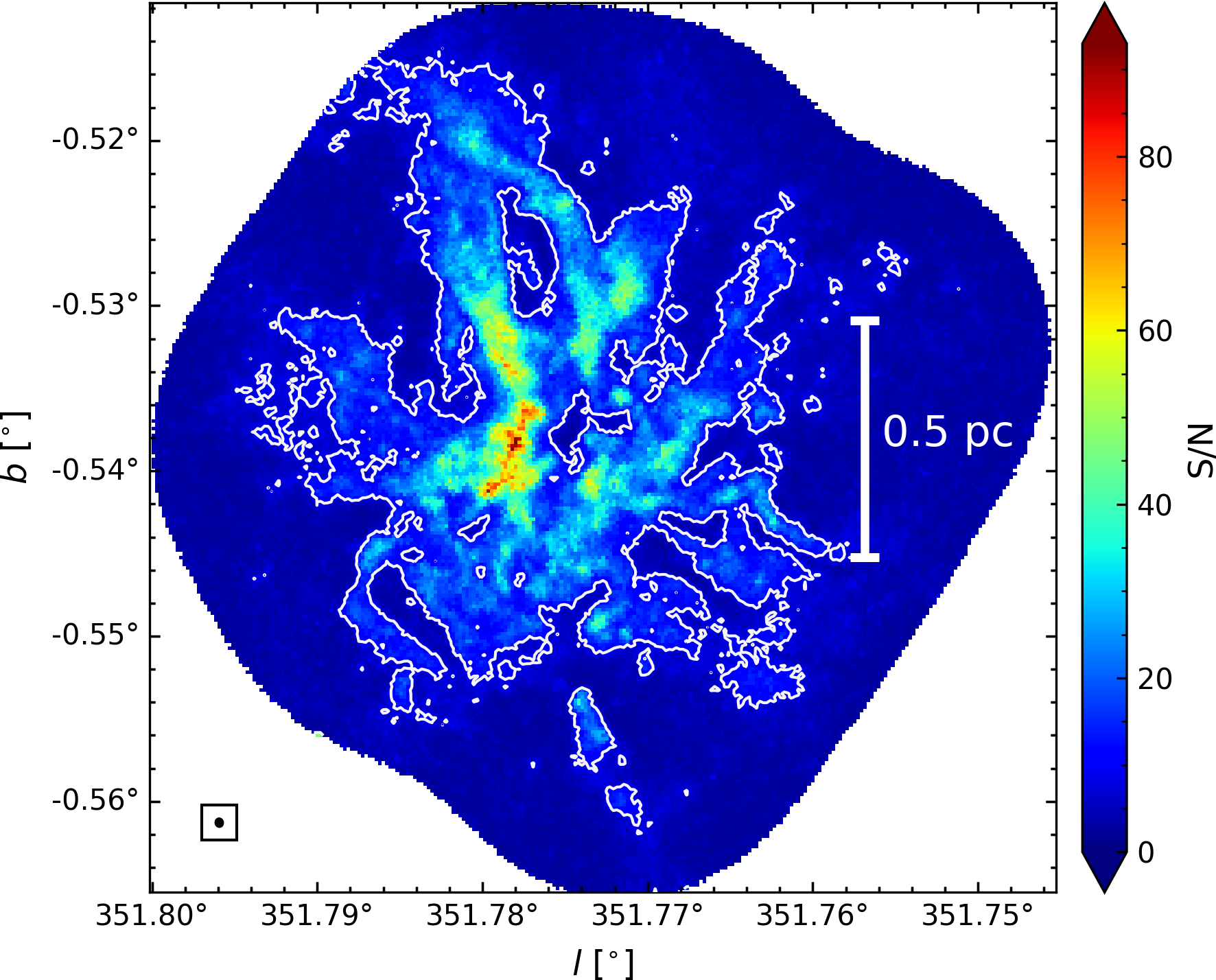}
\caption{Top: Noise map of the fully combined \nhp image (see text) in 
  the G351.77 protocluster. The spectra with the highest noise values 
  are located near the edges, as expected. The mean rms value across the entire map is $\sim$~0.66~K, whereas the mean value in regions with S/N > 9 is $\sim$~0.40~K. Bottom: Signal-to-Noise ratio 
  (S/N) map. Spectra with high S/N are distributed following filamentary 
  structures. The contours show the areas where spectra with S/N~$\geq\,9$ are found (see 
  \S~\autoref{sec:experiment}). The
  ellipse in the bottom-left corner represents the final beam 
  size of the \nhp data of $2.3\arcsec \times 2.1\arcsec$, or 4.6~kau~$\times$~4.2~kau.}
\vspace{-0.3cm}
\label{fig:snrmap}
\end{figure}  

Focusing on the most prominent clump, which we will henceforth refer to as the G351.77 protocluster, it is classified as being in an Intermediate evolutionary stage compared to other ALMA-IMF protoclusters (see above). An UC\HII region is located at its center \citep[][]{motte22, Roberto}, and the presence of clear outflows, sometimes bipolar, indicating vigorous and ongoing star formation
\citep{Leurini2008, Zapata2008, Leurini2009, Leurini2011_2, Leurini2013, Leurini2014, Klaassen, Antyufeyev, towner}. 
These outflows are closely linked to Young 
Stellar Objects (YSOs). Additionally, an evident velocity gradient is observed 
in several tracers, revealing two different velocity components 
\citep{Leurini2019}. This gradient is also observed in some tracers from 
the ALMA-IMF survey such as DCN~(3-2), and C$^{18}$O~(2-1) \citep[e.g.][Koley et al. 
in prep]{Nichol}. Furthermore, eighteen cores have been cataloged within the G351.77 protocluster, with masses from 0.5~\msun to 
36.5~\msun \citep{Fabien}. An analysis of CH$_3$OCHO has identified 
five sources in the central part of the protocluster, 3 of which have 
been classified as hot cores \citep[2 detected in the 1.3~mm band,][]{Melisse}. These hot cores may explain 
the lack of \nhp~(1-0) emission in the center of the protocluster (see 
Fig.~\ref{fig:snrmap}). Moreover, kinematic analysis at scales $< 1$~kau in 
the central zone of the protocluster reveals the existence of disks with 
signs of outflows, infall, and rotation \citep{Zapata2008, Beuther}.
  
In this paper, we focus on the kinematics of dense and cold gas within 
the massive G351.77 protocluster using the \nhp~(1-0) spectral line
emission. To achieve this, we image the observations obtained from 
the 12m-array and 7m-array configurations for the ALMA-IMF Large Program, 
feathering them with TP observations, as explained in \S~\ref{sec:data}. 
We employ specialized software to fit the individual spectra, enabling precise 
and meticulous characterization of the kinematics and spectral properties, a process detailed 
in \S~\ref{sec:line-fitting-process}. We estimate column densities, relative
abundances, and masses of \nhp~(1-0) in \S~\ref{sec:column-density-and-masses}. 
Additionally, we analyze PV diagrams to characterize structures at both small
and large scales in order to comprehend the physical processes occurring within
the protocluster in \S~\ref{sec:kinematics-analysis}. We present
the discussion of the results, proposing potential physical scenarios
in \S~\ref{sec:discussion}. Finally, in \S~\ref{sec:conclusion} we present our conclusions.

\vspace{-0.1cm}
\section{Data}\label{sec:data}

We analyze the \nhp~(1-0) observations of the G351.77 protocluster in
the 3~mm spectral band, observed by the 12m-array, 7m-array and TP
configurations obtained from the ALMA-IMF Large Program
\citep{motte22}. Our data has a maximum spatial resolution of $\sim$~4000~au
and a spectral resolution of 0.23~$\kms$. The beam sizes are (2.3\arcsec
~$\times$~2.1\arcsec), (16.9\arcsec~$\times$~10.1\arcsec), and 
(69.6\arcsec~$\times$~69.6\arcsec) for the 12m-array, 7m-array, and TP, 
respectively.

\subsection{\nhp data imaging}\label{sec:data-reduction}

The data imaging process was performed using CASA 6.2 \citep{CASA} along 
with the ALMA-IMF data
pipeline\footnote{\url{https://github.com/ALMA-IMF/reduction}}. Initially,
we imaged the 12m-array data to determine the optimal parameter 
settings to achieve the highest quality results. Among the different 
parameters tested within the
\textit{tclean}\footnote{\url{https://casadocs.readthedocs.io/en/stable/api/tt/casatasks.imaging.tclean.html}}
CASA task, we explored different imaging results by varying:
\textit{threshold}, \textit{deconvolver}, \textit{pbmask},
\textit{pblimit} and \textit{scales}. These parameter values
\footnote{\url{https://github.com/ALMA-IMF/reduction/blob/master/reduction/imaging_parameters.py}} 
were selected based on the analysis of the residuals, model, 
and  the image returned after each imaging process, in a way 
that minimized the residuals in the final product. 
Subsequently, we applied the same parameter values to combine 
the 12m-array and 7m-array data with \textit{tclean} to 
increase the Fourier coverage (7M12M, henceforth). This combination produces artifacts such as peaks and bowls of intensity around the edges. These appear to be due to the differing UV-plane coverage of the 12m-array versus 7m-array, as well as the low signal near the edges. However, by increasing the pblimit value from 0.05 to 0.2 for this product – the later of which is a commonly used threshold in similar ALMA datasets – these artifacts are no longer produced, as the edge regions with lower sensitivity are effectively excluded.

Continuum subtraction was accomplished using the CASA task
\textit{imcontsub}. This task uses the channels free of line emission to generate a
continuum model, which is subtracted from the line emission
channels. This process yields the continuum-subtracted 7M12M data.

As a final step, we employed the CASA task
\textit{feather} to combine the 7M12M continuum subtracted data with the TP data to
recover the extended line emission. Here we define the 7M12M continuum
subtracted data as the high resolution data, and the TP as the low
resolution data. Finally, we obtain a fully combined spectral cube (data-cube, see Fig.~\ref{fig:channelmap}), with a spectral resolution of 0.23~$\kms$, and whose size final synthesized beam is 2.3\arcsec~$\times$~2.1\arcsec and a Beam-Position-Angle (BPA) of 89$^{\circ}$. The noise and signal-to-noise ratio (S/N) map of our imaged data are shown in Fig.~\ref{fig:snrmap}. The mean rms value across the entire map is $\sim$~0.66~K, whereas the mean value in regions with S/N > 9 is $\sim$~0.40~K.

\subsection{Other ALMA-IMF data}

In our analysis, we use the smoothed \textit{getsf} \citep{Menshchikov}
core catalog from \citet{Fabien}, derived from the continuum 
images of the 1.3~mm band \citep{adam22}. We also utilize the core kinematics,
integrated intensity map, and mean velocity map obtained from 
the DCN~(3-2) spectral line fits from the 12m-array \citep{Nichol}.
Further, we make use of the integrated intensity map and
mean velocity map of H$_2$CO~(3-2) derived from the 12m-array observations
as part of the spectral setup of the ALMA-IMF Large Program \citep{motte22}. 
Additionally, we use an intermediate product of the \hdos column density 
map of G351.77 at 6\arcsec resolution, derived from a combination of 
1.3~mm band, SOFIA/HAWC+ (53~$\mu$m, 89~$\mu$m, and 214~$\mu$m), 
APEX/SABOCA (350~$\mu$m) and APEX/LABOCA (870~$\mu$m) observations 
\citep{Pierre}. Although the resolution of the final product of the \hdos 
column density is 2.5\arcsec, we use this intermediate product in order 
to reduce the uncertainties of our estimations that could be produced by the
combination of the data of \hdos and from our fits. Moreover, this approach ensures a more uniform and reliable analysis by minimizing artifacts that may appear in maps at
higher resolution and allows the calculation of one representative
value of the relative abundance for the protocluster (Dell'Ova,
private communication 2024).

\vspace{-0.1cm}
\section{Line fitting procedure}\label{sec:line-fitting-process}

\begin{figure}[h]
\centering
\includegraphics[width = \columnwidth]{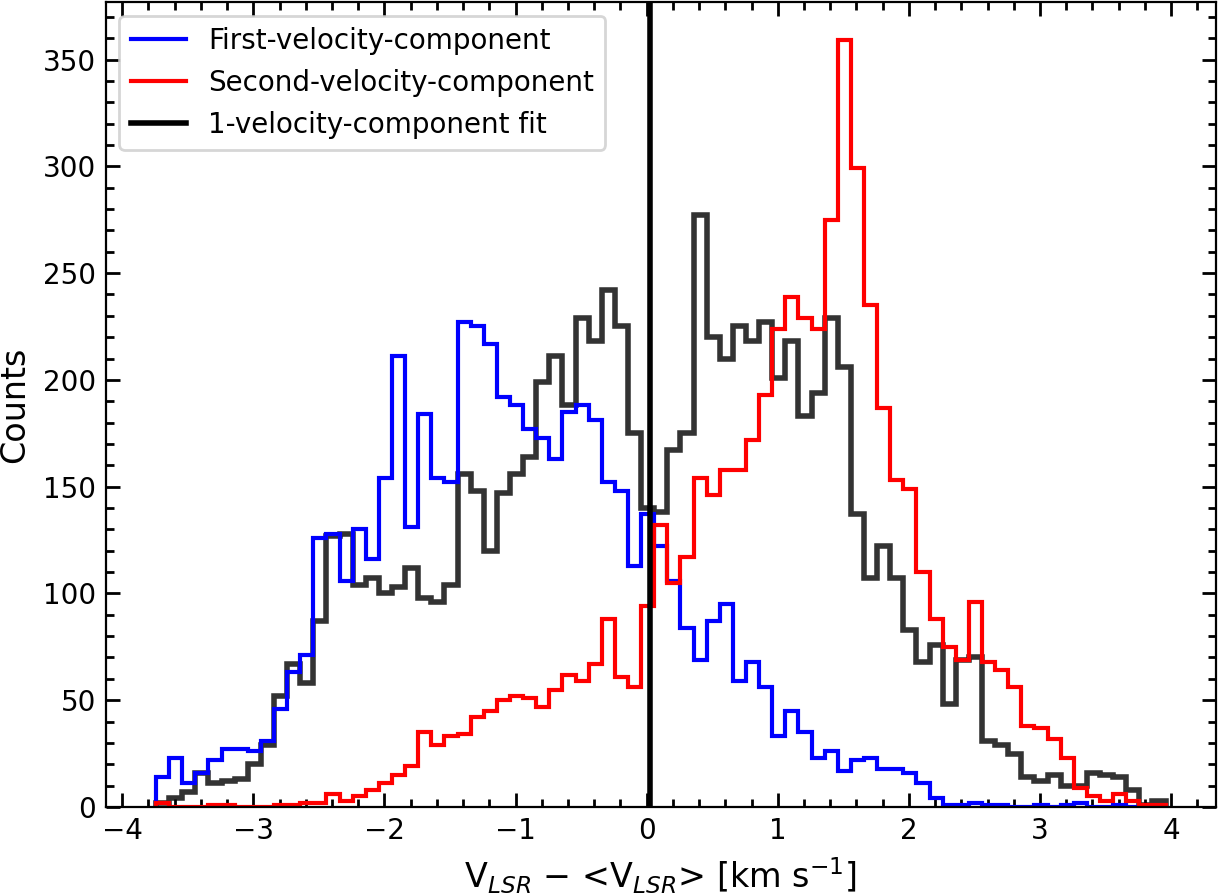}
\caption{Centroid velocity distributions of the 1- and
  2-velocity-components fits. The black histogram represents the
  centroid velocity distribution of the spectra fitted by the
  1-velocity-component model. The blue and red histograms show the centroid
  velocity distributions of the spectra fitted with the
  2-velocity-components model, where blue corresponds to the 
  First-velocity component and red to the Second-velocity-component. The black solid line
  represents the middle point $X$ located at ~0.025~$\kms$, measured via
  Eq.~\ref{eq:vel_limit_eq} (see text). This divides the black histogram
  into two velocities, as explained in
  \S~\ref{sec:bluest-and-reddest-velocity-component}.}
\vspace{-0.3cm}
\label{fig:histogram_veldistribution}
\end{figure}

Since our main goal in this paper is to analyze the fine and large
scale protocluster kinematics of dense gas, we require the use of
spectral line fitting to retrieve the radial velocity field. Moreover, the
line fitting of \nhp~(1-0) hyperfine structure provides additional parameters of
interest, i.e.\ the optical depth ($\tau$) and excitation temperatures
(T$_{\rm ex}$, see below). The first examination of the data-cube immediately 
reveals that multiple velocity components exist over a
significant number of spectra, consistent with previous research 
\citep[e.g.][see Fig.~\ref{fig:histogram_veldistribution}]{Leurini2019}. Hence, we adopt an iterative
approach to line fitting, as described in detail below.

We only fit spectra with S/N~$>$~9. This approach stems from
our experimental findings, which reveal that spectra with a S/N~$<$~9
are inadequately fitted, yielding high uncertainties in the returned parameters (see
Fig.~\ref{fig:snrmap} and \S~\autoref{sec:experiment}). We ultimately
fit the spectral cube with 2-velocity-components when possible (driven
mainly by S/N considerations, see \S~\ref{sec:two-component-fit}) and 
1-velocity-component when the spectra are either relatively simple or 
the noise precludes more detailed velocity decomposition. To accomplish 
this fitting, we begin with a 1-velocity-component fit (see 
Fig.~\ref{fig:specwellfited1comp}) and then with the 2-velocity-components fit
(see Fig.~\ref{fig:specwellfited2comp}), which are independent. 
We refer to these two relatively ``raw'' fits as the First Fitting 
Procedures (FFPs). We then analyze the S/N of the 
decomposed spectra to identify where we have reliable 
2-velocity-components fits; where we do not, we adopt the
1-velocity-component fit for the spectrum being analyzed 
(see \S~\ref{sec:best-fit-and-final-model} and
Fig.~\ref{fig:map_components}).

\begin{table*}[h]
\caption{Guess and limit values entered for 1- and 2-velocity-components fits.}
\vspace{-0.4cm}
\begin{center}
	\scalebox{0.96}{
\begin{tabular}{ccccccccc}
\hline\hline
\\[-3mm]
Parameters  &  T$_{\rm {ex,1}}$ [K] & $\tau_{1}$ & V$_{\rm {LSR,1}}$ [$\kms$] & $\sigma_{\rm {v,1}}$ [$\kms$]  & T$_{\rm {ex,2}}$ [K] & $\tau_2$ & V$_{\rm {LSR,2}}$ [$\kms$] & $\sigma_{\rm {v,2}}$ [$\kms$] \\ 
\hline
\\[-3mm]
 & \multicolumn{4}{c}{Single velocity component} & \multicolumn{4}{c}{} \\
               
\hline
\\[-3mm]

FFP guesses & 20.0 & 1.0 & -3.0 & 1.0 & \novel & \novel & \novel & \novel \\

Final guesses& 12.4 & 3.64 & -3.77 & 1.0 & \novel & \novel & \novel & \novel    \\

Limits& (2.73, 100) & (0, 100) & (-13, 3) & (0, 6) & \novel & \novel & \novel & \novel \\    

\hline
\\[-3mm]

 & \multicolumn{4}{c}{First-velocity-component} & \multicolumn{4}{c}{Second-velocity-component} \\

\hline
\\[-3mm]

FFP guesses & 20.0 & 1.0 & -5.0 & 1.0 & 20.0 & 1.0 & -2.0 & 1.0 \\

Final guesses & 9.22 & 4.60 & -5.0 & 0.83 & 9.0 & 7.4 & -2.6 & 0.66 \\

Limits  & (2.73, 100) & (0, 100) & (-13, 3) & (0, 6) & (2.73, 100) & (0, 100) & (-13, 3) & (0, 6) \\
\hline

\end{tabular}
}
\tablefoot{The ``Single velocity component'' shows the guesses and limits 
used in the 1-velocity-component fit. The ``First-velocity-component'' 
and ``Second-velocity-component'' show the guesses and limits used in the 
2-velocity-component fit. The FFP guesses represent the initial guesses 
used in the first fitting process for both 1- and 2-velocity-component 
fits. The Final guesses are averaged of the parameters obtained from 
previous fits, used to improve the models in both 1- and 
2-velocity-component fits. The Limits represent the range of values 
within which the returned parameters will lie.}
\vspace{-0.5cm}
\label{tab:guesses_table}
\end{center}
\end{table*}

Here we use the \nhp~(1-0) line model
\textit{n2hp\_vtau}\footnote{\url{https://pyspeckit.readthedocs.io/en/latest/example_n2hp_cube.html}}
from
PySpecKit\footnote{\url{https://pyspeckit.readthedocs.io/en/latest/index.html\#}}
\citep{PySpecKit11, PySpecKit22} to fit the cube
\citep[e.g.,][]{ Elena, Valentina19, Rodrigo21}. This model
is an LTE model, whose spectroscopic predictions come from known molecular data
under the LTE assumption, such as 
rest frequencies and hyperfine structures. This procedure returns
4 parameters per velocity component: excitation temperature
(T$_{\rm ex}$, see Fig.~\ref{fig:temp_map}), optical depth ($\tau$, see Fig.~\ref{fig:tau_map}), centroid velocity (V$_{\text{LSR}}$), and line
width ($\sigma_{\rm v}$), each representing a measurement for the entire spectrum.

To fit the data-cube, four different guesses must
be entered to initialize PySpecKit. In addition to the
guesses, each parameter is assigned both a lower and upper limit
value, defining the range within which the final fitted values will
lie. In Table~\ref{tab:guesses_table}, we show the guess and limit
values for each parameter used in the FFP and in the final fit. After
the fitting process, PySpecKit returns the model spectra, parameter
values (see above), and associated parameter errors. Finally, applying 
methods in order to define the best fit for each spectra  we merge the 
1- and 2-velocity-components fits into one model cube (see 
\S~\ref{sec:best-fit-and-final-model}, Fig.~\ref{fig:momentmaps0}, and Fig.~\ref{fig:velocity_map} 
for example maps of the spectral fits, including the integrated 
intensity, mean velocity and the line width, and also Fig.~\ref{fig:temp_map} and Fig.~\ref{fig:tau_map} for examples of excitation temperature maps and optical depth maps). In the analysis in 
\S~\ref{sec:bluest-and-reddest-velocity-component}, we define the two 
main velocity components of the protocluster (see 
Fig.~\ref{fig:momentmaps0} and Fig.~\ref{fig:velocity_map}), 
where we also perform a careful measurement of the V$_{\text{LSR}}$ of 
G351.77 where $<$V$_{\rm LSR}>~=~-3.843~\pm~0.001~\kms$. In the text 
that follows, we present more details of the procedures outlined above, and we will show in
\S~\ref{sec:bluest-and-reddest-velocity-component} that there is a continuity between the 1-
and 2-velocity-component fits except in some locations presenting jumps. 

\subsection{One-velocity-component fit}\label{sec:one-component-fit}

As an initial step, we employed a straightforward approach by fitting
a 1-velocity-component across the entire spectral cube (see 
Fig.~\ref{fig:specwellfited1comp}), entering the ``FFP guesses'' and 
``Limits'' listed in Table~\ref{tab:guesses_table}. Since our goal is 
to achieve the best possible fit, we utilize the average of the 
parameters returned throughout the cube from the FFP to refine the 
input guesses, which are then applied in a new fitting process. New 
guesses are then estimated from these new fitting process. This is 
repeated until no variations are observed in the averages obtained
for each parameter compared to previous fits. We define these 
parameter averages as ``Final guesses'' 
(see Table~\ref{tab:guesses_table}). In this way, we provide 
PySpecKit values that are more representative of the data in order 
to produce the final model with a 1-velocity-component fit.

The FFP reveals an issue produced by spectra with $\tau\,<\,1$,
which generates misleading estimations of T$_{\rm ex}$, whose values lie
between 80~K\,$\lesssim$\,T$_{\rm ex}\,<\,10^4$~K. This issue also occurs in
the 2-velocity-components fitting. This problem arises due to the 
intrinsic degeneracy when fitting a single transition, which makes it 
impossible to measure both T$_{\rm ex}$ and $\tau$ from the same transition when $\tau~<~1$.  
Since we aimed to preserve most of the data and use reliable
values for each parameter, we address this issue by refitting the
spectra with $\tau\,<\,1$ and assigning them a constant T$_{\rm ex}$ value
\citep[e.g.,][]{Caselli1, Caselli2}.

Thus, spectra with $\tau\,<\,1$ are selected and separated from the
cube to be re-fitted by a 1-velocity-component with a constant
T$_{\rm ex}$ value, whose value comes from averaging T$_{\rm ex}$ from spectra
with $\tau\,>\,1$. The re-fitted spectra use the ``Final guesses'' listed
in Table~\ref{tab:guesses_table}. However, the T$_{\rm ex}$ limits are
now 2.73~K to 12.4~K.  At the end of this process, we obtain two
separate spectral models and parameters from the spectra with $\tau \geq
1$, and from the spectra with $\tau\,<\,1$ for the 1-velocity-component
fit, which correspond to the 64\% and 36\% of the spectra, respectively. As a final step in the fitting process, we merge these two
spectral models and parameters, creating a final spectral model
entirely fitted by the 1-velocity-component, with its corresponding
parameters and associated errors.

Although this does not occur in every case, some spectra with $\tau~\gg~1$ 
($\tau$~$\gtrsim~40$) exhibit  a ``flat-top'' effect at the intensity peak 
of the hyperfine structures in the model spectra. These spectra have
errors in $\tau$ exceeding 100\% and are excluded from the analysis. 
However, this occurs in less than 1\% of the spectra, where the 
median value of $\tau\,\sim~3$. The same issue arises in the 
2-velocity-component fits with similar frequency.

\subsection{Two-velocity-component fit}\label{sec:two-component-fit}

\begin{figure}[h]
\centering
\includegraphics[width = \columnwidth]{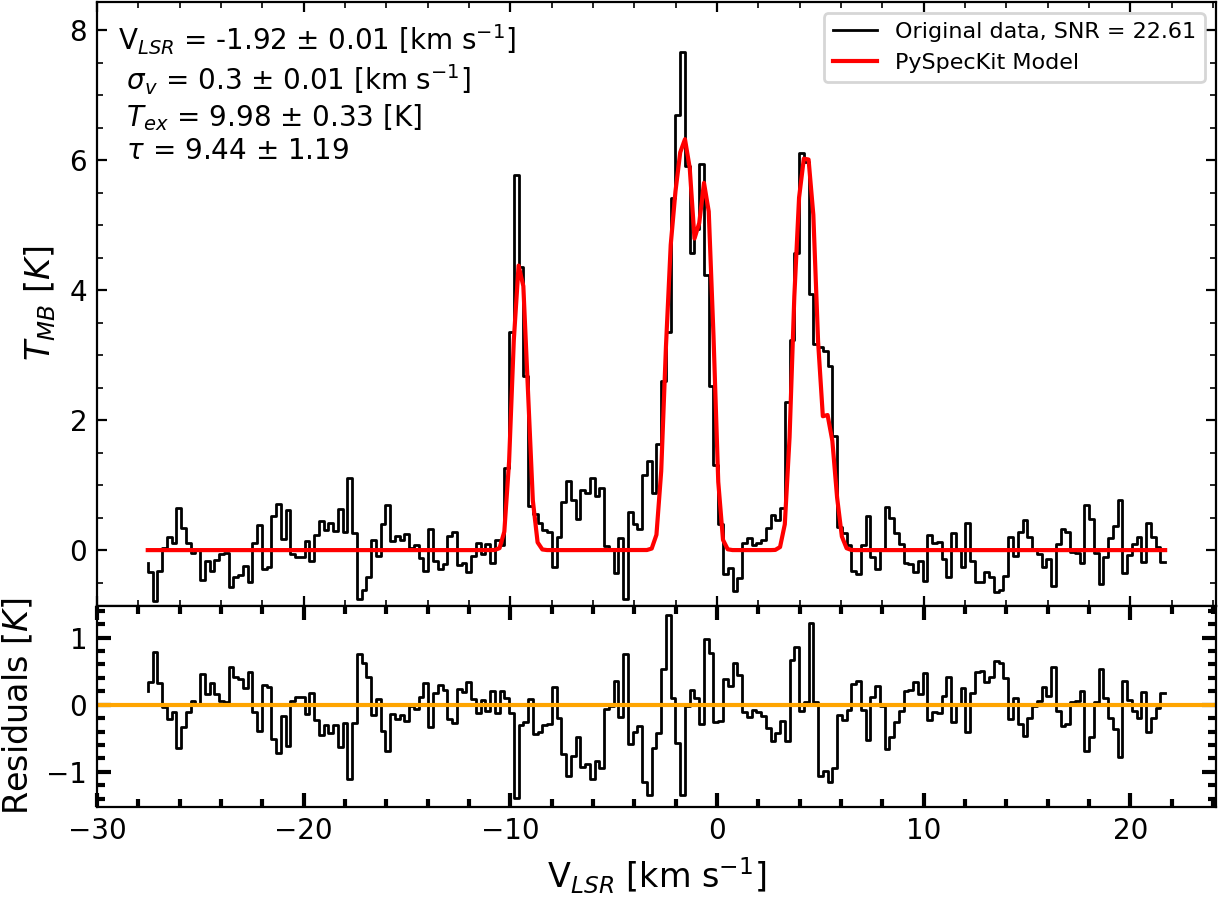}
\includegraphics[width = \columnwidth]{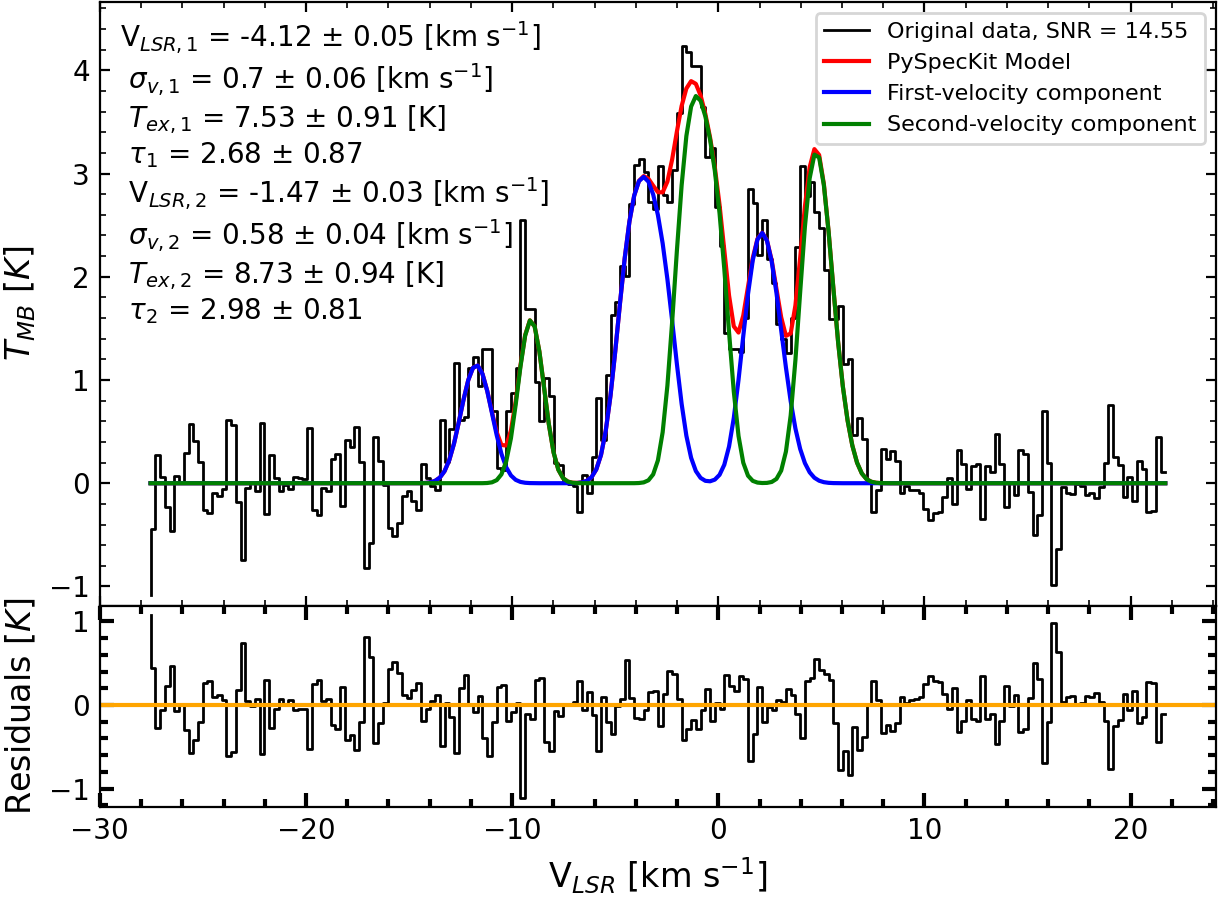}
\caption{Example of 1-velocity (top) and 2-velocity (bottom) component fits
to two different spectra; the former (later) corresponds to a spectrum
extracted from the blue (orange) areas in Fig. 5. In both panels: the
data are represented on the top (solid black histograms) with their
corresponding fits (colored curves), and the residuals are shown on
the bottom, here the orange line refers to the nil value. The
S/N of each spectrum, as well as the best-fit parameter values, are
listed in each panel.  In the bottom panel, the red curve represents
the sum over the two velocity components.
}
\vspace{-0.2cm}
\label{fig:specwellfited1comp}
\label{fig:specwellfited2comp}
\end{figure}

Upon examining the spectral cube, we notice the presence of two distinct 
velocity components in several spectra. Consequently, a 
1-velocity-component fit is not sufficient to describe the global kinematics of
the data. This forces us to fit 2-velocity-components in the spectral 
cube (see Fig.~\ref{fig:specwellfited2comp}), entering the 
``FFP guesses'' and ``Limits'' for each velocity component, listed in 
Table~\ref{tab:guesses_table}. In a similar way as we made in 
\S~\ref{sec:one-component-fit}, and in order to get the best fit, we 
derive new guesses measuring the average of the values returned for each 
parameter of each component throughout the cube, until we obtain the 
``Final guesses'' (see Table~\ref{tab:guesses_table}), which are used to 
generate the final model with a 2-velocity-components fit.

As we mentioned in \S~\ref{sec:one-component-fit}, spectra with $\tau\,
<\,1$ produce bad estimations of the T$_{\rm ex}$. Therefore, spectra with
$\tau\,<\,1$ are selected and separated to be re-fitted. However,
because we are fitting two-velocity-components, it is necessary to
separate the spectra into three different conditions (see
Table~\ref{tab:new_tex_limits}) to assign them a constant T$_{\rm ex}$
value. These constant values come from averaging T$_{\rm ex}$ of spectra
with $\tau\,\geq\,1$ of each component. The re-fitted spectra will use the
same ``Final guesses'' listed in Table~\ref{tab:guesses_table}, while
the T$_{\rm ex}$ limits depend on the the opacity values (see
Table~\ref{tab:new_tex_limits}).

\begin{table}[h]
		\caption{Conditions and T$_{\rm ex}$ limits for the re-fit of 2-velocity-components.}
\vspace{-0.4cm}		
\begin{center}
	\begin{tabular}{cccc}
	\hline\hline 
	\\[-3mm]	

		Number & Condition & T$_{ex,1}$ [K] & T$_{ex,2}$ [K] \\
	\hline
	\\[-3mm]
		
		1 & $\tau_{1} < 1$ \& $\tau_{2} \geq 1$ & (2.73, 9.22) & (2.73, 100) \\

		2 & $\tau_{1} \geq 1$ \& $\tau_{2} < 1$ & (2.73, 100) & (2.73, 9.0) \\ 
		
		3 & $\tau_{1} < 1$ \& $\tau_{2} < 1$. & (2.73, 9.22) & (2.73, 9.0) \\
		\hline

	\end{tabular}
	\tablefoot{New T$_{\rm ex}$ limits applied under each condition for the re-fit of the 2-velocity-components, based on $\tau$ values, in order to avoid poor T$_{\rm ex}$ estimations.}
\end{center}
\vspace{-0.3cm}
\label{tab:new_tex_limits}
\end{table}

The refitting process produces new $\tau$ estimations for spectra with $\tau$~$\approx$~1 where the value can drop below 1. Specifically, this occurs in 5\% of the total spectra when applying conditions 1 and 2 (see Table~\ref{tab:new_tex_limits}). As we already 
mentioned, these $\tau$ values generate bad estimations of 
T$_{\rm ex}$. It is necessary to 
select and separate the re-fitted model spectra under two new 
conditions:
\begin{enumerate}
\item[a.] Model returned from the condition number 1 with $\tau_{2}\,<\,1$.
\item[b.] Model returned from the condition number 2 with $\tau_{1}\,<\,1$.
\end{enumerate}

These spectra are re-fitted again, using the ``Final guesses'' and
``Limits'' of the Table~\ref{tab:guesses_table}. However, the T$_{\rm ex}$
limits are now T$_{\rm ex}$~=~(2.73~K, 9.22~K) for the condition ``a'' and
T$_{\rm ex}$~=~(2.73~K, 9.0~K) for the condition ``b''.  The separation of
the cube into different conditions to re-fit the spectra with $\tau\,<\,
1$ provides us with the security that all spectra have greater freedom
in order to estimate $\tau$ values and obtain good or more reliable
T$_{\rm ex}$ values in each spectrum.

At the end of this process, we obtain six separate model spectra and
parameter cubes fitted by 2-velocity-components. One of these 
correspond to the model spectra with $\tau~\geq~1$, which account for 38\% of the total spectra. The remaining five models spectra are derived from refitting process: conditions 1, 2, and 3 (see Table~\ref{tab:new_tex_limits}) represent 30\%, 15\%, and 11\% of the final spectra, respectively, while the condition ``a'' and ``b'' (see above) contribute 3\% and 2\% to the final spectra, respectively. 
As a final step in this process, we 
merge these six spectral models and parameter cubes, creating a final 
spectral model entirely fitted by 2-velocity-components 
(see Fig.~\ref{fig:momentmaps0}). We refer to each component as the
``First-velocity-component'' and the ``Second-velocity-component'', with
their centroid velocities being the lowest and highest, respectively,
and with their parameters and associated errors 
(see Fig.~\ref{fig:specwellfited2comp}).

\subsection{Best fit and merging models}\label{sec:best-fit-and-final-model}

In \S~\ref{sec:one-component-fit} and
\S~\ref{sec:two-component-fit}, we described the complete fitting process
of the data, fitting the whole cube with 1- and
2-velocity-components. However, our spectral cube shows spectra where 
we can identify just 1-velocity-component (see
Fig.~\ref{fig:specwellfited1comp}), spectra where we can identify
2-velocity-components (see Fig.~\ref{fig:specwellfited2comp}), and
spectra where it is not possible to make a clear identification by
eye. Furthermore, it is necessary to determine which fit (1- or
2-velocity-components) is better for each spectra to simplify our
analysis and to recover reliable kinematic information about the
protocluster. Thus, we devised a method to determine if 1- or
2-velocity-components is appropriate for each spectrum based on 
S/N criteria and the parameters associated with each
spectrum, specifically the line width.  

\begin{figure}[h]
\centering
\includegraphics[width = \columnwidth]{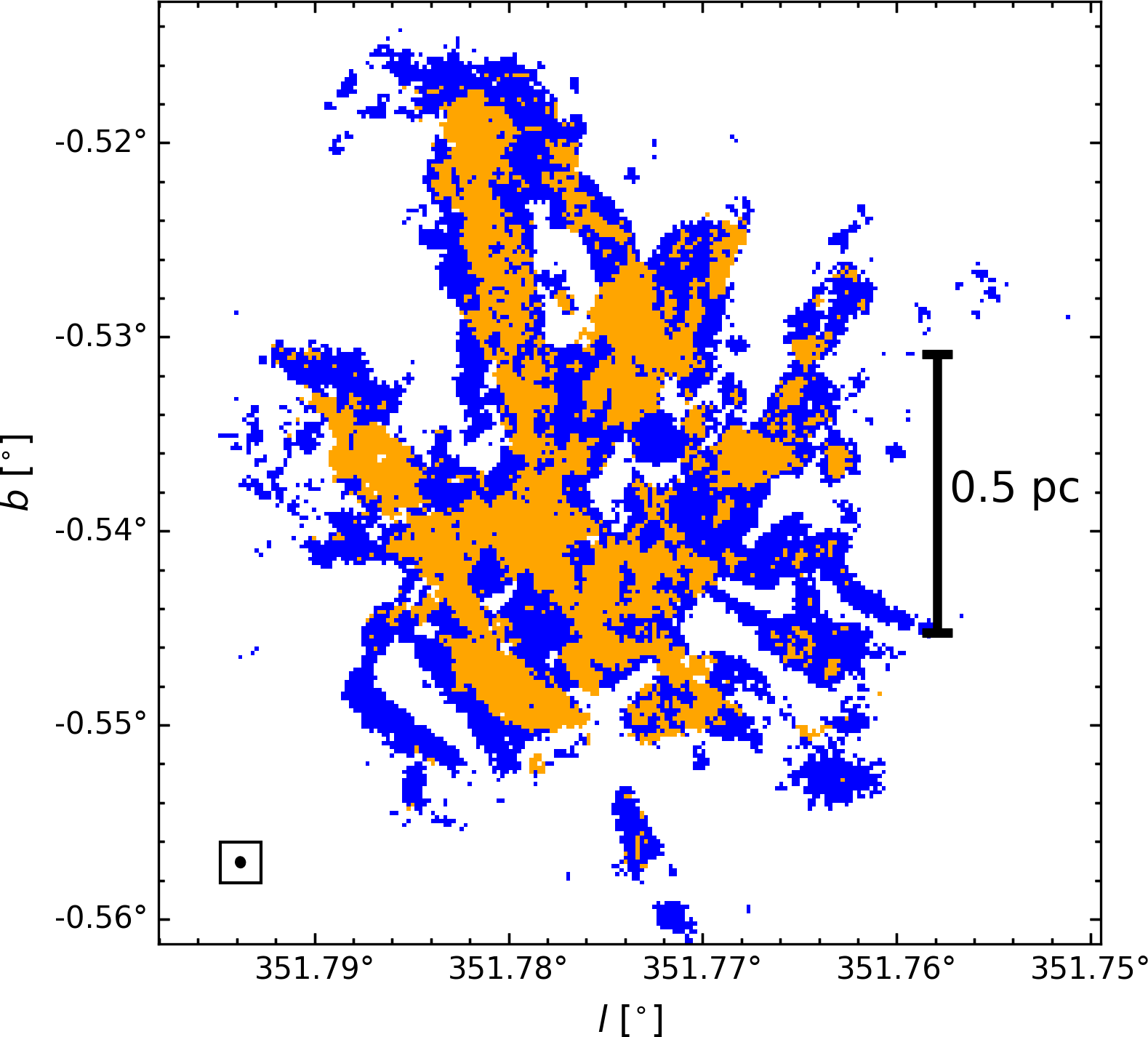}
\caption{Map of the number of velocity components in the \nhp~(1-0) 
  spectral fits in the. The blue (orange) pixels represent 59\% 
  (41\%) of the spectra where we adopt a 1-velocity-component fit 
  (2-velocity-component fit), see 
  \S~\ref{sec:best-fit-and-final-model}. Most of the orange pixels are
  located in central regions, where we observe the highest S/N and
  integrated intensity values, similar to the results for G353.41 
  \citep{rodrigo24}. The blue areas are located preferentially near the
  edges. The ellipse in the bottom-left corner represents the beam
  size of the \nhp data.}
\vspace{-0.3cm}
\label{fig:map_components}
\end{figure}

In order to determine if the spectrum is better fitted by 1- or 2-velocity-components we:
\begin{enumerate}
\item Check the $\sigma_v$ value in the spectral model fitted
  by 2-velocity-components. From  inspection of the spectral
  model fitted by 2-velocity-components, we find some 
  ``false component''
  spectra without emission whose 
  $\sigma_v$ is equal to 0 (see 
  Fig.~\ref{fig:specdef1comp_1}). This implies that
  all spectra with these characteristics are
  better characterized by 1-velocity-component fit (see
  Fig.~\ref{fig:specdef1comp_1}).

\begin{figure}[H]
\centering
\includegraphics[width = \columnwidth]{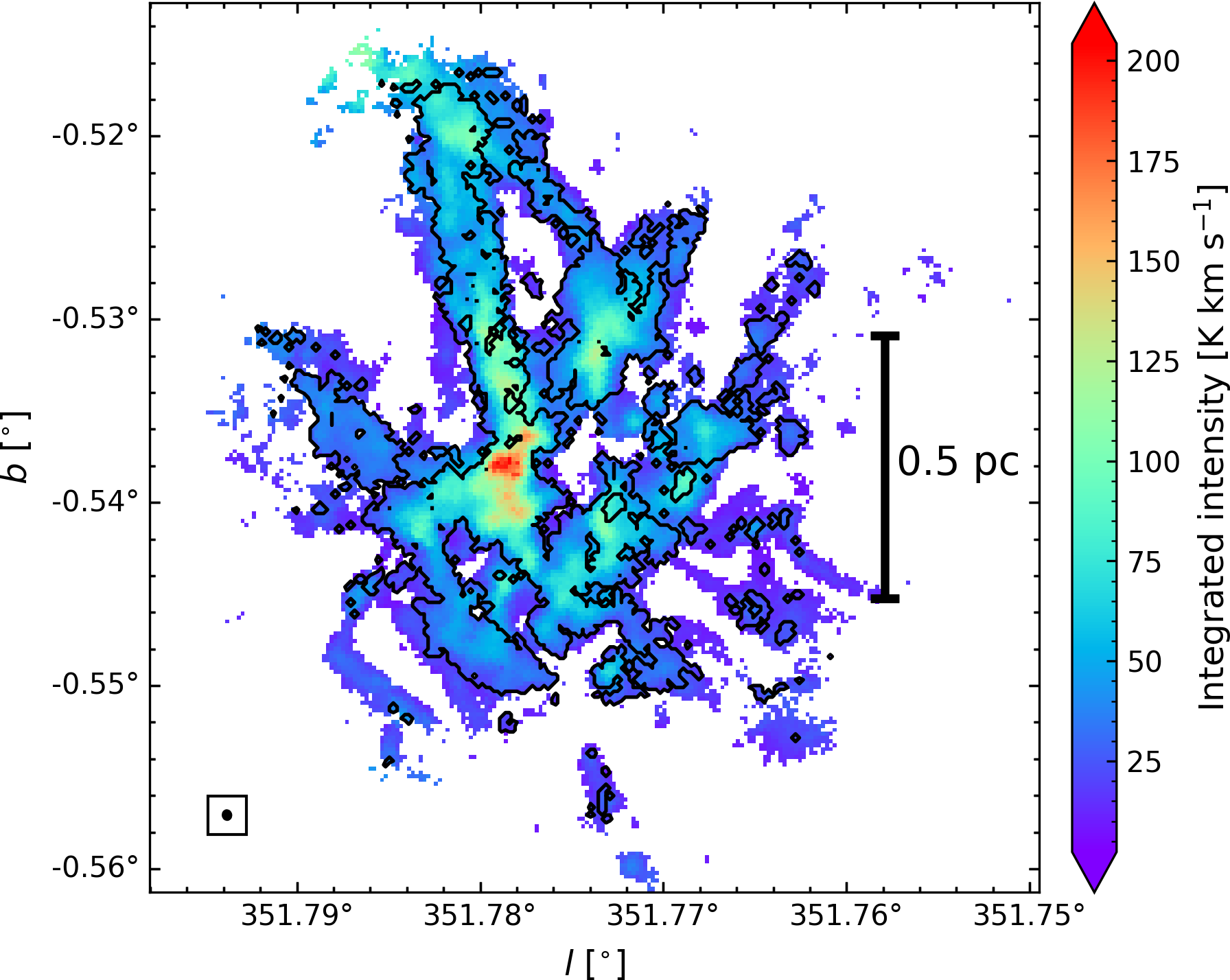}
\vspace{0.2cm}
\includegraphics[width = \columnwidth]{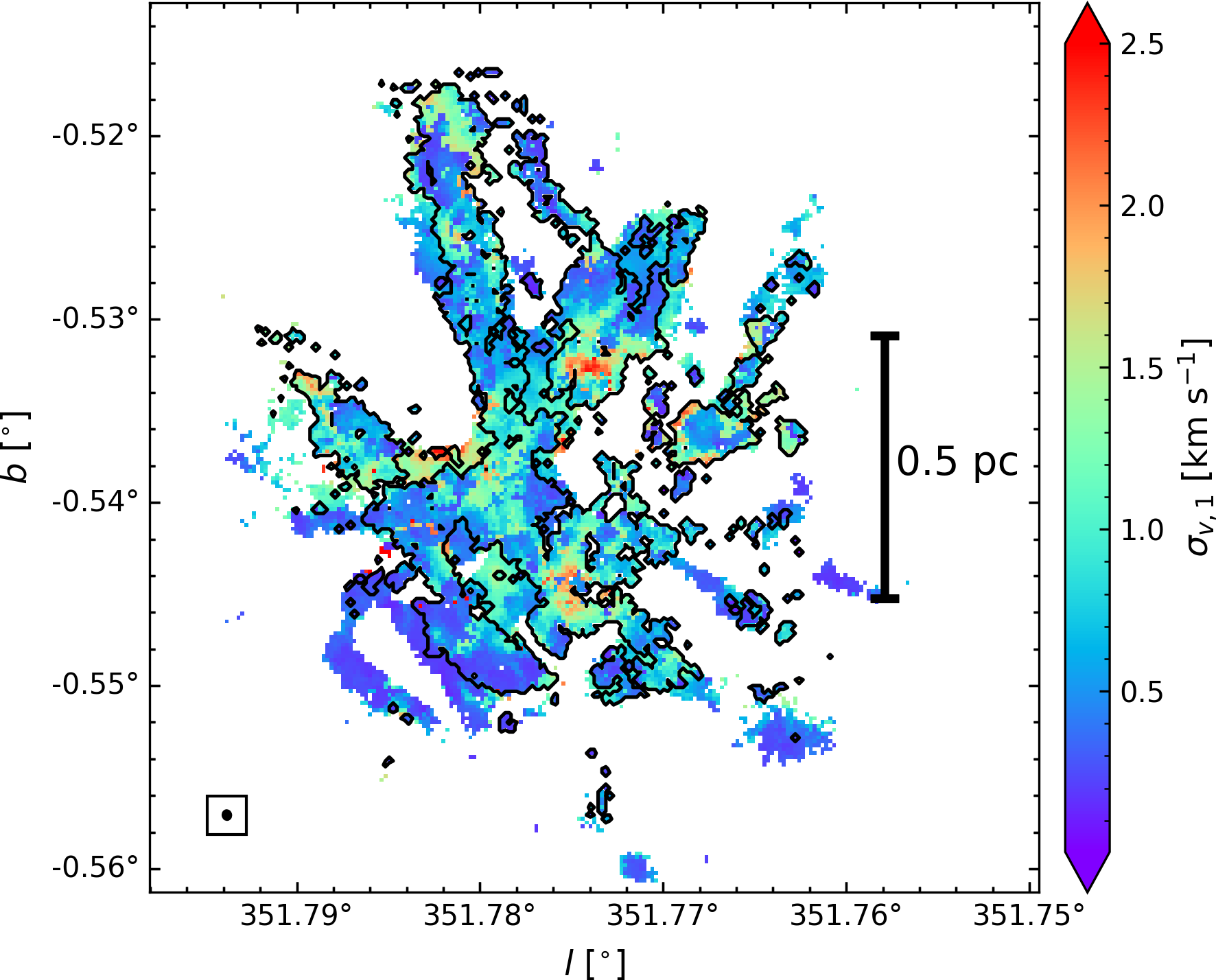}
\includegraphics[width = \columnwidth]{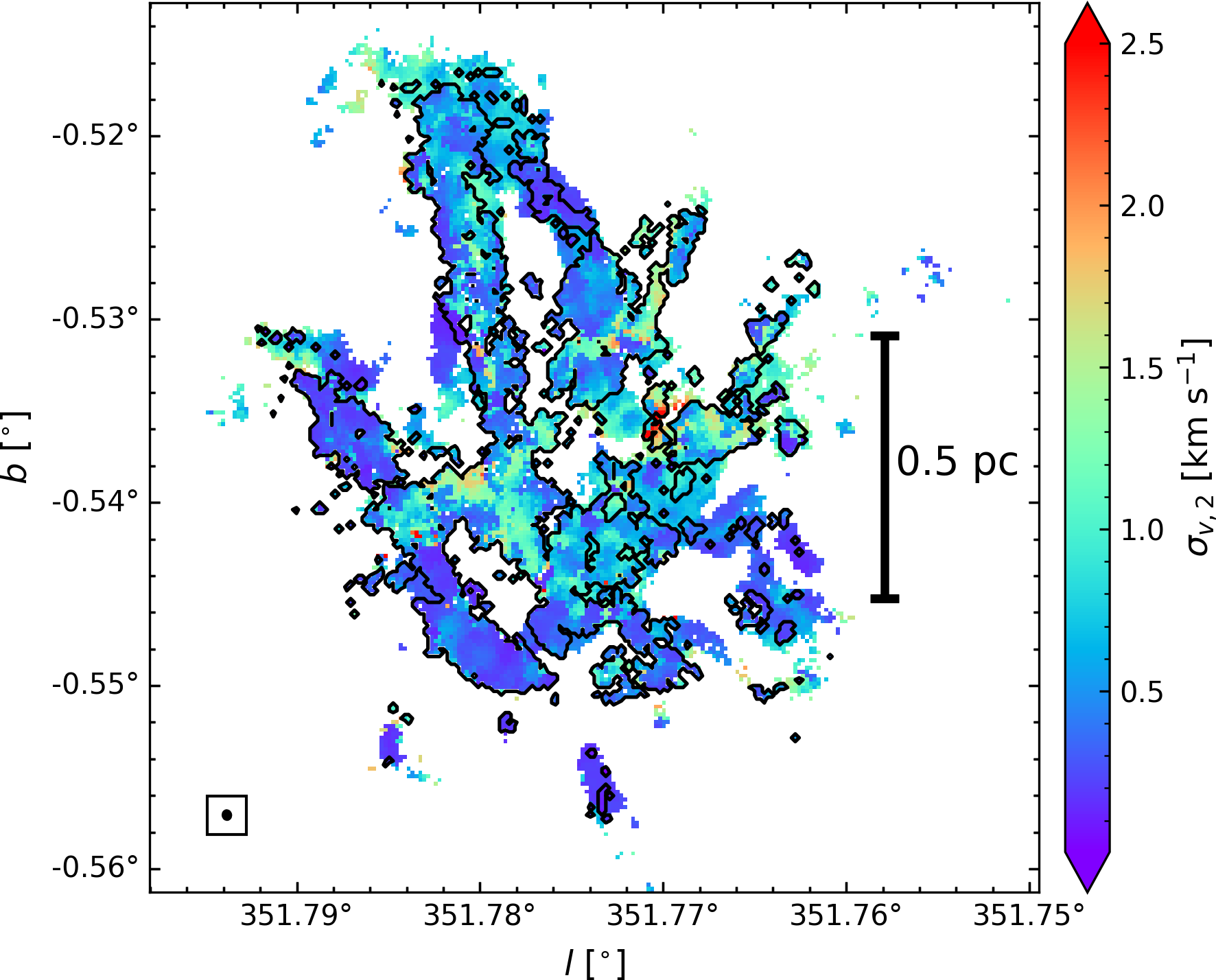}
\caption{ Top: Integrated intensity map from the final spectral model
  composed of both the 1- and 2-velocity-components. Middle: Line width map of 
  the Blue-velocity component. Bottom: Line width map
  of the Red-velocity component. The spectra inside the
  black contour are fit with 2-velocity-components. The ellipse in the
  bottom-left corner represents the beam size of the \nhp data.}
\label{fig:momentmaps0}
\end{figure}

\item Measure the S/N of the velocity component with the lower intensity 
  in the 2-velocity-component fit. Deeper inspection
  shows that some 
  additional velocity components have intensities similar to noise values.
  So, if the peak intensity is lower than five times the noise measured in that
  spectrum from the data-cube, the 1-velocity-component fit is
  adopted. Thus, we ensure the reliability of the additional velocity
  component (see Fig.~\ref{fig:specdef1comp_2}).
\end{enumerate}

Additionally, all fitted spectra with errors higher than five times the median error estimated for the centroid velocity ($\sim$~0.2~$\kms$) are removed for the kinematic analysis (see \S~\ref{sec:kinematics-analysis}). For the estimation of column density (see \S~\ref{sec:column-density-and-masses}), spectra with errors exceeding 10\% for the line width, 30\% for T$_{\rm ex}$, and 30\% for $\tau$ are also excluded. These thresholds were estimated considering twice the median error measured for each parameter. Furthermore, we find that 95\% of the spectra have error within the following ranges: [0.05 K, 1.78 K] for T$_{\rm ex}$, [0.4, 9.7] for $\tau$, [0.007 $\kms$, 0.26 $\kms$] for the centroid velocity, and [0.1 $\kms$, 0.17 $\kms$] for the line width.

After applying these criteria, we obtain spectra well-fitted by
1- and 2-velocity-components (see
Fig.~\ref{fig:specwellfited2comp}). All these spectra are merged in
order to create a final spectral model composed of both 1- and
2-velocity-components (see Fig.~\ref{fig:map_components}). Most of the 
spectra fitted by 2-velocity-components are located in internal regions toward highest integrated intensities (see
Fig.~\ref{fig:momentmaps0}).

\subsection{Blue and Red velocity component}\label{sec:bluest-and-reddest-velocity-component}

From the four parameters returned by PySpecKit of the \nhp~(1-0) spectra, the
centroid velocity is the one that has the lowest
uncertainties with median values of 0.034~$\kms$ (cases with higher errors than five times the median value are excluded, see \S~\ref{sec:best-fit-and-final-model}), which gives 
us information about the radial velocity distributions of the regions where we
observe line emission. The general centroid velocity distributions that we
derive are plotted in Fig.~\ref{fig:histogram_veldistribution}, which
we discuss in detail below. 

\begin{figure}[H]
\centering
\includegraphics[width = 0.98\columnwidth]{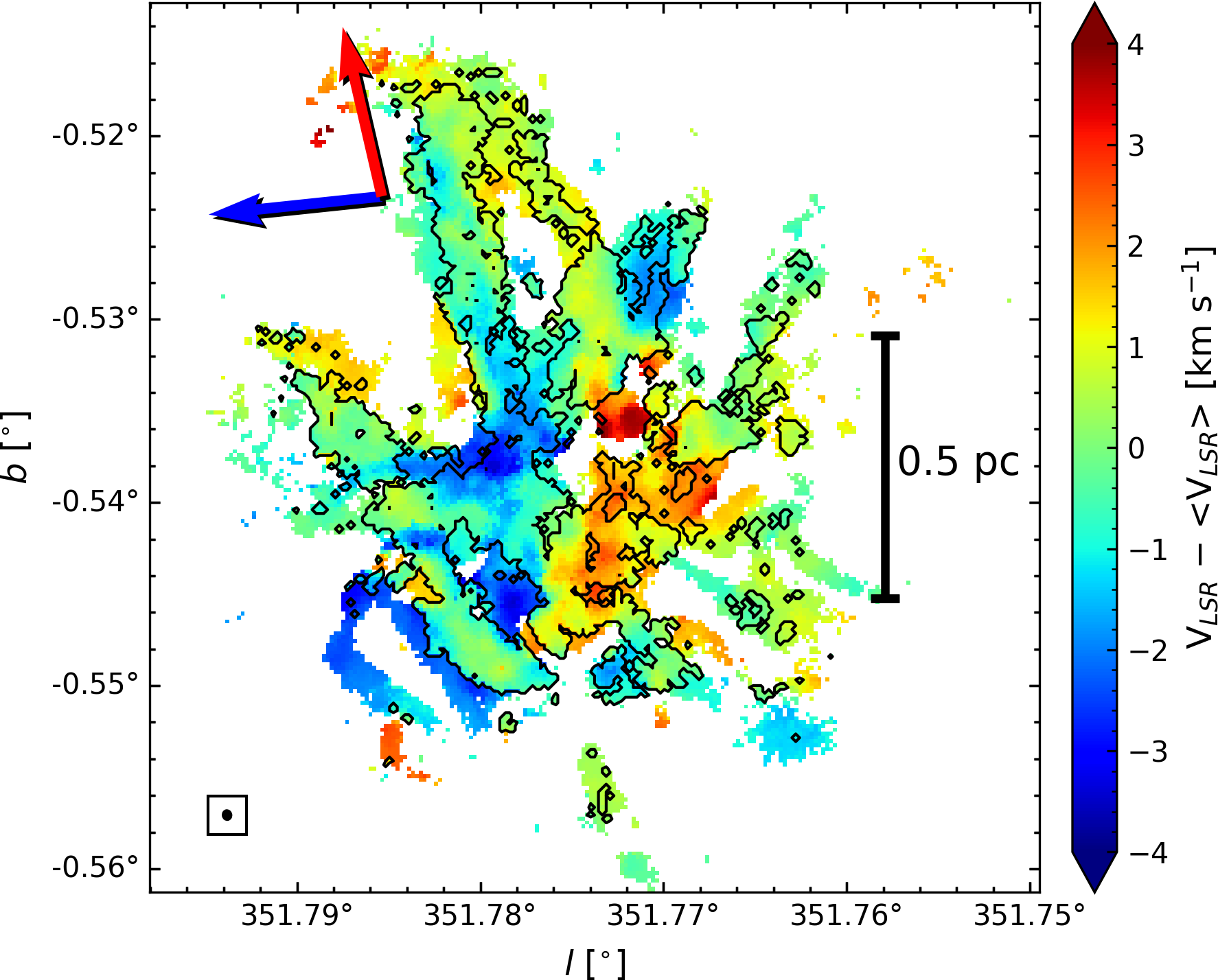}
\vspace{0.02cm}
\includegraphics[width = 0.98\columnwidth]{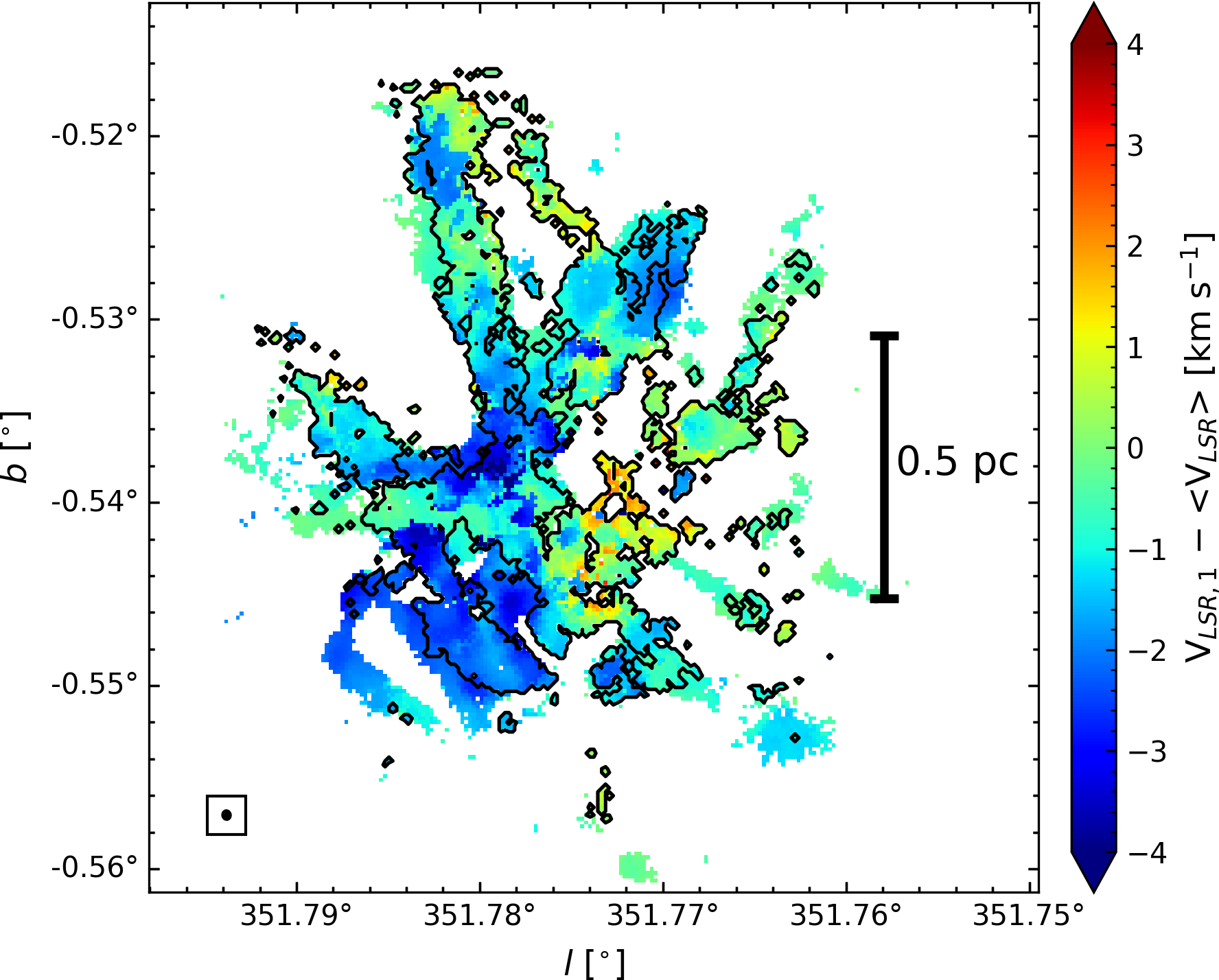}
\includegraphics[width = 0.98\columnwidth]{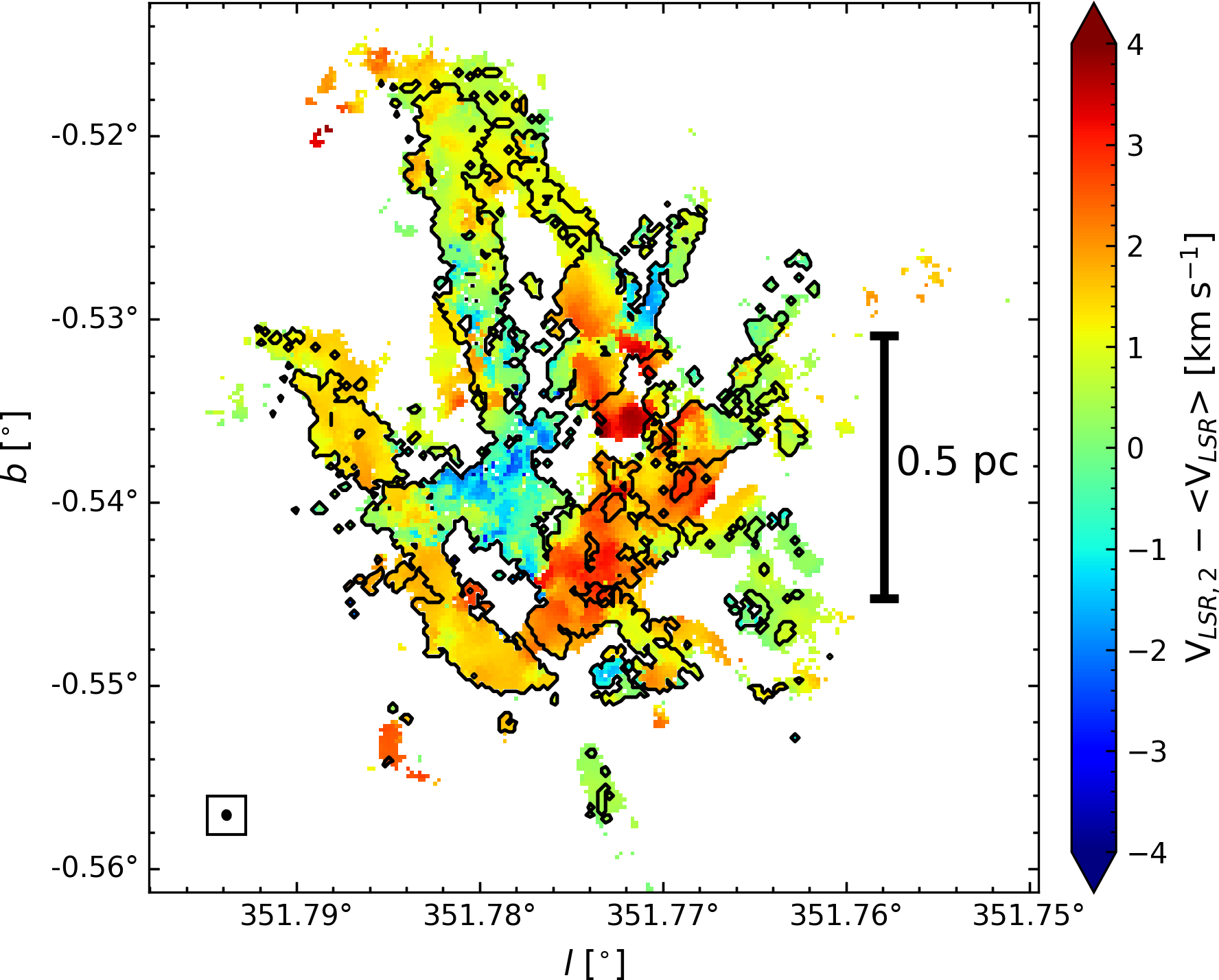}
\caption{Top: Mean velocity map from the final spectral model
  composed of both the 1- and 2-velocity-components. The red arrow display
  the direction of the Mother Filament. The blue arrow indicates
  the direction of the large-scale velocity gradient measured from the centroid velocities. 
  Middle: Centroid velocity 
  map of the Blue-velocity component. Bottom: Centroid velocity map of the 
  Red-velocity component. The spectra inside the black contour are fit 
  with 2-velocity-components. The ellipse in the bottom-left corner represents the beam
  size of the \nhp data.}
\label{fig:velocity_map}
\end{figure}

The mean velocity map from the model in the top
panel of Fig.~\ref{fig:velocity_map}, shows various interesting features. One of these includes indications of a 
large-scale velocity gradient (VG) that is almost perpendicular ($\sim$\,83\,$^{\circ}$) to the direction in which the molecular cloud filament that hosts the
protocluster extends (which we call the Mother Filament, see
Fig.~\ref{fig:rgbimage}).

On the other hand, we observe jumps 
between the velocities of the spectra with 1- and 2-velocity-components 
that reaches velocities of up to $\sim$ 4~$\kms$ (see top panel of 
Fig.~\ref{fig:velocity_map}).  This is an effect produced by averaging 
the two centroid velocities in spectra with 2-velocity-components, 
so is an artificial velocity averaging effect rooted in the complex motions in G351.77.
Indeed, in the bottom two panels of Fig.~\ref{fig:velocity_map}, we
separate the measured velocities and assign them to either a Blue-
or Red-velocity component. To make this somewhat arbitrary but
necessary velocity separation, we define the cut-off velocity as the
midpoint based on the velocity distributions of the
First-velocity-component and the Second-velocity-component (see
Fig.~\ref{fig:histogram_veldistribution}). We achieve this by
applying the following simple relation:

\begin{equation}
<\text{V}_{\rm LSR,1}> + \hspace{0.1cm} X \times \sigma_{\rm v_{\rm LSR,1}} = \hspace{0.1cm} <\text{V}_{\rm LSR,2}> - \hspace{0.1cm} X \times \sigma_{\rm v_{\rm LSR,2}}\text{,}
\label{eq:vel_limit_eq}
\end{equation}

\noindent where $<\text{V}_{\rm LSR,1}>$ and $<\text{V}_{\rm LSR,2}>$ are
the averaged centroid velocities of the First- and
Second-velocity-component, respectively, $\sigma_{\rm v_{\rm LSR,1}}$ and
$\sigma_{\rm v_{\rm LSR,2}}$ are the standard deviations of the centroid
velocities in the First- and Second-velocity-component, respectively,
and $X$ represents the cut-off value used as the middle \linebreak
\noindent point, normalized by
the standard deviations of the centroid velocity distributions.

Applying Eq.~\ref{eq:vel_limit_eq} we obtain a $X=0.025$. 
This value allows us to separate the centroid velocity
distribution of the 1-velocity-component fit into two parts and 
combine the measurements with the First- and the
Second-velocity-component. We define ``Blue-velocity component'' as the merger
of all the spectra from the First-velocity-component and the spectra from the
1-velocity-component fit with centroid velocity $< 0.025~\kms$, see
middle panel of Fig.~\ref{fig:velocity_map}. Similarly, the ``Red-velocity
component'' is the merger of all the spectra from the Second-velocity-component and the
spectra from the 1-velocity-component fit with centroid velocity 
$>\,0.025~\kms$ (see bottom panel of Fig.~\ref{fig:velocity_map}).

\section{Column Density, mass, and \nhp relative abundance}\label{sec:column-density-and-masses}

The spectral line fitting of \nhp~(1-0) shown in
\S~\ref{sec:line-fitting-process} provides relevant information about the \nhp~(1-0) spectra, specifically excitation temperature (T$_{\rm ex}$),
optical depth ($\tau$), and line width ($\sigma_v$). These parameters enable the estimation of column density maps of \nhp, 
N(\nhp), calculated by Eq.~\ref{eq:column-density} (see Fig.~\ref{fig:columndensity}),
and the mass of \nhp in the protocluster, M(\nhp), 
calculated by Eq.~\ref{eq:n2hp-mass}, which can be
determined for both the Blue- and the Red-velocity component
(see \S~\ref{sec:n2h+_column_density}). Additionally, the column density map of \hdos, N(\hdos), 
provided by \citet{Pierre}, shown in Fig.~\ref{fig:h2map}, gives 
us the opportunity to estimate the \nhp relative
abundances, X(\nhp), 
in the protocluster (see Fig.~\ref{fig:xfactor})
covering ten out of eighteen
1.3~mm dust continuum cores (hereafter referred to as ``dense cores'') from the core catalog of \citet{Fabien}, see Table~\ref{tab:core_parameters}.

\begin{table*}[b]
		\caption{Physical parameters of the dense cores in G351.77 protocluster.}
\vspace{-0.4cm}
\begin{center}
	\scalebox{0.88}{
	\begin{tabular}{cccccccccc}
		\hline\hline
		\\[-3mm]

		Core  &  Int $\times$ $10^3$ & V$_{\text{LSR}} -$ <V$_{\text{LSR}}$>$^1$ & $\sigma_v$ $^1$ & T$_{\rm ex}$ $^1$ & $\tau$ $^1$ & N(\nhp) & X(\nhp)$^1$ & \ntot $^2$ & \mtot $^2$ \\
		ID  &  [K $\kms$] & [$\kms$] & [$\kms$] & [K] &  & $\times$ $10^{14}$ [cm$^{-2}$] & $\times$ $10^{-10}$ & $\times$ $10^{24}$ [cm$^{-2}$] & [\msun] \\
		\hline
		\\[-3mm]
		
		L10 &  0.358 & -1.41 $\pm$ 0.03 & 0.92 $\pm$ 0.02 & 11.2 $\pm$ 0.7 & 5.44 $\pm$ 0.5 &  2.44 $\pm$ 0.26 & 2.91 $\pm$ 0.84 & 0.65 $\pm$ 0.08 & 0.56 $\pm$ 0.07  \\

		L11 &  0.205 &  2.04 $\pm$ 0.03 & 1.51 $\pm$ 0.03 & 12.3 $\pm$ 1.1 &  1.97 $\pm$ 0.3 &  0.88 $\pm$ 0.14 & 0.34 $\pm$ 0.09 & 0.23 $\pm$ 0.02 & 0.20 $\pm$ 0.02 \\ 
		
		L12 &  0.176 & -0.48 $\pm$ 0.06 & 1.08 $\pm$ 0.04 & 11.3 $\pm$ 1.4 &  1.51 $\pm$ 0.2 &  0.72 $\pm$ 0.18 & 0.45 $\pm$ 0.13 & 0.19 $\pm$ 0.02 & 0.16 $\pm$ 0.02 \\ 
		
		L13 &  0.254 & -1.84 $\pm$ 0.10 & 1.19 $\pm$ 0.06 & 10.3 $\pm$ 1.2 &  2.64 $\pm$ 0.2 &  1.24 $\pm$ 0.26 & 1.15 $\pm$ 0.33 & 0.33 $\pm$ 0.03 & 0.28 $\pm$ 0.02 \\
		
		L14 &  0.042 & -0.55 $\pm$ 0.04 & 0.70 $\pm$ 0.05 & 12.4 $\pm$ 1.1 &  0.71 $\pm$ 0.1 &  0.15 $\pm$ 0.04 & 0.06 $\pm$ 0.01 & 0.04 $\pm$ 0.01 & 0.03 $\pm$ 0.01\\
		
		L16 &  0.370 & -0.72 $\pm$ 0.01 & 0.56 $\pm$ 0.01 & 10.0 $\pm$ 0.2 & 22.8 $\pm$ 1.1 &  2.66 $\pm$ 0.25 & 2.90 $\pm$ 0.84 & 0.70 $\pm$ 0.08 & 0.61 $\pm$ 0.07 \\
		
		L18 &  0.409 & -1.74 $\pm$ 0.01 & 1.01 $\pm$ 0.01 & 17.8 $\pm$ 0.3 &  4.74 $\pm$ 0.2 &  2.09 $\pm$ 0.08 & 0.83 $\pm$ 0.24 & 0.55 $\pm$ 0.06 & 0.47 $\pm$ 0.05 \\
		
		L19 &  0.940 & -2.87 $\pm$ 0.01 & 1.18 $\pm$ 0.01 & 17.5 $\pm$ 0.2 &  5.21 $\pm$ 0.2 &  4.80 $\pm$ 0.22 & 1.44 $\pm$ 0.42 & 1.27 $\pm$ 0.15 & 1.10 $\pm$ 0.13 \\
		
		L20 &  0.396 & -2.36 $\pm$ 0.02 & 0.82 $\pm$ 0.01 & 9.1 $\pm$ 0.9 &  8.08 $\pm$ 0.4 &  2.78 $\pm$ 0.51 & 0.24 $\pm$ 0.07 & 0.74 $\pm$ 0.09 & 0.64 $\pm$ 0.08 \\
		
		L23 &  0.785 & -1.61 $\pm$ 0.03 & 0.74 $\pm$ 0.02 & 5.8 $\pm$ 1.4 &  2.61 $\pm$ 0.2 &  2.92 $\pm$ 0.36 & 0.33 $\pm$ 0.09 & 0.77 $\pm$ 0.01 & 0.67 $\pm$ 0.01 \\
		\hline

	\end{tabular}}
	\tablefoot{The number in the first column indicate the cores number in 
	Table E.11 in \citet{Fabien}, and in Table F.9 in \citet{Nichol}. 
	The ``L'' is used to distinguish between the PV features associated with 
	the dense cores from \citet{Fabien} and the PV features of 
	possible cores under the 1.3~mm band detection limit 
	(see \S~\ref{sec:kinematics-analysis-at-large-scales} and Table~\ref{tab:VGcores}). The errors of N(\nhp) and \ntot are 
	estimated based on the distribution of the percentage error in column 
	densities, to prevent underestimations. \\
	$^1$ These quantities represent average measurements inside core regions. \\
	$^2$ Total column density and masses estimated from the N(\nhp) and relative abundance (see \S~\ref{sec:relative-abundances}).} 
\end{center}
\label{tab:core_parameters}
\end{table*}

\subsection{N$_2$H$^+$ column density and  mass}\label{sec:n2h+_column_density}

In order to estimate the column density of \nhp, and given that T$_{\rm ex}$ is not well constrained due to the low optical depth of some spectrum, we use the optically 
thick transitions approximation given by the following expression 
\citep{Caselli2, Elena}:

\begin{equation}
    \text{N(N}_2\text{H}^+) = \frac{4\pi^{3/2}}{\sqrt{\ln(2)}} \frac{\nu^3 \cdot Q \cdot \sigma_v}{c^3 \cdot A_{ul} \cdot g_u} \frac{\tau}{e^{h\nu/k_bT_{ex}} - 1} \cdot e^{E_u/k_bT_{ex}}, 
\label{eq:column-density}
\end{equation}

\begin{figure}[h]
\includegraphics[width = \columnwidth]{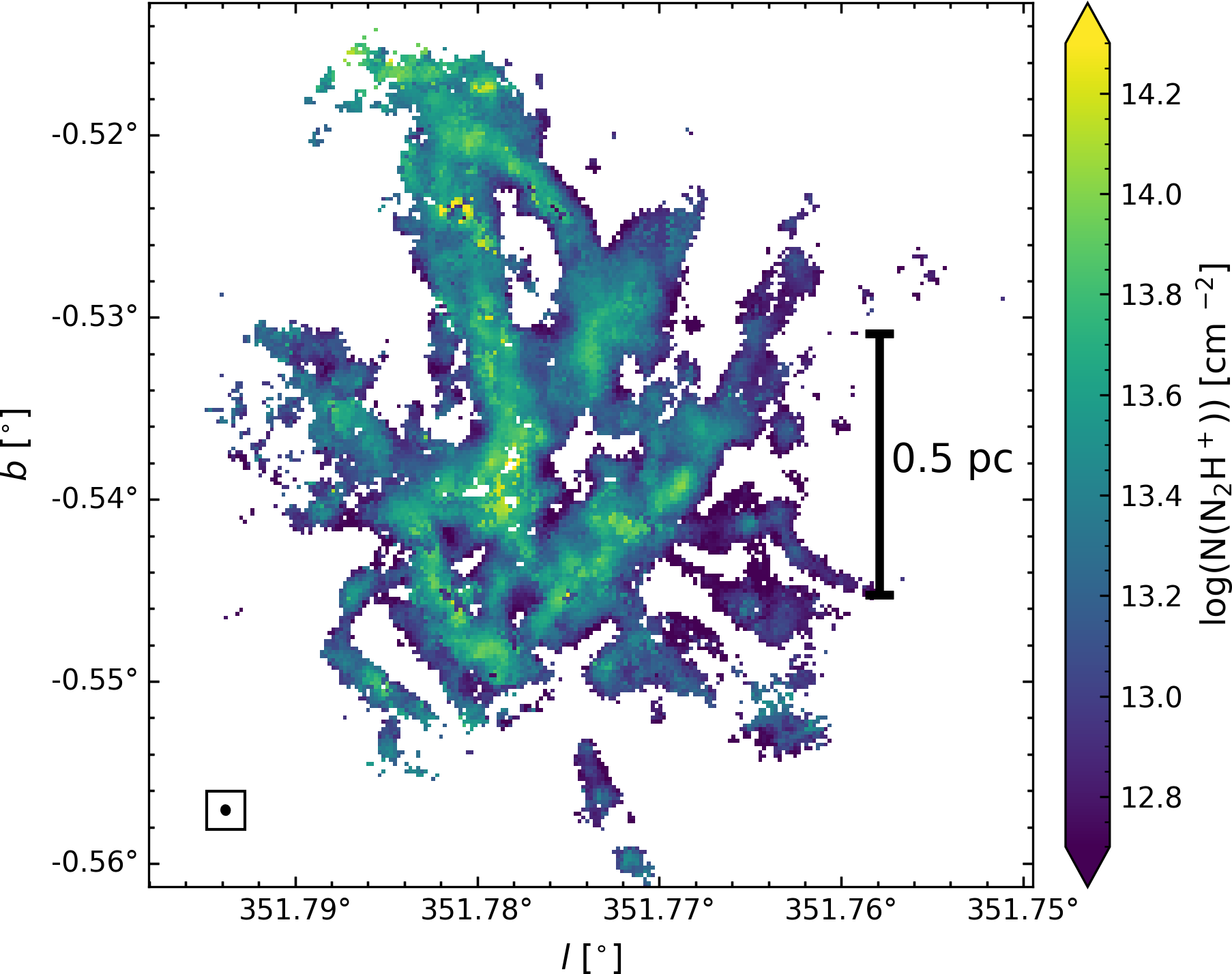}
\caption{\nhp total column density map derived
  from Eq.~\ref{eq:column-density}. The ellipse in the bottom-left
  corner represents the beam size of the \nhp data.}
\label{fig:columndensity}

\end{figure}

\noindent where $\sigma_v$, $\tau$, and $T_{ex}$ correspond to the
parameters returned by our fits, while $\nu$ is the frequency, $Q$ is
the partition function, $c$ is the speed of light, $A_{ul}$ is the
Einstein coefficient, $g_u$ is the statistical weight, $h$ is the
Planck's constant, $k_B$ is the Boltzmann's constant, and $E_u$ is the
energy of the upper limit \citep{Laurent, Jeffrey, Elena}. The error 
of N(\nhp) is estimated making the error propagation over the 
Eq.~\ref{eq:column-density}. We apply
this expression over the Blue- and the Red-velocity component to
obtain the column density of each one, N$_\text{B}$(\nhp) and
N$_\text{R}$(\nhp), respectively. The total column density
N$_\text{T}$(\nhp) is obtained by summing these (see
Fig.~\ref{fig:columndensity} and
Table~\ref{tab:columndensity_table}). Then, we can measure the mass per pixel of
\nhp by applying the following expression:

\begin{equation}
    \text{M(N}_2\text{H}^+) = \text{N(N}_2\text{H}^+) \cdot \text{A}_{\rm pixel} \cdot m_{\text{N}_2\text{H}^+}, 
\label{eq:n2hp-mass}
\end{equation}

\noindent where N(\nhp) is given by the Eq.~\ref{eq:column-density},
A$_{\rm pixel}$ corresponds to the area of the pixel, and
$m_{\text{N}_2\text{H}^+}$ corresponds to the mass of \nhp molecule,
where $m_{\text{N}_2\text{H}^+} = 4.817 \times 10^{-23}$~g. 
Table~\ref{tab:columndensity_table} shows the mass for the Blue-
and the Red-velocity components as well as the total \nhp mass in the
protocluster.

\begin{table}[h]
		\caption{Column densities and masses of \nhp in G351.77 protocluster.}
\vspace{-0.4cm}
\begin{center}
	\begin{tabular}{ccc}
	\hline\hline 
	\\[-3mm]	
		N$_\text{col}$ & <N(\nhp)>  & M(\nhp) \\
		               & $\times$ $10^{13}$ [$\text{cm}^{-2}$] & $\times$ $10^{-6}$ [\msun] \\
		\hline 
		\\[-3mm]
		
		N$_\text{B}$(\nhp)  & 1.49 $\pm$ 0.41 & 1.77 $\pm$ 0.49 \\

		N$_\text{R}$(\nhp)  & 1.79 $\pm$ 0.49 & 2.19 $\pm$ 0.61 \\ 
		
		N$_\text{T}$(\nhp)  & 2.28 $\pm$ 0.56 & 3.96 $\pm$ 0.78 \\
		\hline

	\end{tabular}
	\tablefoot{Averages of \nhp column density estimations and total \nhp masses. N$_\text{B}$, N$_\text{R}$ and N$_\text{T}$ correspond to the Blue-velocity component, the Red-velocity component and N$_\text{B}$ $+$ N$_\text{R}$, respectively. The errors are estimated based on the distribution of the percentage errors in column densities, to prevent underestimations.}
	\vspace{-0.5cm}
\end{center}
\label{tab:columndensity_table}
\end{table}

\subsection{N$_2$H$^+$ relative abundance}\label{sec:relative-abundances}

We expect to estimate a new column density map of \hdos (\ntot) and mass map of 
\hdos (\mtot) for both small 
and large scale structures (see \S~\ref{sec:kinematics-analysis}), 
with at smaller spatial scale (see below) than the N(\hdos) provided by 
\citet{Pierre}. This estimation can not be made with the ALMA-IMF dust 
emission given its small spatial coverage. However, the \nhp emission 
provides us with a better angular resolution and a larger spatial 
coverage. For this is necessary to know the X(\nhp) in the 
protocluster, which is given by the following expression:

\begin{equation}
\text{X(N}_2\text{H}^+) = \frac{\text{N(N}_2\text{H}^+)}{\text{N(H}_2)}.  
\label{eq:relativeabundance_equation}
\end{equation} 

\noindent where the error of the relative abundance is estimated making
the error propagation of the Eq.~\ref{eq:relativeabundance_equation} 
utilizing the error maps of N(\nhp) and N(\hdos) at the same resolution. 

\begin{figure}[h]
\includegraphics[width = \columnwidth]{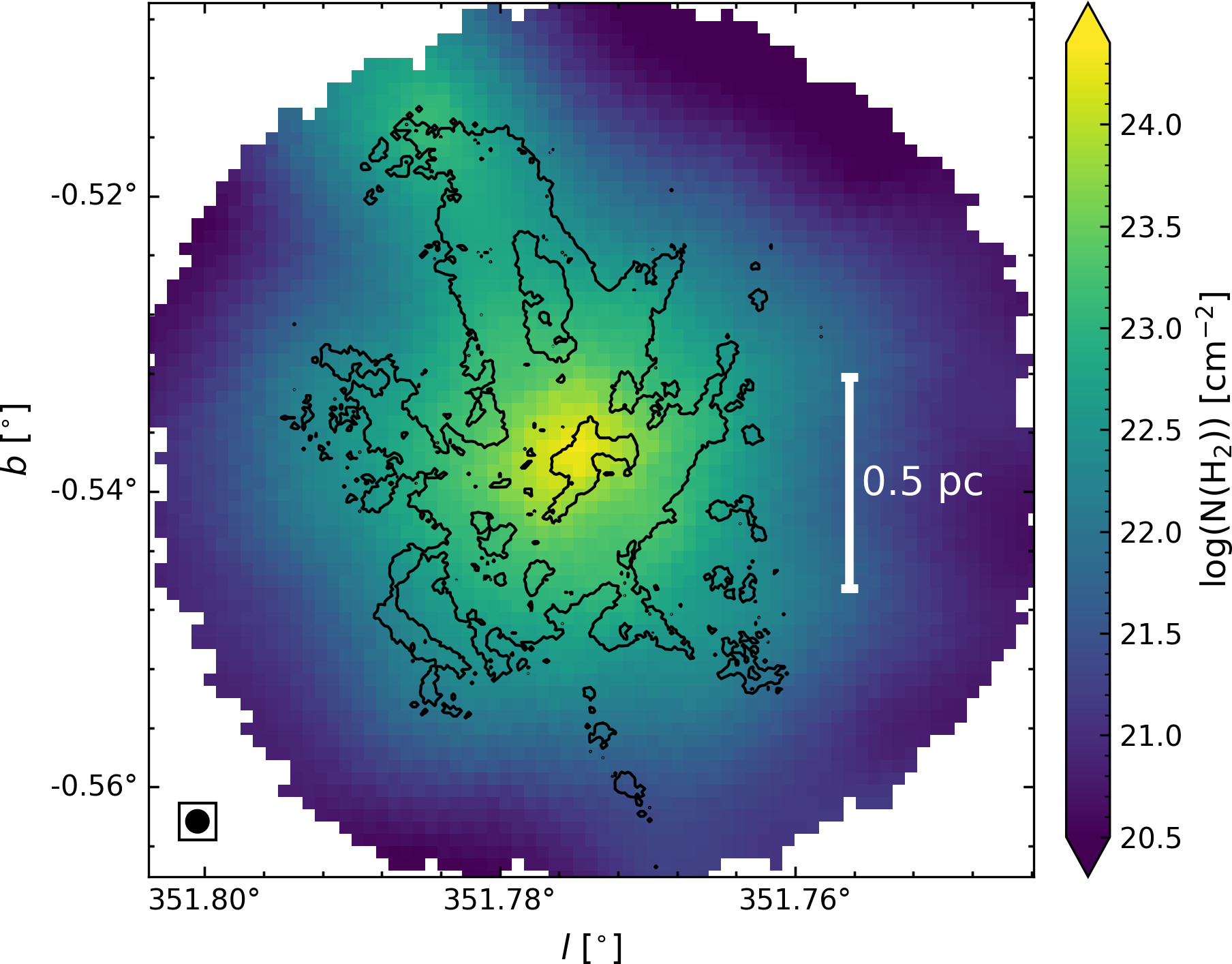}
\caption{\hdos column density map from \citet{Pierre} of the G351.77 protocluster. The black contour represents
  the spectra with N$_\text{T}$(\nhp) $>$ 1 $\times$ 10$^{13}$
  cm$^{-2}$. The circle on the bottom left represents the beam size of
  6\arcsec $\times$ 6\arcsec.}
\label{fig:h2map}
\end{figure}

Given the lower resolution of the N(\hdos) data (see Fig.~\ref{fig:h2map}), 
we perform a new fitting 
process of the \nhp cube starting from the original spectral cube. In the 
first step, we convolved to a matched resolution and pixel scales the 
N(\hdos) data. Subsequently, we implement the fitting processes shown in 
\S~\ref{sec:line-fitting-process}. In this way, it is feasible to 
analyze both data-sets together in order to estimate the X(\nhp) 
correctly. 

\begin{figure}[h]
\includegraphics[width = \columnwidth]{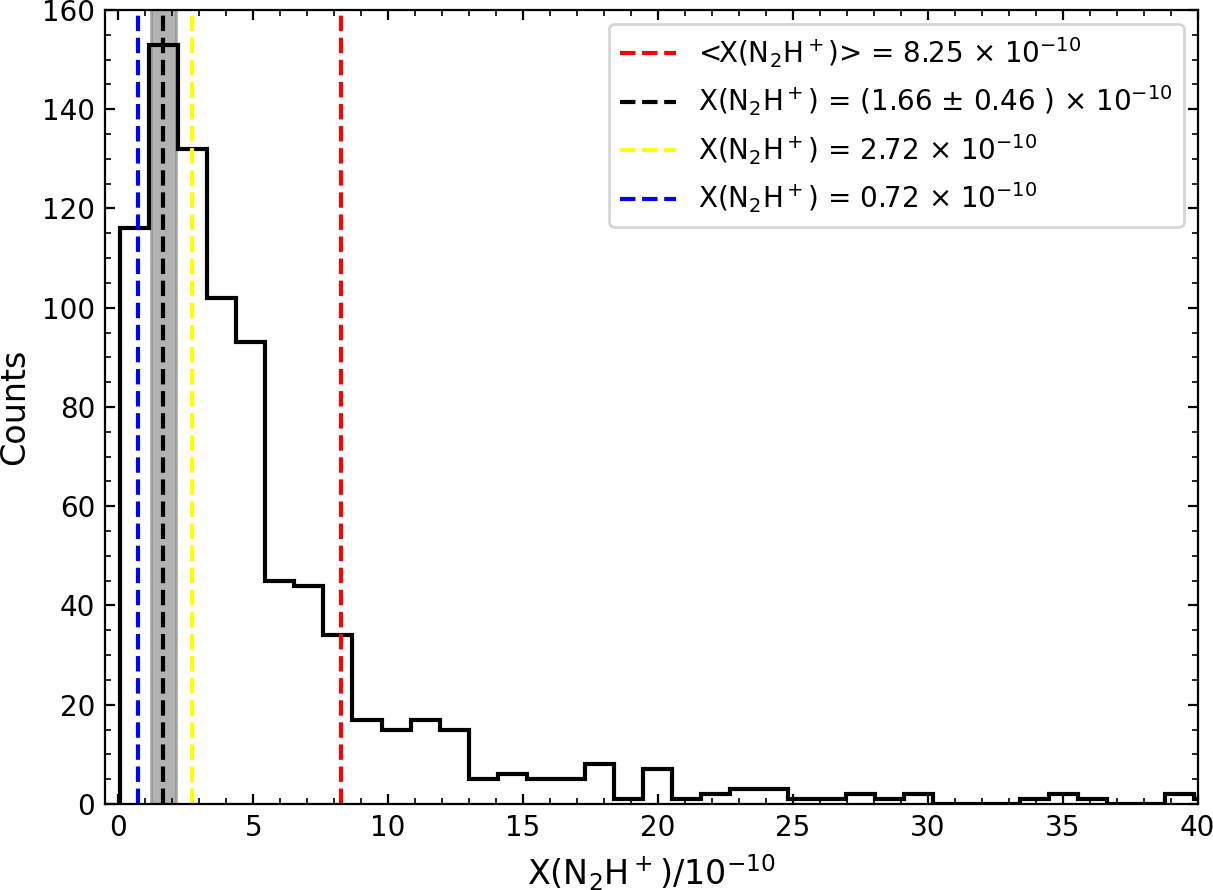}
\caption{Relative abundance X(\nhp) distribution in the G351.77
  protocluster zone. The red dashed line represents the mean of the
  distribution. The black dashed line
  represents the mean inside the most
  prominent bin of the distribution. The shaded region
  represents the estimated error around the mode.}
\vspace{-0.3cm}
\label{fig:xfactor}
\end{figure}

To ensure a more robust measurement of X(\nhp), we restrict our
analysis to N(\nhp) and N(\hdos) regions with S/N in column densities
$\,>\,4$.  Applying
Eq.~\ref{eq:relativeabundance_equation} we obtain the relative abundance
values for each pixel, which appears to follow a log-normal distribution
(see Fig.~\ref{fig:xfactor}). Based on these findings, our
objective is to determine a representative value of X(\nhp), whose value
will be equal to the mode of this distribution (the mean within the
largest bin of the histogram, see Fig.~\ref{fig:xfactor}). This allow us to
compare structures across the whole protocluster and between the velocity components.

However, taking into account the significant scatter of 
X(\nhp), with values that ranging from $\sim\,10^{-11}$ to $\sim\,10^{-9}$,
and considering that the mode may vary depending on the bin width used, 
we implement the \textit{Freedman Diaconis Estimator} method in 
order to derive the most representative value for the protocluster.
This method\footnote{\url{https://numpy.org/devdocs/reference/generated/numpy.histogram_bin_edges.html}}
provides the optimal bin width for a histogram, dividing the data sample 
into interquartile ranges (IQR), in order to equilibrate the 
distribution type, dispersion, and the size of the data sample.   

\begin{table}[h]
		\caption{Total column densities and masses in G351.77 protocluster.}
\vspace{-0.4cm}
\begin{center}
	\begin{tabular}{ccc}
		\hline\hline
		\\[-3mm]

		N$_\text{col}$ & <\ntot >     & \mtot     \\
		               & $\times$ $10^{23}$ [cm$^{-2}$]  & [\msun]   \\
		\hline 
		\\[-3mm]
		
		N$_{\rm tot,B}$     & 0.89 $\pm$ 0.24 & 740 $\pm$ 205 \\

		N$_{\rm tot,R}$     & 1.08 $\pm$ 0.29 & 917 $\pm$ 253 \\ 
		
		N$_{\rm tot,T}$ & 1.370 $\pm$ 0.38 & 1657 $\pm$ 326 \\
		\hline

	\end{tabular}
	\tablefoot{Averages of total column density estimations and
               total masses derived from the \nhp
               emission N$_{\rm tot,B}$, N$_{\rm tot,R}$ and
               N$_{\rm tot,T}$ corresponds to the Blue-velocity component,
               the Red-velocity component and N$_{\rm tot,B}$ $+$
               N$_{\rm tot,R}$, respectively.}
\end{center}
\vspace{-0.5cm}
\label{tab:h2col_table}
\end{table}

Applying this method, the estimated optimal number of bins for our data
distribution is 119 (see Fig.~\ref{fig:xfactor}). With this number of
bins, the mean within the largest bin (mode) corresponds to 
X(\nhp)\,$=\,(1.66\,\pm\,0.46)\,\times\,10^{-10}$. Additionally, by increasing
the number of bins while preserving the log-normal shape of the distribution and averaging all the measured modes, we find that the modes converge to X(\nhp)\,$=\,1.66\,\times\,10^{-10}$. Thus, we 
define the representative relative abundance of G351.77 in the 
protocluster zone as \linebreak X(\nhp)\,$=\,(1.66\,\pm\,0.46)\,\times\,10^{-10}$, 
whose value is within the ranges and magnitude order previously
reported \citep[e.g.][]{Caselli2002, fontani2006, Henshaw}.
We use this value to determine \ntot and the \mtot at a higher resolution 
based on the N(\nhp) emission at its original resolution (before convolving it).
This facilitates the measurement of the \mtot, at smaller scales, 
allowing to estimate masses related to cores and velocity 
gradients (see Table~\ref{tab:h2col_table} and \S~\ref{sec:kinematics-analysis}).

We estimate a total mass \mtot in the G351.77 protocluster
$\sim\,1660\,\pm\,326$\,~\msun, which we consider a lower limit
since this corresponds to the mass derived only from the regions with
\nhp emission. Additionally, the total mass estimated directly from
the dust-based \hdos column density map from \citet{Pierre} 
is M(\hdos)\,~$\sim$\,~2820\,~\msun. However, if we consider the
mass inside the \nhp coverage (at S/N~$>$~9), we estimate a M(\hdos)\,~$\sim$\,~2000\,~\msun.
This mass is within our uncertainties and hence represent a good
agreement considering the different tracers and techniques involved in
this comparison. Given that the N(\hdos) map from \citet{Pierre} is 
not affected by the interferometric filtering, like the \nhp data of 
this research, the column densities, and then the masses, are 
correctly comparable. Our \mtot estimation is also consistent with the 
total mass measured in \citet{Simon}, which represents $\sim$~16\% of 
the total mass measured in the Mother Filament $\sim$~10200\,~\msun 
in \citet{Simon}. Aside from the global protocluster properties, we 
provide column densities, masses, and \nhp relative abundances,
in addition to the spectral fitting parameters 
(see \S~\ref{sec:line-fitting-process} and 
Table~\ref{tab:core_parameters}) of the dense cores.
.

\section{Kinematic analysis}\label{sec:kinematics-analysis}

\begin{figure*}[h]
\includegraphics[width = 0.98\textwidth]{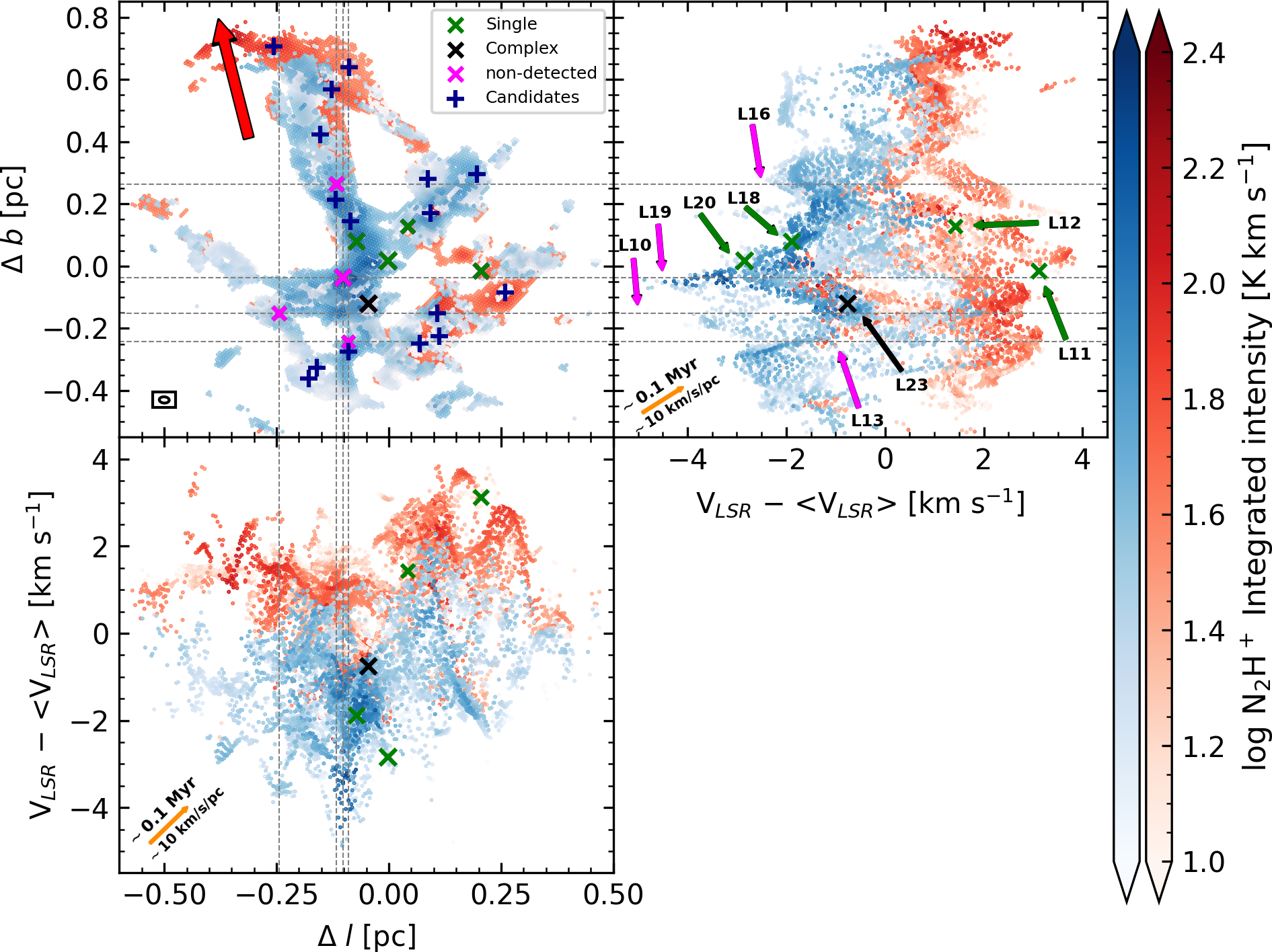}
\caption{Integrated intensity and position-velocity (PV) diagrams of
  the Blue- (blue colorbar) and Red- (red color bar) velocity components
  seen in \nhp~(1-0). Top left: Spatial distribution of \nhp~(1-0) emission in
  G351.77.  The green, black, and magenta $\times$ markers indicate
  the positions of the 9 out of the 18 (see \S~\ref{sec:kinematic-analysis-at-small-scales})      
  dense cores \citep{Fabien},
  where each color represents the DCN spectral
  classification: single, complex, and non-detected, respectively \citep{Nichol}.
  The $+$ markers indicate the position of the 16 core candidates, proposed on the basis
   on the \nhp PV features observed at scales of 
  $\sim$~0.1~pc (see \S~\ref{sec:kinematic-analysis-at-small-scales}
  and Table~\ref{tab:VGcores}). Dashed lines indicate the 
  positions of the four cores that do not have measurable velocities. The red arrow
  indicates the direction of the Mother Filament.
  The ellipse in the bottom-left represents the
  beam size of the \nhp data. Top right and bottom left: PV diagrams along
  the two perpendicular directions. Top right: The colored arrows indicate the position of the 
  dense cores and their ID in Table~\ref{tab:core_parameters}. 
  We observe multiple structures,
  such as V-shapes and Straight structures (see text) along the filaments,
  some associated with cores in both position and
  velocity. The orange arrows represent a velocity gradient of 
  10~$\kms$~pc$^{-1}$\,$\approx$\,0.1\,Myr.}
\label{fig:pv_diagram1}
\end{figure*}

The main goal is to analyze the
kinematics of the dense gas within the G351.77 protocluster
traced by \nhp~(1-0). The spectral line fitting process described in
\S~\ref{sec:line-fitting-process} provides us with the most reliable
parameter measurable from each spectrum, the centroid velocity,
whose median error estimations are $\sim\,0.034~\kms$. This centroid
velocity, combined with the angular resolution of our data, enables us
to analyze the kinematics of the protocluster on different scales (see below). To
achieve this, we utilize PV diagrams (see Fig.~\ref{fig:pv_diagram1}), 
which correspond to projections of Position-Position-Velocity (PPV) diagrams. 
These PV diagrams allow us to characterize PV features, which may be produced 
by inflow, rotation, outflow, or cloud-cloud collisions 
\citep[][]{Tobin, Henshaw, Haworth, Mori}, and access to the physical
information occurring at both $\sim$ small and large scales. 
To simplify the analysis, we therefore define small scales as scales from
the resolution limit of the observations, or the ``core'' scales, of
about 4~kau to about 10$\times$ larger.  Conversely, larger scales
encompass approximately from filament scales to the protocluster
scale, so $\sim 0.2$~pc to 1.2~pc.

In addition, we create and analyze PPV diagrams\footnote{\url{https://nicolas11399.github.io/Nicolas.Sandoval-Garrido_PVV.github.io/}},
in order to inspect and distinguish PPV structures. 
This way, it is possible to extract structures at both small and large, potentially 
coarser, scales. In doing so, we isolate 
structures based on PPV distributions combined with their integrated intensity. We then analyze these while avoiding degeneracies that might arise from the use of projected PV diagrams alone.

We utilize three parameters to generate the intensity-weighted PV diagrams shown in e.g.,
Fig.~\ref{fig:pv_diagram1}. These parameters are the centroid
velocity, integrated intensity (excluding points with emission below 11~K~$\kms$ to avoid introducing noise and to better highlight the dominant structures in the diagrams), and $b$ or $l$ coordinates
\citep[e.g.][]{Henshaw, Valentina19, Rodrigo21, Elena22,
rodrigo24}. The top left panel of the Fig.~\ref{fig:pv_diagram1}
shows the spatial distribution of the \nhp~(1-0) emission, where the blue
and red colorbars represent the Blue- and Red-velocity
components described in \S~\ref{sec:bluest-and-reddest-velocity-component}.
The top right and bottom left panels display PV diagrams, 
where the color of each point represents the integrated intensity
weighted.  

\subsection{Kinematic analysis on small scales}\label{sec:kinematic-analysis-at-small-scales}

A cursory inspection of
Fig.~\ref{fig:pv_diagram1} reveals distinct PV features
along the filamentary structures present in both the Blue- and the Red-velocity
component, some of which appear to be
consistent in position both with the dense cores \citep{Fabien} 
and with their estimated DCN~(3-2) velocities (see Table F.9
in \citealt{Nichol}).

\begin{figure}[H]
\centering
\includegraphics[width = \columnwidth]{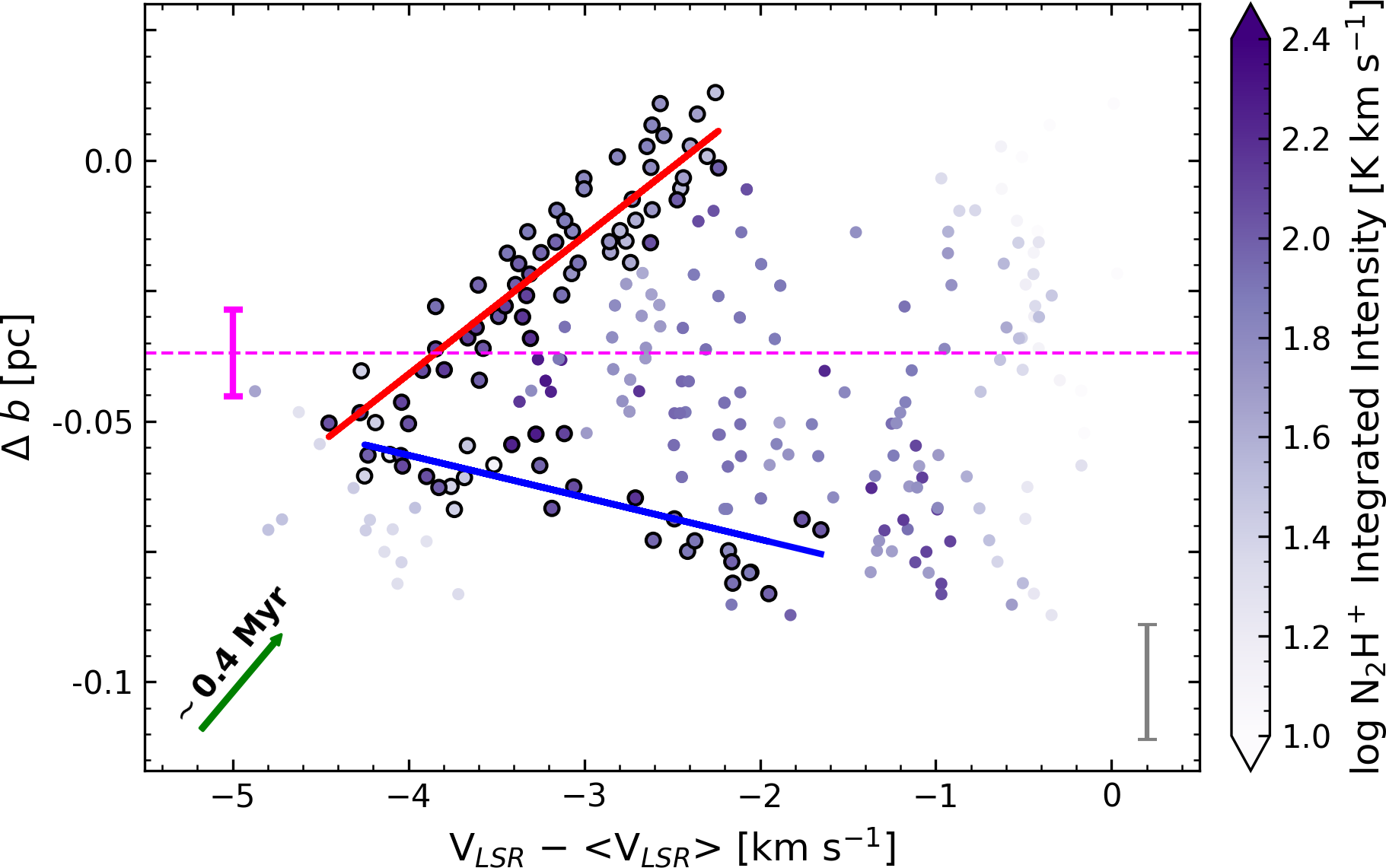}
\includegraphics[width = \columnwidth]{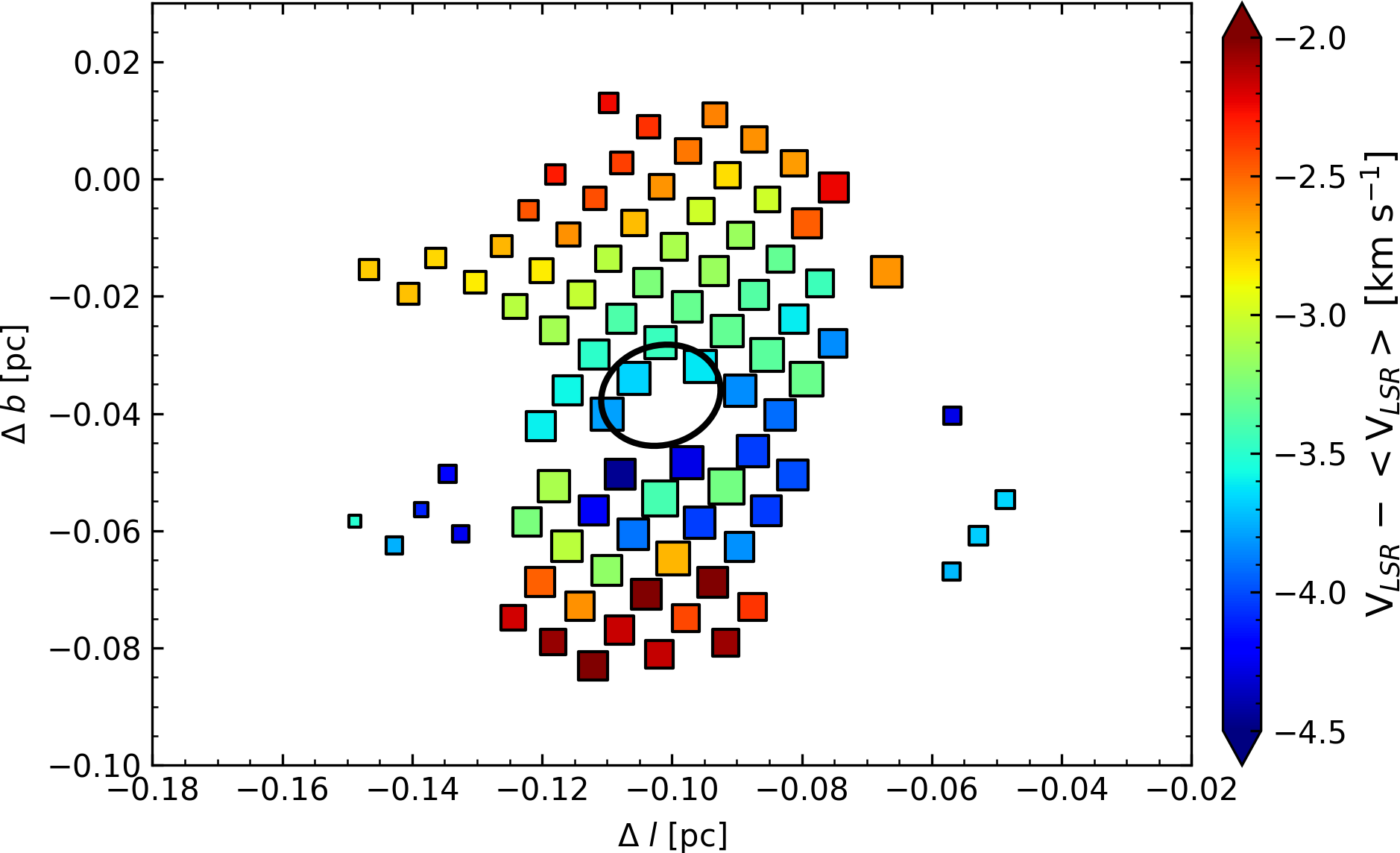}
\includegraphics[width = \columnwidth]{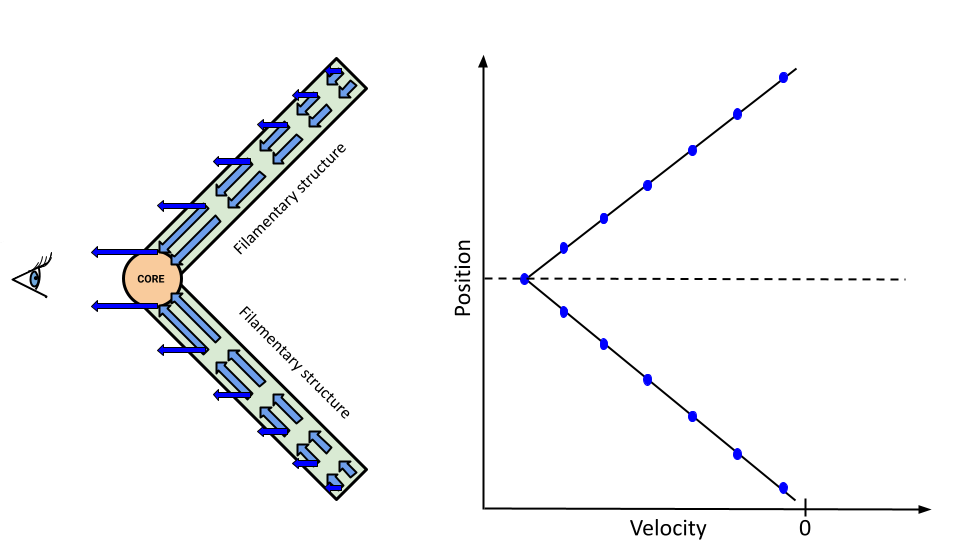}
\caption{Top: PV diagram of the most prominent V-shape (L19 in Fig.~\ref{fig:pv_diagram1}). The black 
  outlined points show the points used in order to
  generate the linear fits of the velocity gradients. The red and blue
  lines represent the upper and lower VG, respectively, which converge
  at $\sim$ 0.015 pc below the central position of the dense
  core. The magenta 
  errorbar represents the projected size of the core \citep{Fabien}, while the horizontal dashed 
  line
  represents the position of the core. The colorbar shows the integrated
  intensity of \nhp~(1-0) levels. The gray errorbar represents the beam size of the
  \nhp data. The green arrow represents a velocity gradient of 25~$\kms$~pc$^{-1}$, 
  which corresponds to a timescale of $\sim$~0.4~Myr. Middle: Position-position diagram of the most prominent V-shape. The squares represent the points used for the linear fit (top figure). The size of the squares correspond to the integrated intensity of each spectrum, while the ellipse indicates the position and size of the core (L19 in Fig.~\ref{fig:pv_diagram1}). Bottom right: Proposed model 
  for the observed PV features as V-shapes (see also Fig.\ 12 in \citealt{Henshaw} 
  and \citealt{rodrigo24}). Bottom left: inflowing filamentary structure. The
  light blue and blue arrows represent the velocity magnitude of the
  gas and the radial component of the gas velocity,
  respectively. Right: PV diagram of the model.}
\vspace{-0.3cm}
\label{fig:core6_2}
\label{fig:cartoon1}
\end{figure}

\begin{figure}[h]
\centering
\includegraphics[width = \columnwidth]{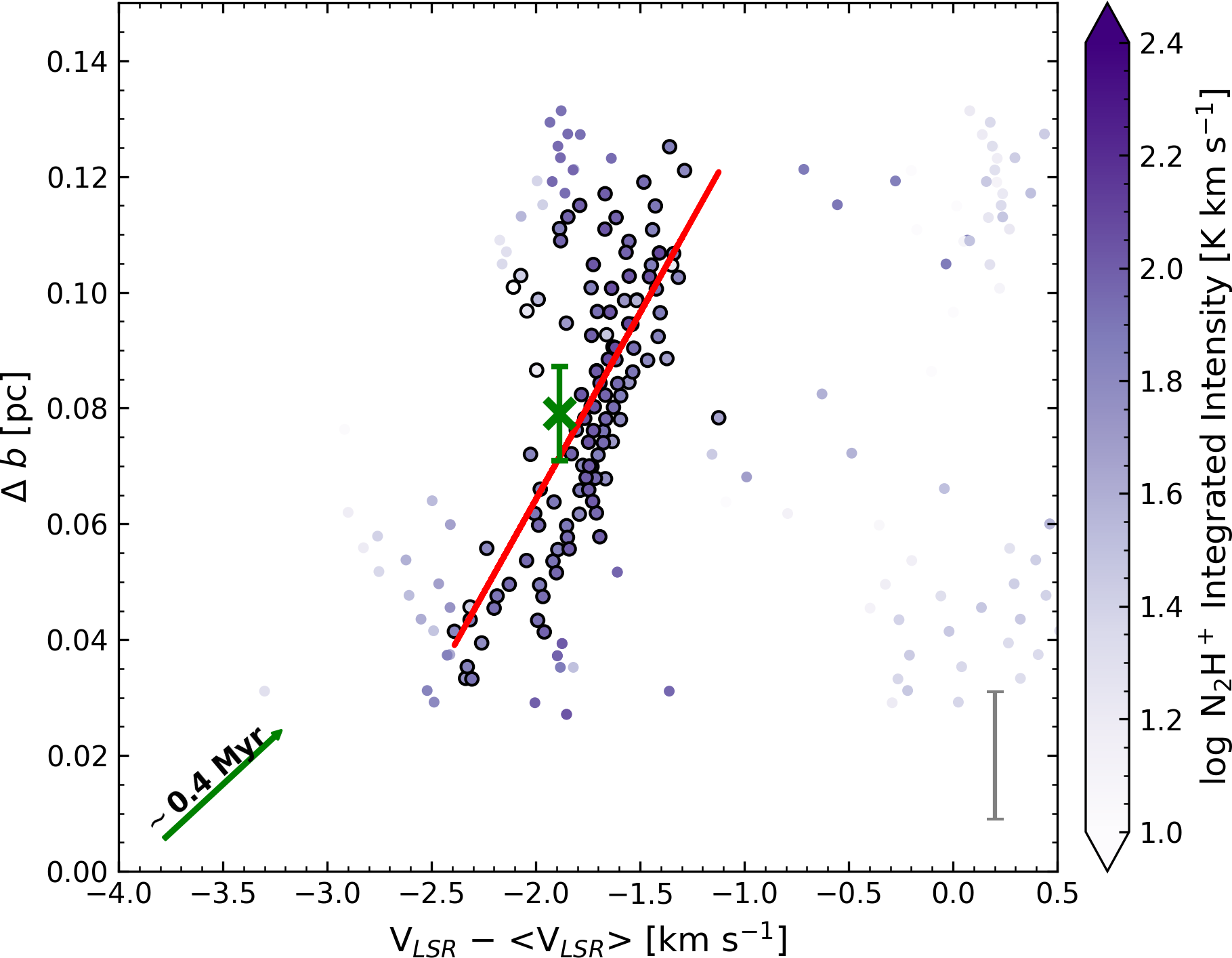}
\caption{PV diagram of a single core (L18 in Table~\ref{tab:VGcores}) from the catalog of \citet{Fabien}. 
  The PV feature outlines what we define
  as a Straight-shape. The black outlined points show the points used
  to generate the linear fit of the velocity gradient. The red line
  represents the linear fit of the velocity gradient. The green $\times$
  marker indicates the position \citep{Fabien} and velocity of the core measured in
  DCN (see Table F.9 in \citealt{Nichol}), while the green bar
  indicates the projected core size. The gray errorbar represents the beam size 
  of the \nhp data. The green arrow represents a velocity gradient of 25~$\kms$~pc~$^{-1}$, which correspond to a 
  timescale of $\sim$~0.4~Myr.}
\label{fig:strigth-shape}
\end{figure}

We analyze the \nhp~(1-0) velocities around the dense cores within a
radius 6 times the size of the core ($\sim$ 0.1 pc, see Table E.11
\citealt{Fabien}). This radius criteria enables us to examine
the behavior of the surrounding gas while avoiding the overlap of gas
around other cataloged cores. Subsequently, we re-create the
PV diagrams of \nhp~(1-0) emission in the vicinity of the cores, which
permits the identification of two distinct types of features:
\begin{enumerate} 
\item V-shapes: where the core is positioned at or near the point
  where we measure the inflection point in the velocity field of \nhp~(1-0),
  and the surrounding gas transitions towards lower or higher
  velocities (see Fig.~\ref{fig:cartoon1} and \citealt{rodrigo24}).
\item Straight-shapes: where the gas around the core transitions from
  high to low velocities, while the core is located $\sim$ halfway
  along this Straight-shape (see
  Fig.~\ref{fig:strigth-shape}; see also Figure 1 in \citealt{Tobin}).
\end{enumerate} 

From the 1.3~mm continuum data,
18 dense cores have been cataloged (see Table E.11 in \citealt{Fabien}), 
where 9 are located in zones with \nhp~(1-0) integrated intensity higher than 11~K~$\kms$ and S/N~$>$~9
(see Fig.~\ref{fig:pv_diagram1}). However, only six dense cores have identifiable
kinematic patterns in the PV diagram (3 V-shapes and 3
Straight-shapes, see Table~\ref{tab:VGcores}).

In order to characterize the V- and Straight-shapes we apply a linear fit to the
observed VGs, encompassing the upper
and lower VG for the V-shapes (see Fig.~\ref{fig:core6_2} and Fig.~\ref{fig:strigth-shape}). The selection criteria for the points used in the fit are based on their integrated intensity and location along the V-shape structure. This approach allows us to specifically characterize the enveloping points of the V-shape structure, effectively providing a boundary or limit for the measurements derived from the characterization. For the linear fitting process the integrated
intensity is used as a statistical weight. After characterizing the VGs,
we estimate their associated timescales as
$t_{\text{VG}} = \frac{1}{\text{VG}}$,
\citep[e.g.,][]{Rodrigo21,rodrigo24}.  Furthermore, we measure the
\mtot (\S~\ref{sec:column-density-and-masses}) associated with
the points used to quantify the VGs in order to estimate the mass
inflow rate \minflow following:

\begin{equation}
\label{eq:massinflowrate}
\dot{\textup{M}}_{\text{tot,in}} = \frac{\text{M}_{\text{tot}}}{t_{\text{VG}}}.
\end{equation}

We note that the above analysis of the VG and timescales is affected by the inclination angle $\theta$ of the filament relative to
the plane of the sky. The relationship is given by $t_{\rm VG}\,=\,l/v_r\tan(\theta)$, where $l$ represents
the projected length of the filament in the plane of the sky, $v_r$ denotes the radial velocity, and $\theta$ is the
inclination angle of the filament. For our purposes, we can not measure the inclination of any given structure. The assumptions here are equivalent to assuming a 45$^{\circ}$ average inclination angle for all timescales and $\dot{\textup{M}}_{\text{tot,in}}$. Besides, \citet{rodrigo24} showed in their Fig. 11 that
the estimated gradients (which they measure in a similar fashion as
here) do not depend on the orientation angle of the spatial axis used in
the PV diagrams.  That is, V-shapes extracted along the $b$ coordinate
still appear as clear V-shapes at other angles with gradients (and so timescales). 

\begin{figure}[H]
\centering
\includegraphics[width = \columnwidth]{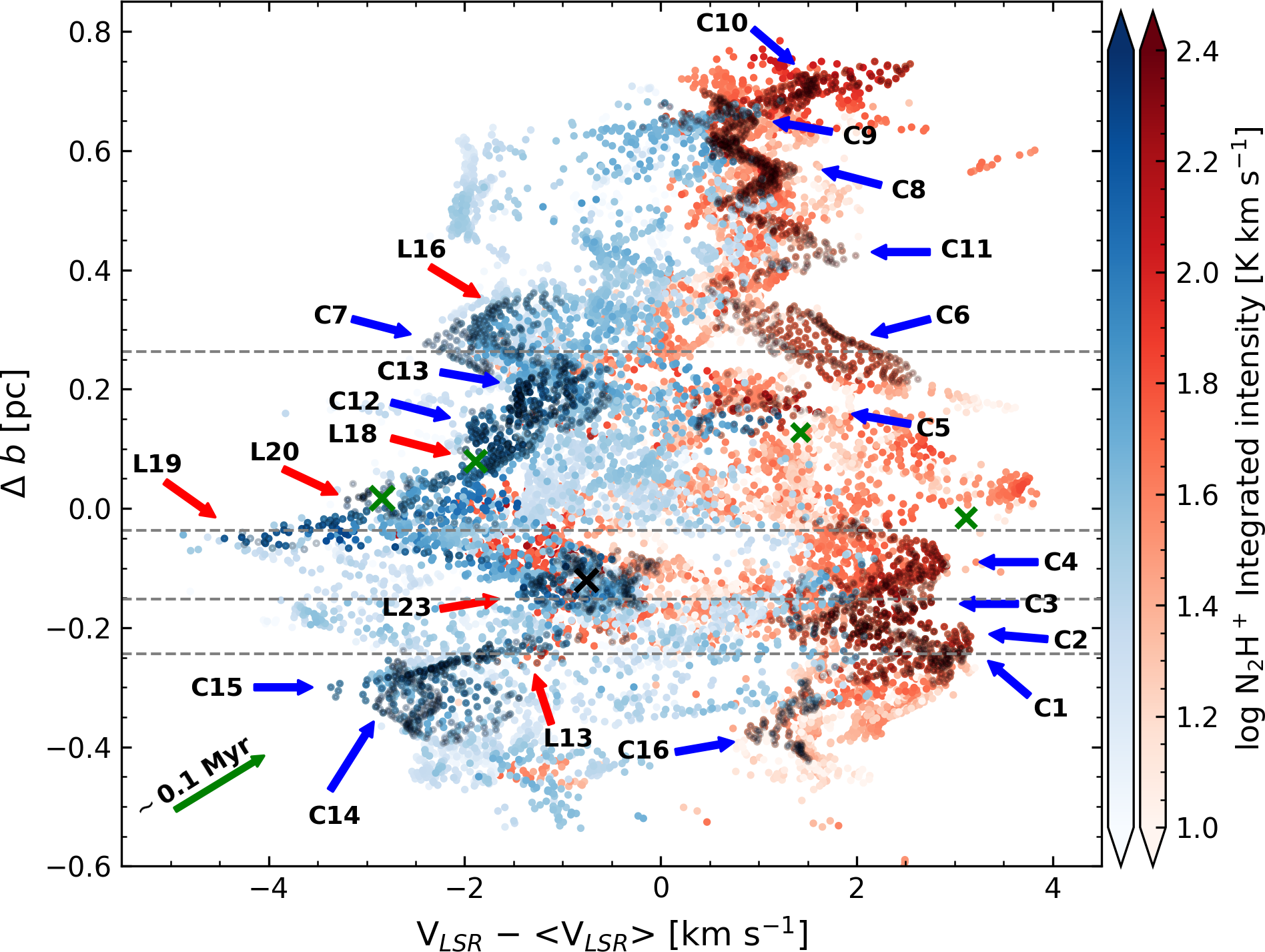}
\caption{PV diagram of the Blue- (blue colorbar) and Red- 
(red colorbar) velocity component seen in \nhp~(1-0). The dark points 
highlight the PV features observed. The blue arrows represent the 
location of the 16 PV features of candidates 
cores under the 1.3~mm band detection limit 
(see Table~\ref{tab:VGcores}). The red arrows indicate the location 
in the PV diagram of PV features associated with the dense cores 
(\citealt{Fabien}, see Table~\ref{tab:VGcores}). The markers and dashed lines are the same as in Fig.~\ref{fig:pv_diagram1}.
The green arrow represents a velocity gradient of 10~$\kms$~pc~$^{-1}$, which correspond to a timescale of 0.1~Myr.}
\vspace{-0.3cm}
\label{fig:vshapes_location}
\end{figure}

Focusing on the most prominent V-shape, we identify two clear VGs
associated with a dense core (L19 in Fig.~\ref{fig:vshapes_location} and Table~\ref{tab:VGcores}). 
This core is located in the region with the
highest \nhp~(1-0) emission, where the spectra are fitted by
2-velocity-components. 
We measure the VGs, timescales, mass, and mass inflow rate of the 
six PV features associated with dense cores, and we present 
these parameters in Table~\ref{tab:VGcores}.

\citet{Henshaw} proposed two physical processes related to the
V-shapes structures in PV diagrams; inflows and outflows. However, we do not
observe agreement between the position of outflow catalogs \citep[e.g.][]{towner}
and the position of cores associated with V-shapes in PV diagrams. 
We  \linebreak 


\noindent propose that the points utilized for the characterization of the VG in V-shapes
represent the inflowing gas with the highest velocity along a
filamentary structure, where the core may be located in a ``knee'' in the
filament (see Fig.~\ref{fig:cartoon1} and \citealt{rodrigo24} for
similar interpretation of the V-shapes). On the other hand, the Straight-shapes may be explained as 
filamentary structures with embedded cores, where the
filaments inflow toward the inner and denser regions and the cores
follow the stream of the surrounding gas, flowing along with the filaments
\citep[][]{hacar2011, Lu, anualmotte, Kim, Arzoumanian,  Yang, Pan}.

\begin{table*}[t]
		\caption{Characterizations of the \nhp PV features at small scale observed
                  PV diagram.}
\vspace{-0.4cm}
\begin{center}
	\begin{tabular}{ccccccccc}
		\hline\hline
		\\[-3mm]
		PV-feature & RA & DEC & Shape &       VG         & $t_{\rm VG}$ & \mtot & \minflow \\
		ID  &    \multicolumn{2}{c}{[ICRS (J2000)]}     &       & [km s$^{-1}$pc$^{-1}$] &   [kyr]   & [\msun]  & $\times$ $10^{-4}$  [\msunyr] \\

		\hline
		\\[-3mm]

		L13 & 17:26:45.089 & -36:09:19.42 & Straight & 26.90 $\pm$ 0.95 & 36.35 $\pm$ 1.29 & 9.74 $\pm$ 2.69 & 2.68 $\pm$ 0.74 \\
		\\[-3mm]

		\multirow{2}{*}{L16}  & \multirow{2}{*}{17:26:41.635} & \multirow{2}{*}{-36:08:48.00} & \multirow{2}{*}{V-shape} & 22.61 $\pm$ 2.65 & 43.25 $\pm$ 5.07 & 1.71 $\pm$ 0.47 & 0.40 $\pm$ 0.11 \\
		
		 & & & & 15.46 $\pm$ 0.89 & 63.24 $\pm$ 3.65 & 5.44 $\pm$ 1.50 & 0.86 $\pm$ 0.24 \\
		\\[-3mm]

		L18 & 17:26:42.725 & -36:09:02.42 & Straight & 15.50 $\pm$ 1.43 & 63.09 $\pm$ 5.81 & 18.36 $\pm$ 5.07 & 2.91 $\pm$ 0.80 \\
		\\[-3mm]

		\multirow{2}{*}{L19} & \multirow{2}{*}{17:26:43.687} & \multirow{2}{*}{-36:09:06.51} & \multirow{2}{*}{V-shape} & 37.78 $\pm$ 2.08 & 25.88 $\pm$ 1.43 & 10.01 $\pm$ 2.76 & 3.87 $\pm$ 1.07 \\

		 &  & & & 133.60 $\pm$ 16.96 & 7.32 $\pm$ 2.02 & 5.35 $\pm$ 1.48 & 7.31 $\pm$ 2.02 \\	
		\\[-3mm]

		\multirow{2}{*}{L20} & \multirow{2}{*}{17:26:42.830} & \multirow{2}{*}{-36:09:11.97} & \multirow{2}{*}{V-shape} & 24.8 $\pm$ 2.07 & 39.43 $\pm$ 3.30 & 1.02 $\pm$ 0.28 & 0.26 $\pm$ 0.07 \\	

		 & & &  & 29.42 $\pm$ 3.35 & 33.24 $\pm$ 3.78 & 2.39 $\pm$ 0.66 & 0.75 $\pm$ 0.20 \\
		\\[-3mm]

		L23 & 17:26:44.011 & -36:09:16.18 & Straight & 2.46 $\pm$ 0.14 & 397.63 $\pm$ 23.10 & 38.21 $\pm$ 10.56 & 0.96 $\pm$ 0.27 \\

		\hline
		\\[-3mm]

		\multirow{2}{*}{C1} & \multirow{2}{*}{17:26:44.369} & \multirow{2}{*}{-36:09:33.05} & \multirow{2}{*}{V-shape} & 33.04 $\pm$ 2.41 & 29.6 $\pm$ 2.16 & 15.29 $\pm$ 4.22 & 5.17 $\pm$ 1.43 \\
		                   &                               &                               &                          & 48.05 $\pm$ 6.21 & 20.35 $\pm$ 2.63 & 12.44 $\pm$ 3.44 & 6.11 $\pm$ 1.69 \\
		\\[-3mm]

		\multirow{2}{*}{C2} & \multirow{2}{*}{17:26:44.003} & \multirow{2}{*}{-36:09:35.61} & \multirow{2}{*}{V-shape} & 33.34 $\pm$ 2.19 & 29.32 $\pm$ 1.93 & 17.97 $\pm$ 4.96 & 6.13 $\pm$ 1.69 \\
		                   &                               &                               &                          & 44.40 $\pm$ 3.53 & 22.02 $\pm$ 1.75 & 17.11 $\pm$ 4.72 & 7.77 $\pm$ 2.14 \\
		\\[-3mm]
		
		\multirow{2}{*}{C3} & \multirow{2}{*}{17:26:43.501} & \multirow{2}{*}{-36:09:31.06} & \multirow{2}{*}{V-shape} & 58.51 $\pm$ 10.24 & 16.71 $\pm$ 2.92 & 6.58 $\pm$ 1.82 & 3.94 $\pm$ 1.09 \\
                           &                               &                               &                          & 22.52 $\pm$ 1.03 & 43.42 $\pm$ 1.99 & 23.65 $\pm$ 6.53 & 5.45 $\pm$ 1.50 \\
		\\[-3mm]
		
		\multirow{2}{*}{C4} & \multirow{2}{*}{17:26:41.743} & \multirow{2}{*}{-36:09:35.57} & \multirow{2}{*}{V-shape} & 19.11 $\pm$ 1.65 & 51.17 $\pm$ 4.41 & 13.21 $\pm$ 3.64 & 2.58 $\pm$ 0.71 \\
		                   &                               &                               &                          & 22.49 $\pm$ 2.32 & 43.47 $\pm$ 4.49 & 12.45 $\pm$ 3.44 & 2.86 $\pm$ 0.79 \\
		\\[-3mm]
		
		\multirow{2}{*}{C5} & \multirow{2}{*}{17:26:41.304} & \multirow{2}{*}{-36:09:11.29} & \multirow{2}{*}{V-shape} & 24.35 $\pm$ 2.73 & 40.16 $\pm$ 4.50 & 4.81 $\pm$ 1.32 & 1.20 $\pm$ 0.33 \\
		                   &                               &                               &                          & 37.54 $\pm$ 4.13 & 26.05 $\pm$ 2.86 & 7.59 $\pm$ 2.09 & 2.91 $\pm$ 0.80 \\
		\\[-3mm]
		
		C6 		           &  17:26:40.551                 &    -36:09:04.45               & Straight                 & 24.35 $\pm$ 2.73 & 40.16 $\pm$ 4.50 & 18.58 $\pm$ 5.13 & 4.63 $\pm$ 1.28 \\
		\\[-3mm]
		
		\multirow{2}{*}{C7} & \multirow{2}{*}{17:26:39.924} & \multirow{2}{*}{-36:09:12.81} & \multirow{2}{*}{V-shape} & 16.10 $\pm$ 1.38 & 60.75 $\pm$ 5.19 & 7.16 $\pm$ 1.97 & 1.18 $\pm$ 0.32 \\
		                   &                               &                               &                          & 15.30 $\pm$ 2.81 & 63.92 $\pm$ 11.73 & 2.41 $\pm$ 0.66 & 0.38 $\pm$ 0.10 \\	
		\\[-3mm]
		
		\multirow{2}{*}{C8} & \multirow{2}{*}{17:26:39.548} & \multirow{2}{*}{-36:08:29.49} & \multirow{2}{*}{V-shape} & 17.77 $\pm$ 1.19 & 55.01 $\pm$ 3.68 & 22.79 $\pm$ 6.29 & 4.14 $\pm$ 1.14 \\	
                           &                               &                               &                          & 11.18 $\pm$ 0.78 & 87.43 $\pm$ 6.13 & 23.45 $\pm$ 6.48 & 2.68 $\pm$ 0.74 \\
		\\[-3mm]
		
		\multirow{2}{*}{C9} & \multirow{2}{*}{17:26:38.858} & \multirow{2}{*}{-36:08:28.73} & \multirow{2}{*}{V-shape} & 11.06 $\pm$ 0.90 & 88.40 $\pm$ 7.16 & 13.51 $\pm$ 3.73 & 1.53 $\pm$ 0.42 \\
		                   &                               &                               &                          & 17.23 $\pm$ 3.55 & 56.76 $\pm$ 11.70 & 6.14 $\pm$ 1.69 & 1.08 $\pm$ 0.30 \\
		\\[-3mm]
		
		C10                  &   17:26:39.172                &   -36:08:10.49               &                Straight  & 25.32 $\pm$ 1.13 & 38.62 $\pm$ 1.72 & 45.64 $\pm$ 5.08 & 11.82 $\pm$ 1.32 \\
		\\[-3mm]
		
		\multirow{2}{*}{C11} & \multirow{2}{*}{17:26:40.677} & \multirow{2}{*}{-36:08:35.57} & \multirow{2}{*}{V-shape} & 17.75 $\pm$ 1.61 & 55.08 $\pm$ 4.99 & 5.08 $\pm$ 1.41 & 0.92 $\pm$ 0.26 \\
                            &                               &                               &                          & 20.91 $\pm$ 1.88 & 46.77 $\pm$ 4.20 & 2.33 $\pm$ 0.64 & 0.50 $\pm$ 0.14 \\
		\\[-3mm]
		
		\multirow{2}{*}{C12} & \multirow{2}{*}{17:26:42.190} & \multirow{2}{*}{-36:08:53.34} & \multirow{2}{*}{V-shape} & 20.47 $\pm$ 1.23 & 47.77 $\pm$ 2.86 & 12.76 $\pm$ 3.52 & 2.67 $\pm$ 0.74 \\	
		                    &                               &                               &                          & 33.46 $\pm$ 2.60 & 29.23 $\pm$ 2.27 & 10.65 $\pm$ 2.94 & 3.64 $\pm$ 1.01 \\
		\\[-3mm]
		
		\multirow{2}{*}{C13} & \multirow{2}{*}{17:26:42.185} & \multirow{2}{*}{-36:08:55.85} & \multirow{2}{*}{V-shape} & 20.49 $\pm$ 2.41 & 47.72 $\pm$ 5.60 & 10.53 $\pm$ 2.91 & 2.21 $\pm$ 0.61 \\	
		                    &                               &                               &                          & 35.59 $\pm$ 4.61 & 29.23 $\pm$ 3.56 & 9.17 $\pm$ 2.53 & 3.14 $\pm$ 0.87 \\
		\\[-3mm]
		
		\multirow{2}{*}{C14} & \multirow{2}{*}{17:26:46.011} & \multirow{2}{*}{-36:09:18.13} & \multirow{2}{*}{V-shape} & 12.55 $\pm$ 1.04 & 77.90 $\pm$ 6.43 & 3.18 $\pm$ 0.88 & 0.41 $\pm$ 0.11 \\
                            &                               &                               &                          & 34.67 $\pm$ 2.44 & 28.20 $\pm$ 1.98 & 3.62 $\pm$ 1.01 & 1.28 $\pm$ 0.36 \\
		\\[-3mm]
		
		\multirow{2}{*}{C15} & \multirow{2}{*}{17:26:45.320} & \multirow{2}{*}{-36:09:21.17} & \multirow{2}{*}{V-shape} & 48.40 $\pm$ 3.80 & 20.20 $\pm$ 1.58 & 8.81 $\pm$ 2.43 & 4.36 $\pm$ 1.20 \\
                            &                               &                               &                          & 42.72 $\pm$ 4.74 & 22.89 $\pm$ 2.54 & 5.01 $\pm$ 1.38 & 2.19 $\pm$ 0.60 \\
		\\[-3mm]
		
		\multirow{2}{*}{C16} & \multirow{2}{*}{17:26:46.340} & \multirow{2}{*}{-36:09:18.52} & \multirow{2}{*}{V-shape} & 12.27 $\pm$ 1.35 & 79.70 $\pm$ 8.74 & 9.08 $\pm$ 2.51 & 1.14 $\pm$ 0.31 \\
                            &                               &                               &                          & 16.56 $\pm$ 2.11 & 59.03 $\pm$ 7.53 & 8.22 $\pm$ 2.27 & 1.39 $\pm$ 0.38 \\
		\hline

	\end{tabular}
	\tablefoot{The ``L'' indicates the PV features associated with the dense 
	cores from \citet{Fabien}. The ``C'' indicates the PV 
	features associated with core candidates under the 
	1.3~mm band detection limits. We provide the velocity gradient, 
	timescale, mass and mass inflow rate for each PV feature, in addition 
	to the potential position of the vertex.}
	\vspace{-0.5cm}
\end{center}
\label{tab:VGcores}
\end{table*}

Continuing the PV analysis at scales of 6 times the average radius 
of the dense cores, some of these PV features are 
even found without any detected dense cores 
(see Fig.~\ref{fig:vshapes_location} and Table~\ref{tab:VGcores}).

\begin{table*}[h]
		\caption{Averaged characterization of the \nhp V-shapes observed
                  PV diagram.}
\vspace{-0.4cm}
\begin{center}
	\begin{tabular}{ccccccc}
		\hline\hline
		\\[-3mm]
		V-shape &     RA & DEC &   VG         & $t_{\rm VG}$ & \mtot & \minflow \\
		 ID &   \multicolumn{2}{c}{[ICRS (J2000)]}   & [km s$^{-1}$pc$^{-1}$]  & [kyr]   & [\msun]  & $\times$ $10^{-4}$  [\msunyr] \\

		\hline
		\\[-3mm]

		 L16 & 17:26:41.635 & -36:08:48.00 &  19.04 $\pm$ 1.40 & 51.39 $\pm$ 3.78 & 7.15  $\pm$ 1.57 & 1.39 $\pm$ 0.32 \\

		 L19 & 17:26:43.687 & -36:09:06.51 &  85.69 $\pm$ 8.54 & 11.42 $\pm$ 1.13 & 15.36 $\pm$ 3.13 & 13.45 $\pm$ 3.04 \\

         L20 & 17:26:42.830 & -36:09:11.97 &  27.11 $\pm$ 1.97 & 36.09 $\pm$ 2.62 & 3.41  $\pm$ 0.71 & 0.94 $\pm$ 0.19 \\

		\hline
		\\[-3mm]
		
		 C1 & 17:26:44.369 & -36:09:33.05 &  40.54 $\pm$ 3.33 & 24.13 $\pm$ 1.98 & 27.72 $\pm$ 5.44 & 11.49 $\pm$ 2.25 \\
		
		 C2 & 17:26:44.003 & -36:09:35.61 &  38.87 $\pm$ 2.07 & 25.17 $\pm$ 1.34 & 35.08 $\pm$ 6.84 & 13.93 $\pm$ 2.72 \\

		 C3 & 17:26:43.501 & -36:09:31.06 &  40.51 $\pm$ 5.14 & 24.15 $\pm$ 3.06 & 30.22 $\pm$ 6.77 & 12.51 $\pm$ 2.90 \\

		 C4 & 17:26:41.743 & -36:09:35.57 &  20.80 $\pm$ 1.42 & 47.04 $\pm$ 3.21 & 25.66 $\pm$ 5.00 & 5.45 $\pm$ 1.06 \\	

         C5 & 17:26:41.304 & -36:09:11.29 &  30.94 $\pm$ 2.47 & 31.61 $\pm$ 2.53 & 12.40 $\pm$ 2.47 & 3.92 $\pm$ 0.78 \\

		 C7 & 17:26:39.924 & -36:09:12.81 &  15.70 $\pm$ 1.56 & 62.32 $\pm$ 6.21 & 9.57 $\pm$ 2.07 & 1.53 $\pm$ 0.33 \\
		
		 C8 & 17:26:39.548 & -36:08:29.49 &  14.47 $\pm$ 0.71 & 67.59 $\pm$ 3.32 & 46.23 $\pm$ 9.03 & 6.84 $\pm$ 1.33 \\

         C9 & 17:26:38:858 & -36:08:28.73 &  14.14 $\pm$ 1.83 & 69.17 $\pm$ 8.95 & 19.65 $\pm$ 4.09 & 2.84 $\pm$ 0.59 \\

		 C11 & 17:26:40.677 & -36:08:35.57 &  19.33 $\pm$ 1.23 & 50.61 $\pm$ 3.24 & 7.41 $\pm$ 1.54 & 1.46 $\pm$ 0.30 \\

		 C12 & 17:26:42.190 & -36:08:53.34 &  26.96 $\pm$ 1.43 & 36.28 $\pm$ 1.93 & 23.41 $\pm$ 4.58 & 6.45 $\pm$ 1.26 \\

         C13 & 17:26:42.185 & -36:08:55.85 &  28.04 $\pm$ 2.60 & 34.89 $\pm$ 3.23 & 19.70 $\pm$ 3.85 & 5.64 $\pm$ 1.10 \\

         C14 & 17:26:46.011 & -36:09:18.13 &  23.61 $\pm$ 1.32 & 41.44 $\pm$ 2.32 & 6.80 $\pm$ 1.33 & 1.64 $\pm$ 0.32 \\

         C15 & 17:26:45.320 & -36:09:21.17 &  45.56 $\pm$ 3.03 & 21.47 $\pm$ 1.43 & 13.82 $\pm$ 2.79 & 6.43 $\pm$ 1.30 \\
         
         C16 & 17:26:46.340 & -36:09:18.52 &  14.41 $\pm$ 1.25 & 67.87 $\pm$ 5.89 & 17.30 $\pm$ 3.38 & 2.54 $\pm$ 0.49 \\
		\hline

	\end{tabular}
	\tablefoot{The ``L'' indicates the V-shapes observed in PV diagram, 
	associated with the dense cores from \citet{Fabien}. 
	The ``C'' represent the V-shapes observed in PV diagram associated 
	with the candidates to possible cores under the 1.3~mm band detection 
	limits. We provide the averaged estimations of velocity gradient, and its respective 
	timescale, mass and mass inflow rate for each V-shape.}
	\vspace{-0.5cm}
\end{center}
\label{tab:meanvg}
\end{table*}

Utilizing the PPV diagram, 
we find 16 PV features both
V- and Straight-shapes in 
the protocluster without associated dense cores (see
Fig.~\ref{fig:vshapes_location}). We distinguish between these 
small scale PV features based on the velocity distribution in the 
PPV diagram and increases in integrated intensity. This way, 
we can rule out the possibility that a given PV feature is produced by the 
overlap of points that do not share nearby regions in the 
position-position diagram. We measured the VGs, timescales,
mass, and \minflow of the PV features 
(see Table~\ref{tab:VGcores}). Both the integrated intensities measured
in the regions where these PV features are observed and
the estimated \minflow values appear similar to what we have estimated
for the PV features associated with dense cores.

Furthermore, it is possible to observe some
consistences in position-position between the dense cores candidates and the cores
detected by \textit{GExt2D} \citep{Fabien}. We suggest a relation between these PV
features and possible cores that are under the detection limit on the
1.3~mm data (undetected by \textit{getsf}) or regions where the gas is just beginning to accrete. 
We propose the location of these possible cores (see
Table~\ref{tab:VGcores}) as the spatial locations of the vertexes of the V-shapes. 
Moreover, based on the dense 
cores and the associated PV features, we may expect a
discrepancy of no more than $\sim$\,0.03\,pc between the proposed
positions and the locations of the undetected cores. This because some gas 
flowing through filaments may move toward a local or global maximum of the
gravitational potential, which might not necessarily coincide exactly with the
position of a core \citep{Izquierdo}. Additionally, inflowing gas
along filaments can be disturbed by turbulence, shear, interaction with molecular outflows,
\HII regions, and other dynamical processes \citep{Ruben2020, Enrique2024}. These effects, in addition with projection effects, could explain the discrepancies 
observed between the position of dense cores and the vertex of the V-shapes, which, nevertheless, are always within about the beam size (see Fig.\ref{fig:core6_2}).

The estimation of \minflow of these PV features, particularly
the V-shapes, could be correlated with the core mass buildup, due to the
convergence of VGs at some dense core positions (see
Fig.~\ref{fig:cartoon1} and \citealt{rodrigo24}). Moreover, the
estimated $t_{\rm VG}$ could suggest the timescale for the collapse of
filamentary structures towards cores \citep[][]{rodrigo24}. In total,
we observe 17 V-shapes (see
Table~\ref{tab:meanvg}), where the average values of $t_{\rm VG}\,=\,$~43.73~$\pm$~0.94~kyr 
and \minflow~$=\,5.59\,\pm$~0.34)~$\times~10^{-4}$~\msunyr.

\subsection{Kinematics analysis on large scales}\label{sec:kinematics-analysis-at-large-scales}

\begin{figure*}[h]
\centering
\includegraphics[width = 0.98\textwidth]{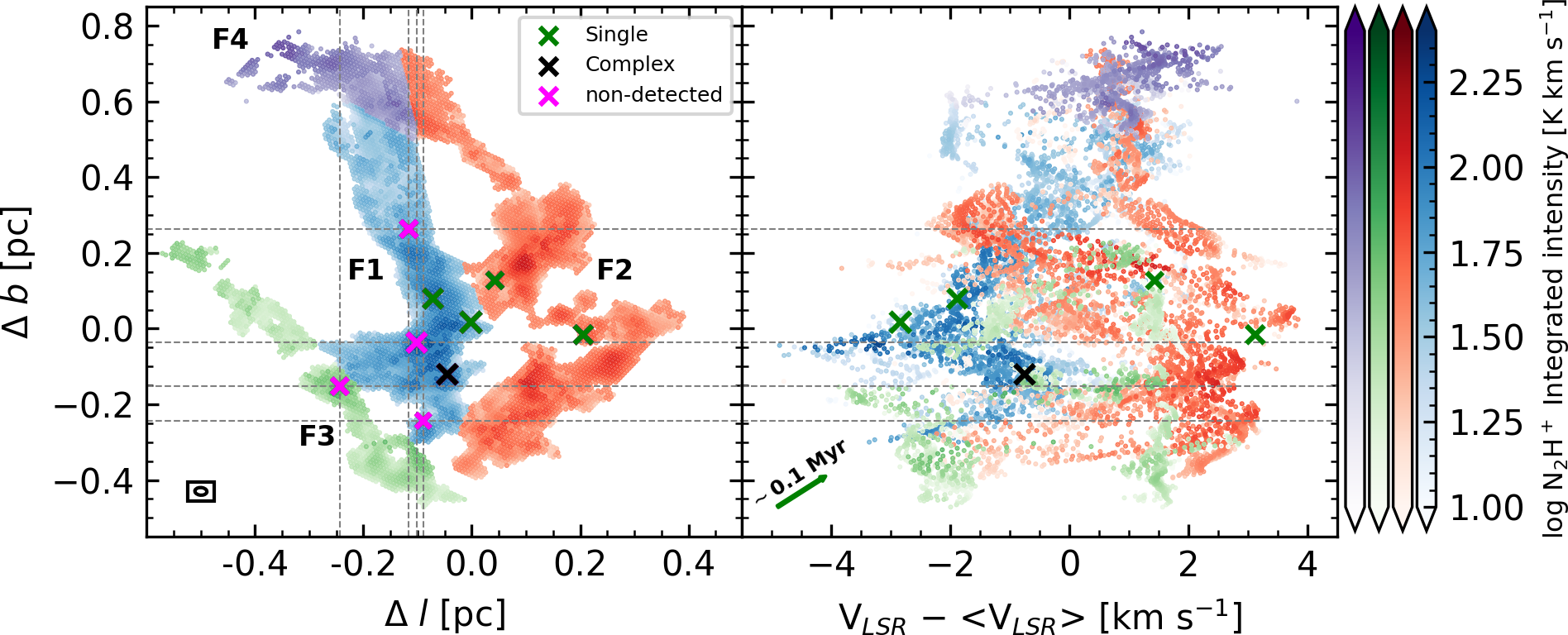}
\caption{Integrated intensity and position-velocity (PV) diagram of 
  the F1 (blue colorbar), F2 (red colorbar), F3 (green colorbar), and
  F4 (purple colorbar) large-scale structures of \nhp~(1-0) into which the 
  protocluster has been divided. Left: Spatial distribution of \nhp~(1-0) 
  emission. Markers and dashed lines are the same as in Fig.~\ref{fig:pv_diagram1}. 
  The ellipse in the bottom-left 
  represents the beam size of the \nhp data. Right: PV diagram. We observe
  multiple large-scale V-shapes and Straight-shapes along the different
  filamentary structures, with some of them associated with cores in
  position and in velocity. The green arrows represent a velocity 
  gradient of 10~$\kms$~pc$^{-1}$, which correspond to a timescale 
  of $\sim$~0.1~Myr.}
  \vspace{-0.3cm}
\label{fig:pvcolores}
\end{figure*}

The PV diagram of the \nhp~(1-0) emission, from filament scales to
protocluster scales ($\sim$~0.2~pc to $\sim$~1.2~pc) reveals two
distinct large-scale structures, separated by
$\sim$~2~$\kms$ (see Fig.~\ref{fig:pv_diagram1}). One appears to be dominated by the 
Blue-velocity component, while the other is characterized by the 
Red-velocity component. Furthermore, these two velocity
structures seem to merge or tighten up towards the Galactic northern region of
the protocluster, in the direction toward the Mother Filament (see top left panel in Fig.~\ref{fig:pv_diagram1}). These large-scale structures present PV
features like a zig-zag along the protocluster, which we suggest are
produced by kinematic effects at smaller scales (see
\S~\ref{sec:kinematic-analysis-at-small-scales}). However, upon deeper
examination of the kinematics and inspection of the PPV diagram, we
realize that these large-scale structures are not only separated
in velocity but also by spatial position (see
Fig.~\ref{fig:pvcolores}).

In order to analyze these structures and identify the large-scale regions contributing to the features observed in the PV-diagram, we divided the emission into four regions (F1, F2, F3, and F4) defining clear boundaries for each. This separation was guided by the integrated intensity map, SNR map, and the column density map, allowing to distinguish between possible filamentary structures. The criteria used include integrated intensities drop below 50~K\,$\kms$ (excluding F3) and velocity differences larger than $\sim$\,0.9\,$\kms$, enabling us to separate structures that exhibit sharp velocity transitions (see Fig.~\ref{fig:pvcolores}):
\begin{itemize}
\item F1: This corresponds to the primary filamentary structure within
  the protocluster, where we observe the highest S/Ns, integrated
  intensities, and column densities. It is connected with the three
  other large-scale structures (F2, F3, and F4) defined within the
  protocluster. Furthermore, six out of the nine dense cores,
  from \citet{Fabien}, located in regions with \nhp emission,
  are situated within this filamentary structure. Notably, 
  this structure exhibits large-scale velocity distributions
  converging near the position of the core (L19 in Table~\ref{tab:core_parameters})
  where we observe the most prominent V-shape (see
  \S~\ref{sec:kinematic-analysis-at-small-scales}) along with three Straight-shapes associated    
  with dense cores. Furthermore, along
  this structure, we find some
  small-scale PV features that we suggest
  represent cores under the detection limit (see
  \S~\ref{sec:kinematic-analysis-at-small-scales}).  Most of the
  spectra in this structure are characterized by
  2-velocity-components.
\item F2: Adjacent to F1, the F2 structure is the next most obvious
  filamentary structure within G351.77. It connects to the F1 structure
  near the protocluster's center through regions with low integrated
  intensity, but it merges with the F1 structure towards the Galactic north of
  the protocluster. This structure appears as a twisted filament, less
  monolithically well-defined than the F1 and not converging to a specific location. 
  Nonetheless, two \citet{Fabien} dense 
  cores are situated inside F2, as well as $\sim$~50\% of the V- and
  Straight-shapes. Multiple spectra within this structure are
  characterized by 2-velocity-components.
\item F3: Located to the south of F1, F3 displays at least two
  clear velocity components along its length with a
  $\Delta$v reaching $\sim~3~\kms$. F3 contains one dense core (L13 in Table~\ref{tab:core_parameters}), coinciding with the position where the filament exhibits
  a large-scale V-shape in the PV diagram (see Fig.~\ref{fig:pvcolores}). Two of the core 
  candidates are located at the Galactic south of this structure.
\item F4: Situated in the northern region of the
  protocluster, where F1 and F2 merge, the F4 structure
  presents well defined V- and Straight-shapes without associated cores. This region is
  connected with the Mother Filament.
\end{itemize}

As a whole, these structures do not appear drastically different from
each other in terms of their internal kinematics; we
observe PV features in all of them.


\section{Discussion}\label{sec:discussion}

\subsection{G351.77 vs G353.41 protoclusters}\label{G351VSG353}

In \S~\ref{sec:kinematic-analysis-at-small-scales} we characterized
the PV features associated with dense cores as well as PV
features without cores, suggesting that the dense gas material 
traced by the \nhp~(1-0) line is inflowing in
filamentary structures into dense cores or in larger scales inflowing towards denser regions
\citep[see also][]{ Hacar2017, Lu, anualmotte, Rodrigo21, Kim, Arzoumanian, Yang, rodrigo24,
Pan}. The recent \nhp analysis in the G353.41 protocluster
\citep[][]{rodrigo24}, also observed by ALMA-IMF at matched physical
resolution to the observations presented here, shows similar kinematic
features although they use the \nhp~(1-0) isolated hyperfine component.  The
comparison to G351.41 is relevant beyond the similarities in the
analysis techniques and the matched sensitivity and spatial resolution
provided by ALMA-IMF. 
 
Both protoclusters have similar masses, sizes, distances
\citep[][]{motte22, Simon, Pierre}, and both exist inside isolated filamentary
structures \citep[see Fig.~\ref{fig:rgbimage} and][]{rodrigo24}.

Moreover, G351.77 and G353.41 have both been classified as existing in
an Intermediate evolutionary stage \citep{motte22}, based
predominantly on H41$\alpha$ emission \citep[see also][]{Roberto}.  In
finer detail, the larger number of dense cores and the smaller
number of PV features in G353.41 compared to G351.77 together suggest
that G351.77 is in a younger evolutionary stage relative to G353.41.
\citet{rodrigo24} characterized multiple V-shapes associated with dense
cores, which they suggest are inflowing material onto or near dense cores,
similar to our results.  Almost all of the V-shapes in G353.41 are
located inside filamentary structures; moreover, the $t_{\rm VG}$ and
\minflow are similar (within the same order of magnitude) as those
measured here for G351.77. This indicates that the V-shapes observed
in PV diagrams represent generic kinematic features in forming
protoclusters and can be used more broadly to access the conditions during the assembly of
stellar clusters and the so called ``initial conditions'' of star 
formation, especially for the kinematic attributes.
  
By averaging the VGs of the V-shapes associated {\it only} with the
dense cores (see Table~\ref{tab:meanvg}), we estimate a timescale
$\sim$~32.96~$\pm$~1.58~kyr. Conversely, by averaging the VGs of the
V-shapes not associated with the dense cores, we estimate a
timescales of $\sim$~43.12~$\pm$~1.10~kyr. When averaging all the VGs
associated with the V-shapes, we estimate a timescale of
$\sim$~41.33~$\pm$~0.95~kyr. We observe that the average timescale
estimated in G353.41 protocluster is $\sim$2$\times$ higher than the
value measured in G351.77.  This may also corroborate the above
suggestion that G351.77 is younger, whereby the cores in the 
protocluster (which are also more numerous) may be accumulating mass 
faster than in G353.41, despite the mass reservoir similarity in both 
protoclusters (see above). 

\subsection{G351.77 on small scales}\label{G351onsmallscales}

The longest timescale we measure for a V-shape
is 69.17~$\pm$~8.95~kyr (see Table~\ref{tab:meanvg}), whose time is lower than the lifetime of the 
prestellar cores (Valeille-Manet et al., submitted). Although we have 
not observed protostellar cores within the G351.77 protocluster, we 
expect that these PV features will disappear before the cores enter the protostellar 
phase, unless these structures are continuously fed with gas in some way. This could 
suggest that more evolved regions should display fewer 
V-shapes features in dense gas, which would be more closely associated with cores rather than 
prestellar and protostellar cores.   

Additionally, utilizing the mass and sizes of the dense cores from Table~D.11 in \citet{Fabien}, associated with the V-shapes, we measure the free-fall time (t$_{ff}$) for each dense core, whose average is $\sim$~13.6~kyr. This value is $\sim$~2.5 times lower than the average timescale for the V-shapes ($\sim$~33~kyr, see \S~\ref{G351VSG353}), suggesting that the dense cores may collapse gravitationally before the V-shapes dissipate. The accumulation of mass into the dense cores via the V-shapes would increases the dense core densities, further reducing t$_{ff}$ and accelerating the gravitational collapse. Consequently, the dense cores could collapse on shorter timescales, limiting the time available for sustained gas accretion from the V-shapes. This implies a dual role for the V-shapes: directly feeding the dense cores while potentially generating new dense regions along the filaments produced by local accumulation of gas. Further, collateral effects from dense core collapse, such as outflows, feedback, and concentration of magnetic fields, could significantly influence the dynamics of these structures. In an evolutionary context, these interactions could affect dense core evolution, and the properties of the stars formed within. The increase in dense core mass could impact the initial mass function (IMF), and the core mass function (CMF) measured in the protocluster \citep[see][]{Fabien}. These effects suggest that the relation between the V-shapes, the dense cores, and the larger filamentary structures regulates gas supply of gas, evolution timescales, and star formation efficiency in the protocluster.
 
Furthermore, we estimate the total mass accretion rate considering both the
V-shape structures associated and not associated with the dense
cores (see Table~\ref{tab:meanvg}). The total mass accretion rate measured is  
$\dot{\text{M}}_{\text{in},\text{T}}\,=\,(9.84\,\pm\,0.63)\,\times\,10^{-3}$~\msunyr.
Considering the total mass of the protocluster is $\sim$~2820~\msun, we can
estimate the depletion timescales given by
$t_{\rm dep}$~=~\mtot/$\dot{\text{M}}_{\text{in},\text{T}}$, obtaining a $t_{\rm dep}\,\sim$~0.3~Myr.



A $t_{\rm dep}$\,$\sim$~0.3~Myr for the V-shapes is consistent with the short lifetime of dense gas traced by \nhp, as described in the simulations of \citet{Priestley}, where \nhp is exclusively associated with dense gas typically linked to material that will collapse to form stars. This suggests that the V-shapes associated with dense cores could represent gas in stages of gravitational collapse. In contrast, the V-shapes without associated dense cores could be material in transit toward higher densities, eventually becoming linked to star formation. Hence, we suggest that the V-shapes may serve as tracers of early SFR and the evolutionary stage of star-forming regions \citep[see e.g.,][for the discussion on the efficacy of YSOs as SFR tracers]{Sami}.

Moreover, the fast $t_{\rm dep}$ measured could represent not only an ongoing star formation in early stages, but also the need for continuous mass feeding from the larger scales, such as the Mother Filament (see below). This scenario maybe consistent with the finding of \citet{Hacar2024}, where they identified dense fibers as structures moving material into forming star clusters. These fibers could continuously feed the protocluster, compensating the fast depletion of the local gas and enabling efficient star cluster formation. We caution that
without direct confirmation of inflow onto a central mass (the cores
in this case), the depletion timescales associated with the total
sample of all V-shapes may be overestimated.

\vspace{-0.3cm}

\subsection{G351.77 on large scales}\label{G351_at_large_scale}

Moving to somewhat larger scales, in
\S~\ref{sec:kinematics-analysis-at-large-scales} we divided the \nhp~(1-0)
emission of the G351.77 protocluster into 4 larger ($\sim$~0.2~pc to
$\sim$~1.2~pc) spatially distinguishable regions or filaments inside
the protocluster. In particular, we show that the F1 filamentary structure
seems to be
inflowing toward a dense core, where we observe the most
prominent V-shape in the PV diagram (see
Fig.~\ref{fig:pv_diagram1} and Fig.~\ref{fig:pvcolores}). 
Given that F1 filamentary structure is
connected with the F4 structure, the region which is directly
connected with the Mother Filament, and its velocity distribution
observed in PPV diagrams, we suggest that the inflowing gas
is coming from the Mother Filament, flowing towards the
protocluster \citep[e.g.][]{Hacar2024}. Furthermore, we find dense cores associated with
Straight-shapes in the G351.77 protocluster, located along of the F1
filament structure (see Fig.~\ref{fig:pv_diagram1}), near the position where we observe the most
prominent V-shape. Considering the velocity distribution of \nhp
displayed in the PV diagrams of Straight-shapes, their location, and
the consistency of the \nhp~(1-0) velocities with those measured for the
cores (see Table F.9 in \citealt{Nichol}), we propose that these cores
are inflowing, along with the \nhp, towards the location of the most
prominent V-shape, potentially entrained in the slow contraction of the
protocluster \citep[][]{Lu, anualmotte, Kim, Arzoumanian, Yang, Pan}. All this,
may suggests that the F1 structure is actively forming cores or stars.

In contrast to F1, the F2 filamentary structure does not show evidence
of inflowing gas towards a specific location but rather presents
inflow towards different locations along the structure. Moreover, most
of the PV features on small scales do not have associated detected dense cores, 
which we propose as cores below the 1.3~mm band detection limit, located within
the filament. This may suggest that this filamentary structure is participating in the 
formation of cores but at an earlier stage than the F1 structure. Additionally, hot cores are located in regions with \nhp emission with S/N~$<$~9, limiting the kinematic information available from them. Regarding the detected SiO outflows \citep[see][]{towner}, we identify four in F1 and two in F2, but none are associated with well-defined structures in the PV-diagrams. Instead, these outflows are located at the edges of the \nhp emission rather than the internal regions of the filaments. This spatial distribution suggest that ongoing star formation activity may not be occurring within the densest regions of the filaments but rather in their outskirts. Furthermore, the more well-defined morphology of F1 compared to F2 supports the idea that these two exist at different evolutionary stages (see also \citet{Nichol} for the proposal of multiple stages inside a given ALMA-IMF protocluster).

The F3 filamentary structure shows little evidence of star
formation activity within it, where we find one dense core, and two
PV features at small scales without associated dense cores. The
PV features and the large-scale velocity distribution of this filamentary 
structure in PPV diagrams, seems to indicate that the gas in
this is inflowing towards the core location.  

The F4 structure presents velocity distributions in the PPV diagram (see
Fig.~\ref{fig:pv_diagram1} and Fig.~\ref{fig:pvcolores}) that supports the
idea that the inflowing gas towards the denser \nhp regions of the
protocluster originate from the Mother Filament.  Additionally, this
region may also be a site where cores or stars could be forming, 
given the presence of PPV features on small scales that do not have detected dense cores associated.

\subsection{G351.77 in a Global Hierarchical Collapse scenario}\label{G351inGHC}

The evidence that we have at small and large scales is consistent with the Global Hierarchical Collapse (GHC) scenario described in \citet{Enrique}. This scenario explains the collapse process occurring on different scales in molecular clouds, where mass and gas flow and accrete through filamentary structures toward clumps and denser regions. It also describes how the timescales of accretion and contraction are shorter on smaller scales. These timescales play an important role in the star formation rate (SFR), with longer timescales allowing most of the mass to be dispersed due to feedback from massive stars, which later results in a low SFR.  

Thus, by comparing the GHC model with our analysis, we suggest that the Mother Filament is undergoing global collapse, with a portion of it accreting and contracting toward the G351.77 protocluster. This implies that the protocluster will increase its mass over time. Although there is evidence of YSOs in the central part \citep{Leurini2011_1, Naiping2018, Leurini2019, Sabatini2019}, the H41$\alpha$ emission is compact in the central region of the protocluster \citep{motte22, Roberto}, where there is no \nhp emission (S/N~$<$~9). This suggests that currently there is insufficient feedback to significantly reduce the SFR in the protocluster \citep{Enrique}. This indicates that the low SFR measured in G351.77 and along the Mother Filament \citep[see][]{Simon} may be due to the fact that the molecular cloud and the protocluster are still in a young stage, consistent with our suggestions and those in \citet{Simon}. Consequently, the SFR is expected to increase with the formation of more massive stars until feedback from these new stars disperses the gas, thereby decreasing the SFR \citep{Enrique}. 

Given that we suggest the G351.77 protocluster is younger than G353.41, and based on the GHC model \citep{Enrique}, we could anticipate a higher SFR in G353.41. This expectation aligns with the H41$\alpha$ emission in G353.41, which is higher than in G351.77 \citep[see Table 4 in][]{motte22}, indicating a larger amount of massive star feedback in the G353.41 protocluster.  However, a larger number of hot cores are observed in G351.77 compared to G353.41 \citep{Melisse}, presenting a scenario that could contradict the notion of G351.77 being younger, as we might expect to find more hot cores in more evolved regions. Nevertheless, considering the various lines of evidence suggesting that G351.77 in indeed younger, we propose that some as of now unknown process may be inhibiting the formation of hot cores in G353.41 protocluster.

\subsection{Multi-tracer modeling}\label{sec:multitracer_model}

\begin{figure*}[h]
\includegraphics[width = 0.98\textwidth]{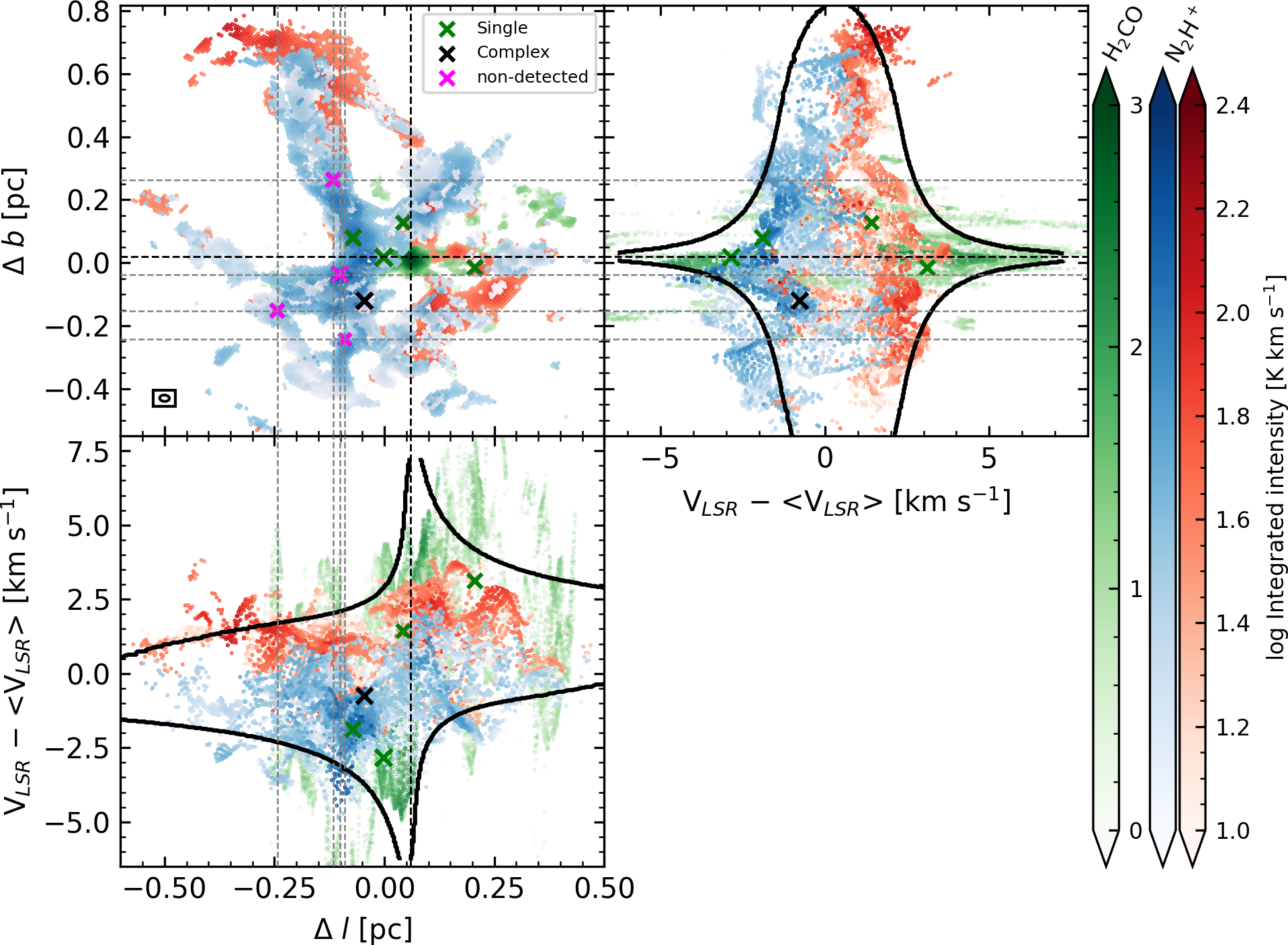}
\caption{Integrated intensity and position-velocity (PV) diagrams of 
  the Blue- (blue colorbar) and Red- (red colorbar) velocity
  components seen in \nhp~(1-0), and of the \hco~(2-1) emission (green colorbar). 
  Top left: Spatial distribution of \nhp and \hco emission in G351.77. 
  Markers and dashed gray lines are the same as in Fig.~\ref{fig:pv_diagram1}.
  The intersection of the dashed black lines represent 
  the center of the rotating and infalling modeled sphere 
  (see \S~\autoref{sec:model}). The ellipse in the bottom-left represents the 
  beam size of the \nhp data. Top right and bottom left: PV diagrams along 
  the two perpendicular directions. The black contours represent the 
  shapes shown by the rotating and infalling modeled sphere in the 
  PV diagram.}
  \vspace{-0.3cm}
\label{fig:tracers_and_models}
\end{figure*}

In order to complement and enhance our understanding of the kinematic
processes occurring in the protocluster at large scales, we utilize
\hco~(2-1) and DCN~(3-2) line emission covering the central part of 
the protocluster without \nhp emission \citep{motte22, Nichol}.
Additionally, we modeled three possible cases for gas
spheres: rotation-only, infall-only, and a combination between
rotation and infall (see \S~\autoref{sec:model}), whose radial
velocities are used to recreate PV diagrams \citep[e.g.,
][]{Tobin, Henshaw, Mori, rodrigo24}. This approach allows us to
analyze the PV features produced by these kinematic processes in the
PV diagrams (see Fig.~\ref{fig:tracers_and_models})

The modeled sphere is formed by \textit{n} numbers of points, which 
are under gravitational infall and keplerian rotation. The velocity
associated to each point is estimated by using the following relations: 
\begin{equation}
V_{\textup{in}} = - \sqrt{\frac{2GM}{r}}, \quad V_{\textup{rot}} = \sqrt{\frac{GM}{r}}
\label{eq:model}
\end{equation}
\noindent where $r$ is the distance of each point to the center of the sphere,
$G$ is the gravitational constant and $M$ is the enclosed mass. The minimum distance
between each point is of 0.007 pc, in order to simulate the pixel width of our data. 
Additionally, the sphere has a density profile that follows a power law described below:
\begin{equation}
\rho(r) = \rho_0\left(\frac{r}{\textup{pc}}\right)^{-\gamma} 
\label{eq:model}
\end{equation}
 
\noindent where $r$ is the distance to the center of the sphere, and $\rho_0$ is a known density. 
To generate an infalling and rotating sphere model, we utilize a $r_{\rm min} = 0.007$pc
and a $r_{\rm max}$ = 0.8~pc, which correspond approximately to the coverage radius of \nhp. 
Also, we define $\gamma$~=~3.1 based on the maximum velocities that the data reach and in
how well are the model fits to the data. The total mass utilized is 270~\msun and 2700~\msun, 
deriving $\rho_0$~=~3.466~\msun~pc$^{-3}$ and $\rho_0$~=~34.66~\msun~pc$^{-3}$, 
respectively. Although the general shape of the PV features of the modeled sphere in PV diagrams is not
affected by the mass, the velocity amplitude does show differences.

Fig.~\ref{fig:tracers_and_models} shows the PV diagram of \nhp~(1-0) and
\hco~(2-1) in comparison with the contours of a rotating and infalling
sphere (see \S~\autoref{sec:model}). We observe that \hco reaches
$\Delta$v~$\sim$~13~$\kms$, with its velocity peaks in the PV diagram
aligning well with the contours of the model. Furthermore, upon
examining the velocity distribution of DCN in the PV diagram, we
observe similar PV features with \hco (see Fig.~\ref{fig:dcn_in+rot} and 
Fig.~\ref{fig:h2co_in+rot}), which appear to match reasonably with
our model of the rotating and infalling sphere and similar models
\citep[][]{Tobin, Mori}.  Based on the spatial locations of \hco and
DCN, along with the PV features in the PV diagrams described by our
model, we suggest that rotation and infall processes are occurring in
the center of the protocluster within a radius of $\sim$~0.2~pc, similar 
scales as e.g., \citet{roberto2009} and \citet{liu2015}, which is
five times smaller than the \nhp spatial coverage.

\begin{figure}[h]
\centering
\includegraphics[width = \columnwidth]{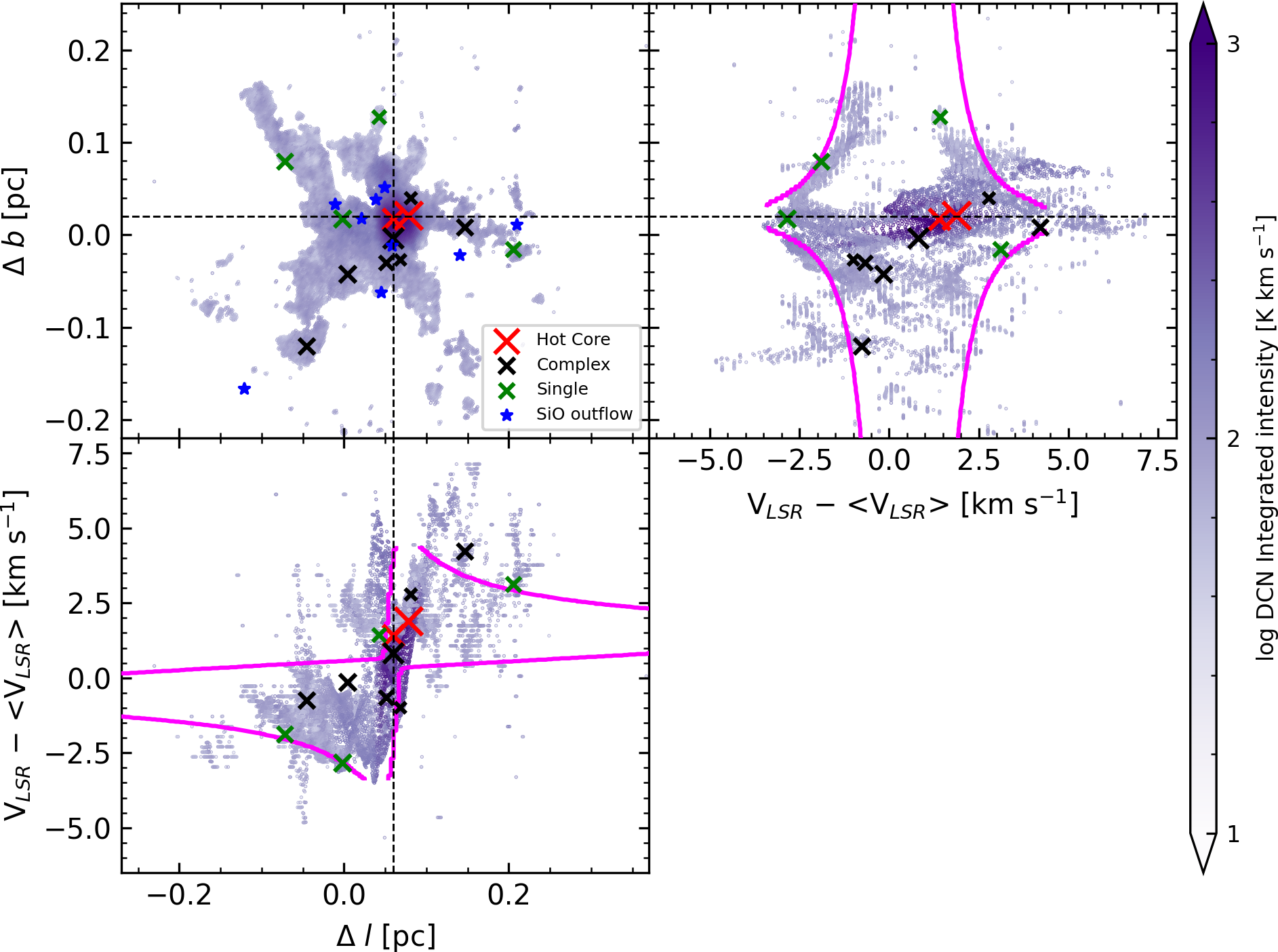}
\caption{Integrated intensity and PV diagrams of 
  the DCN~(3-2) emission. Top left: Spatial distribution of DCN emission in G351.77. 
  The green, black, and red $\times$ markers indicate the position of
  the dense cores \citep{Fabien}, where green and black 
  represent the \citet{Nichol} DCN spectral classification: single, 
  and complex, respectively, and red indicates hot cores \citep[also complex][]{Melisse}.
  The blue stars indicate the position of the cataloged SiO outflows in G351.77, which are
  not associated to observed PV-features neither to range of velocities of DCN \citep{towner}.  
  The intersection of the dashed black curves represent 
  the center of the rotating modeled sphere 
  (see \S~\autoref{sec:model}). Top right and bottom left: PV diagrams along 
  the two perpendicular directions. The magenta contours represent the 
  PV features produced by a rotating modeled sphere in the 
  PV diagram.}
  \vspace{-0.3cm}
\label{fig:dcn_rot}
\end{figure}

This scenario is consistent with the difference in velocity measured
in \nhp emission in the G351.77 protocluster, which suggests that the
two clear \nhp velocity structures observed in
Fig.~\ref{fig:pv_diagram1} are rotating relative to each other with a radial velocity
gradient that is approximately perpendicular to the direction toward
the Mother Filament ($\sim$~83~$^{\circ}$), which extends above the top of
Fig.~\ref{fig:pv_diagram1} (see also Fig.~\ref{fig:velocity_map}). However,
the inflow signature in the radial velocities seems more dominant than
rotation for \nhp on large scales. On the other hand, the
models of rotation-only and infall-only show that the rotating-only
case better fits the shapes observed in \hco and DCN, suggesting that
the central part of the protocluster is more dominated by rotation
than infall (see Fig.~\ref{fig:dcn_rot}, Fig.~\ref{fig:dcn_in}, Fig.~\ref{fig:h2co_rot},
and  Fig.~\ref{fig:h2co_in}). This scenario is
consistent with that proposed in e.g. \citet{anualmotte} for high-mass 
star-forming regions. 

Given that the rotation axis appears parallel to Galactic latitude, in other
words, the velocity gradient points along the Galactic longitude, we may hypothesize 
that the Galactic shear is in some way inducing the rotation in the 
G351.77 protocluster \citep{Braine2018, Braine2020}. This scenario would then imply that the initial angular momentum is set externally, and drains from the outside inward (see also \citet{Rodrigo21} filament rotation scenario and Salinas et al., in prep)

This interplay between infalling and rotating gas might both favor and
disfavor the protocluster star forming activity at different
scales. Despite infall dominating at the larger protocluster scales
(traced by \nhp), in the very center of the protocluster the $-$ now
dominant $-$ rotation might be reducing the rate at which the
protocluster fuels its dense material. This cannot be preventing the
cloud from forming a protocluster, as evidenced by the fact that the
protocluster is there and obviously forming stars.  Yet the Mother
Filament, so on scales beyond the ALMA-IMF coverage, exhibits evidence
for inefficient star forming activity compared to the large mass and
mass-per-unit length profile \citep{Simon}. In this scenario, our
results may be providing an insight to explain this dichotomy observed
on the global cloud scale, because they suggest that rotation of dense
filamentary sub-structures at $\sim$~0.8~pc scales may be ubiquitous
along the entire Mother Filament, but being significantly more
dominant than infall outside the protocluster itself.  Indeed,
correlations between the lack of evidence for star formation and
velocity gradients consistent with rotation have been observed
previously \citep{Rodrigo21}, where they also inferred outside-in
filament evolution as the system sheds its angular momentum and
evolves toward more efficient star formation.  The transition between
these two regimes could be rooted in different angular momentum loss
rates across the entire cloud, potentially driven or influenced by the
magnetic field \citep{amy2016, amy2018}.  Meanwhile, independent of the possible effects of
magnetic fields, the kinematics in the Mother Filament will be
characterized in upcoming work, which will test the rotation
hypothesis.  

In the above modeling, the sphere used to obtain the models employs a
mass of 270~\msun, which is 10 times lower than the 2720~\msun
measured inside a radius of 0.8~pc in G351.77, in the N(\hdos) map
provided by \citet{Pierre}. The larger mass assumption causes our models to exhibit larger velocities than
observed. Specifically, the model would predict velocity spreads of
order $\Delta$v~$\sim$~44~$\kms$, whereas the spreads that we observe 
are $\Delta$v~$\sim$~14~$\kms$. That is, the velocities in the model 
are overestimated by a factor of $\sim$\,3. However, the shapes of the 
model PV diagram compared to the observed one are still very similar, 
which suggests that the qualitative combined infall and rotation 
signature exhibited by the model reproduces the observed PV diagram 
structure \citep[see e.g.,][for models characterizing similar kinematic signatures]{Arzoumanian, Kim, Lu}.

The velocities measured for \nhp, \hco, and DCN seem to be similar,
and the observed PV features of \hco and DCN
seem to describe the same kinematic processes. All of this indicates
that the measurements are reliable. However, in a global analysis the modeled sphere
suggests higher velocities than the observed one given the mass
budget, raising questions such as: Why do we measure lower velocities
in the protocluster? What other physical processes are affecting the
kinematics of the gas? 

The total mass derived from the dense cores located at the center of the protocluster (see Fig.~\ref{fig:dcn_rot}) is $\sim$~100~\msun \citep{Fabien}. Furthermore, using average relative abundances, where X(DCN)~$=2.06~\times~10^{-10}$ and X(H$_2$CO)~$=4.66~\times~10^{-10}$ \citep{Maret, Minh}, and the N(H$_2$) from \citet{Pierre}, we estimate the DCN and H$_2$CO masses to be M(DCN)~$=1.89~\times~10^{-6}$~\msun and M(H$_2$CO)~$=4.56~\times~10^{-6}$~\msun. Given their extremely low masses compared to the total enclosed mass, these components do not significantly contribute to the measured mass of 270~\msun. Nonetheless, they may still play a role in shaping the gravitational potential influencing the observed kinematics. However, Fig.~\ref{fig:dcn_rot} shows possible relations between the positions and velocities of dense cores (including hot cores) and features observed in the PV-diagram of DCN emission. Furthermore, current evidence demonstrates that processes
such as infall and rotation are occurring at smaller scales ($\sim$
0.06\arcsec) in cores located at the central part of the protocluster
\citep[e.g.][]{Beuther}. Additionally, outflows have been cataloged in G351.77 \citep{towner}, whose positions and velocity-ranges not show relations with PV-features (see Fig.~\ref{fig:dcn_rot}). Moreover, the
fragmentation into clumps observed along the Mother Filament has been
attributed to turbulence and magnetic fields produced by the high-mass
star formation \citep{Zapata2008, Leurini2009, Leurini2011_1, Leurini2011_2, Leurini2013, 
Naiping2018, Leurini2019, Sabatini2019}. Considering all these physical processes, we could
suggest that the kinematic behavior of the gas traced by \nhp, \hco,
and DCN may not be as described in our model, which is simple and only considers two
kinematic processes.  However, in broad scopes, the model is consistent
with a comparatively slow gravitational contraction of the
protocluster with inner rotation.  


\section{Conclusions}\label{sec:conclusion}

We conducted an analysis of the kinematics of the dense and cold gas
within the massive G351.77 protocluster, traced by the spectral line
\nhp~(1-0), utilizing the PV diagrams and characterization of physical
processes involved in high-mass star formation. The spatial resolution
of the \nhp data cube is $\sim$~4~kau, making this the first study
of this kind in G351.77.

Through spectral line fitting we obtain the main accessible physical
parameters such as excitation temperature, optical depth, centroid
velocity, and line widths (see \S~\ref{sec:line-fitting-process}). We
have implemented the spectral line fitting over the entire data-cube,
fitting 1- and 2-velocity-components (see
\S~\ref{sec:one-component-fit} and \S~\ref{sec:two-component-fit}). We
developed a method to identify and extract each spectra with 1- or
2-velocity-components based on the line width and S/N of the modeled
spectra (see \S~\ref{sec:best-fit-and-final-model}). We finally
obtained modeled data that possess spectra with 1- or
2-velocity-components. Based on the centroid velocity distribution of
each spectrum, we divided and isolated the data into 2 velocity
components, which we named the Blue-velocity component and
Red-velocity component (see
\S~\ref{sec:bluest-and-reddest-velocity-component}).

The excitation temperature, optical depth, and line widths enable the
estimation of column density maps and masses of \nhp (see
\S~\ref{sec:column-density-and-masses}). Additionally, we utilize the
column density map of \hdos provided by \citet{Pierre}, to estimate
the relative abundance of \nhp, X(\nhp) $\sim$ (1.66 $\pm$ 0.46)
$\times$ 10$^{-10}$, from which we derived an estimated total mass in
the \nthp emitting gas in the protocluster of $\sim$ 1660 $\pm$ 326
\msun.

The characterization of kinematic patterns, our main goal in this
paper, revealed by the scrutiny of both the PV and PPV diagrams cubes
allowed us to identify different small and large-scale features (see
\S~\ref{sec:kinematics-analysis}). Our main results are:
\begin{enumerate}
 
\item The PV diagram on small scales reveals features such as V- and
  Straight-shapes associated in position and in velocity with the
  dense cores \citep[][]{Nichol, Fabien}.  We suggest that the
  V-shapes describe inflowing gas in filamentary structures, where the
  dense cores are located near the vertex of the V-shapes. On the other
  hand, the Straight-shapes describe inflowing filamentary structures
  going to denser \nhp regions with dense cores following the stream
  produced by the gas (see Fig.~\ref{fig:cartoon1}).

 \item We characterize these  PV features by measuring the VGs, timescales, mass, and \minflow.
  The most prominent V-shape reaches velocities $\sim$ -4.5 $\kms$
  relative to the V$_{\rm LSR}$, has an average \minflow $\sim$\,13.45
  $\times$\,10$^{-4}$~\msunyr and timescale $\sim$\,11.42~kyr (see
  \S~\ref{sec:kinematic-analysis-at-small-scales}).
 
\item The above V-shapes have similar properties as those analyzed in
  \citet{rodrigo24} in G353.41 indicating that they may be ubiquitous
  in this stage of protocluster evolution.
 
\item Similarly, we also find V- and Straight-shapes without
  associated dense cores; these have similar properties (timescales
  and \minflow) as the above sample.  We suggest that these arise
  from possible cores or accretion centers that are under the
  detection limit of the 1.3~mm data (see
  \S~\ref{sec:kinematic-analysis-at-small-scales} and
  Table~\ref{tab:VGcores}).
  
\item The averaged t$_{ff}$ of the dense cores is shorter than the timescales of the V-shapes. This suggest mass of cores is increasing over their lifetimes, in turn modifying the t$_{ff}$ itself and likely driving faster dense core collapse.
 
\item The estimated timescales for the V-shapes is shorter than the
  timescale estimates for protostellar core envelope evolution
  (Valeille-Manet et al., submitted and references therein). This
  suggests that the V-shapes are short-lived accretion signatures,
  that are more easily detectable in the most intense mass buildup
  phase of core formation.  Hence we may expect fewer relative numbers
  of V-shapes as star-forming regions evolve (see
  \S~\ref{sec:kinematic-analysis-at-small-scales}).

\item The total V-shape \minflow (summed over the sample) enables us
  to estimate a depletion timescale of $\sim$~0.3~Myr in the
  protocluster. This implies a fast cluster formation timescale and/or
  a continuous mass feeding (see
  \S~\ref{sec:kinematic-analysis-at-small-scales}).
 
\item The PV diagram of \nhp (on the largest scale we study of
  $\sim$~1.2~pc) reveals two main filamentary structures within the
  protocluster (F1 and F2 in Fig.~\ref{fig:pvcolores}). These
  filamentary structures are actively participating in the dense core and
  star formation, but at different levels, so we infer that they are
  at different evolutionary stages (see
  \S~\ref{sec:kinematics-analysis-at-large-scales}).  This is
  consistent with previous findings of indications of different
  episodes of star formation in the ALMA-IMF protoclusters
  \citep[][]{Nichol}.
 
\item The analysis of large-scale VGs reveals that one of the two main
  filamentary structures (F1 in Fig.~\ref{fig:pvcolores}) maybe
  inflowing toward a dense core position, where we observe the most
  prominent V-shape in the PV diagram at small scales (see above and
  Fig.~\ref{fig:core6_2}). Within this filamentary structure we find
  dense cores and PV features that are not associated with detected cores
  (see above).  These seem to be inflowing along the filament (see
  \S~\ref{sec:kinematics-analysis-at-large-scales}).

\item The large-scale PV and PPV features seem to indicate that the
  protocluster is being fed from the the Mother Filament (see
  \S~\ref{sec:kinematics-analysis-at-large-scales}), which the
  protocluster is forming out of.

\end{enumerate}  

The kinematic evidence observed in the dense gas, along with its
characterization, suggests that G351.77 is actively and vigorously
forming dense cores. However, based on the comparison with G354.41
protocluster \citep{rodrigo24}, the low SFR measured along the
filament \citep{Simon}, and its consistency with the Global
Hierarchical Collapse scenario \citep{Enrique}, we conclude that
G351.77 protocluster is still in an early evolutionary stage. As the
protocluster continues to be fed by the Mother Filament, we might
expect to observe an increase in the number of cores, V-shaped
features associated with them, and filamentary structures evolving
into well-defined filaments, along with a rise in the SFR.

We have complemented the PV diagram analysis with the \hco~(2-1) and
DCN~(3-2) tracers, which cover areas without \nhp~(1-0) emission. This
allows us to explore the relation between the large-scale kinematics
described by \nhp and the central region of the protocluster. In
addition, we have implemented a rotating and infalling sphere model in
order to characterize the PV features in the PV diagram produced by
these kinematic processes, which is consistent with previous work
\citep[e.g.,][]{Tobin, Henshaw, Mori}.

Here, in brief, we find that our simple toy model is consistent with
larger scale slow contraction and inner rotation, implying outside-in
evolution as the initial angular momentum, possibly induced by Galactic
shear \citep[]{Braine2018, Braine2020}, is dissipated \citep[see also][]{Rodrigo21}.
The PV features in the PV diagram produced by \hco and DCN suggest
that the central region of the protocluster may be experiencing
simultaneous rotation and inflow, where the rotation appears to be the
most dominant kinematic effect in the center (see
\S~\autoref{sec:multitracer_model}). On the other hand, considering the velocity
distribution and the position-position of the 2 large-scale velocity
structures of \nhp separated by $\sim$\,2\,$\kms$ in the PV diagram,
in addition to the possible rotation features described by \hco and
DCN, we suggest that these 2 large-scale velocity structures could be
rotating relative to each other, with the rotating axis parallel to
the direction of the Mother Filament. However, the most prominent
kinematic process present at the \nhp filamentary structures appears
to be inflow, which seem to be inflowing not towards the center of the
protocluster but rather inflowing onto themselves towards denser \nhp
regions (see Fig.~\ref{fig:tracers_and_models}).

Nevertheless, while the shapes and features presented in the model are
consistent with those observed in the PV diagram by the tracers, the
measured velocities exhibit inconsistency with the velocities returned
by the models by a factor of 3 (see \S~\ref{sec:multitracer_model}). We attribute these differences to the
simplicity of the toy model, which only considers two kinematic
effects.  At the same time, other works suggest the presence of
rotation and infall at small scales, bipolar outflows produced by
YSOs, gravitational instabilities along the filament, and magnetic
fields in the protocluster, whose physical effect could be strongly
affecting the behavior of the gas.

We have carried out a thorough analysis of the \nthp line emission.
One absolute imperative arising from this work is to ascertain the
relationship of the protocluster to the Mother Filament \citep{Simon}
and the kinematic state of this filament as a whole.
This work will be carried out in the near future.

\begin{acknowledgements}
We thank the referee for carefully comments, which helped improve this paper. 
This paper makes use of the following ALMA data: ADS/JAO.ALMA\#2017.1.01355.L, \#2013.1.01365.S, and \#2015.1.01273.S. ALMA is a partnership of ESO (representing its member states), NSF (USA) and NINS (Japan), together with NRC (Canada), MOST and ASIAA (Taiwan), and KASI (Republic of Korea), in cooperation with the Republic of Chile. The Joint ALMA Observatory is operated by ESO, AUI/NRAO and NAOJ.
The project leading to this publication has received support from ORP, which is funded by the European Union's Horizon 2020 research and innovation program under grant agreement No. 101004719 [ORP].
This project has received funding from the European Research Council (ERC) via the ERC Synergy Grant \textsl{ECOGAL} (grant 855130) and from the French Agence Nationale de la Recherche (ANR) through the project \textsl{COSMHIC} (ANR-20-CE31-0009).
A.S. gratefully acknowledges support by the Fondecyt Regular (project code 1220610), and ANID BASAL project FB210003.
R.A. gratefully acknowledges support from ANID Beca Doctorado Nacional 21200897.
R.G.M. acknowledges support from UNAM-PAPIIT project IN108822 and from CONAHCyT Ciencia de Frontera project ID 86372.
Part of this work was performed at the high-performance computers at IRyA-UNAM.
T.Cs. and M.B. have received financial support from the French State in the framework of the IdEx Universit\'e de Bordeaux Investments for the future Program.
A.G. acknowledges support from the NSF under grants AST 2008101 and CAREER 2142300. 
M.B. is a postdoctoral fellow in the University of Virginia’s VICO collaboration and is funded by grants from the NASA Astrophysics Theory Program (grant num- ber 80NSSC18K0558) and the NSF Astronomy \& Astrophysics program (grant number 2206516).
P.S. was partially supported by a Grant-in-Aid for Scientific Research (KAKENHI Number JP22H01271 and JP23H01221) of JSPS.
R.A. gratefully acknowledges support from ANID Beca Doctorado Nacional 21200897.
A.K. and L.B. gratefully acknowledges support from ANID BASAL project FB210003.
S.R. acknowledges the funding and support of ANID-Subdirección de
Capital Humano Magíster/Nacional/2021-22211000.
G.B. acknowledge support from the PID2020-117710GB-I00 grant funded by MCIN/AEI(10.13039/501100011033
A.G. acknowledges the support of the Programme National \lq Physique et Chimie du Milieu Interstellaire' (PCMI) of CNRS/INSU with INC/INP co-funded by CEA and CNES.
H.-L. Liu is supported by Yunnan Fundamental Research Project (grant No. 202301AT070118, 202401AS070121).
This work is supported by the China-Chile Joint Research Fund (CCJRF No. 2312). CCJRF is provided by Chinese Academy of Sciences South America Center for Astronomy (CASSACA) and established by National Astronomical Observatories, Chinese Academy of Sciences (NAOC) and Chilean Astronomy Society (SOCHIAS) to support China-Chile collaborations in astronomy. P. García is sponsored by the Chinese Academy of Sciences (CAS), through a grant to the CAS South America Center for Astronomy (CASSACA).
\end{acknowledgements}


\bibliographystyle{aa}
\bibliography{ref.bib} 

\appendix

\section{\nhp raw data}\label{sec:rawdata}

\begin{figure*}[h]
\centering
\includegraphics[width = \textwidth]{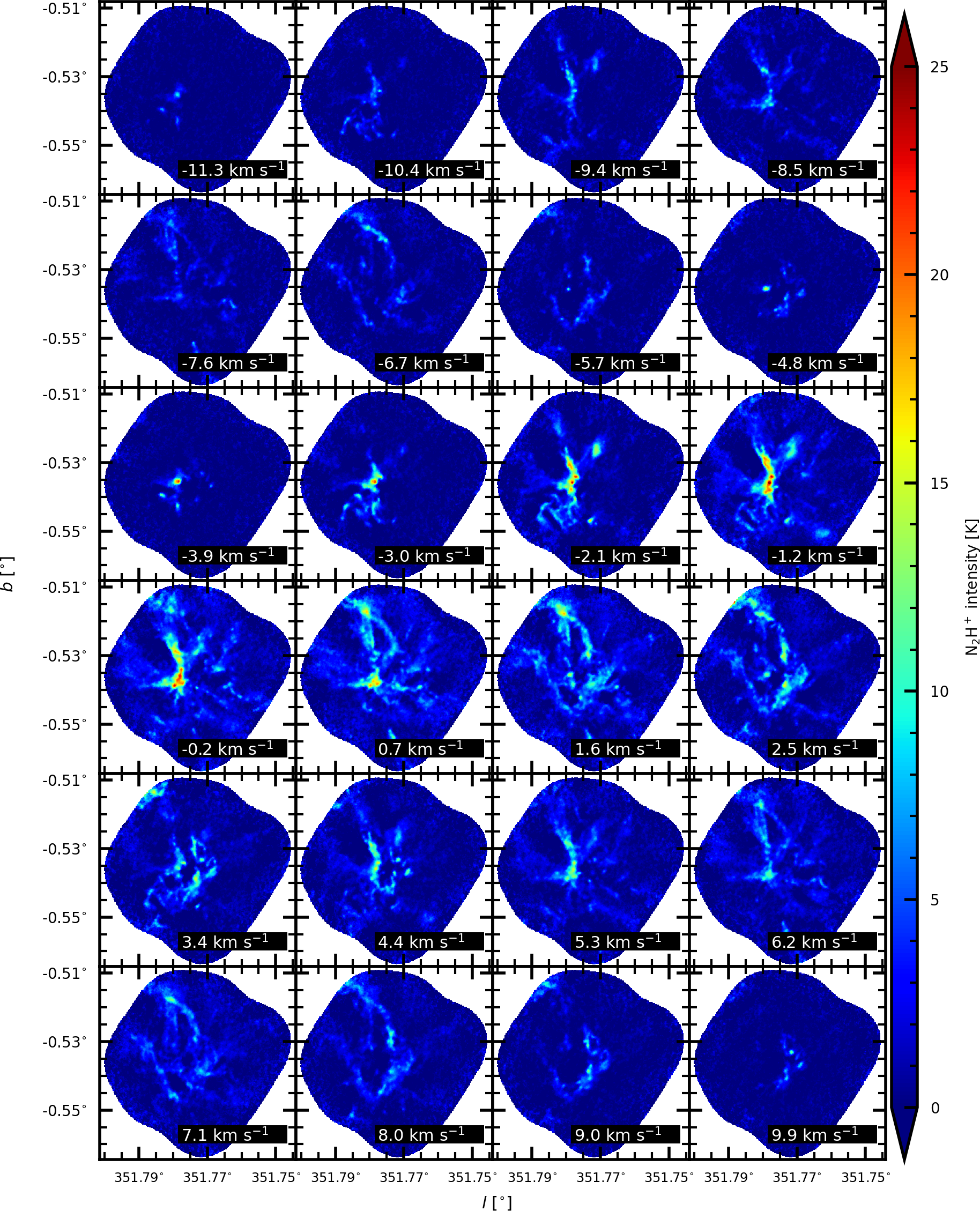}
\caption{Channels map of data-cube of the \nhp~(1-0) spectral line emission. There is a delta of velocity of 0.92~$\kms$ between each channel. The velocities of each panel correspond to V$_{\rm LSR}$ - <V$_{\rm LSR}$>.}
\label{fig:channelmap}
\end{figure*}

Figure~\ref{fig:channelmap} displays different emission channels of the \linebreak \nhp~(1-0) spectral line through the data-cube observed with ALMA (see \S~\ref{sec:data}). Out of the 215 channels of the data-cube, there are 110 that correspond to emission channels, which goes from -13.35 to 11.72~$\kms$, with a spectral line resolution of 0.23~$\kms$. 

\section{PySpecKit experiment}\label{sec:experiment}

\begin{figure}[h]
	\centering
	\includegraphics[width= \columnwidth]{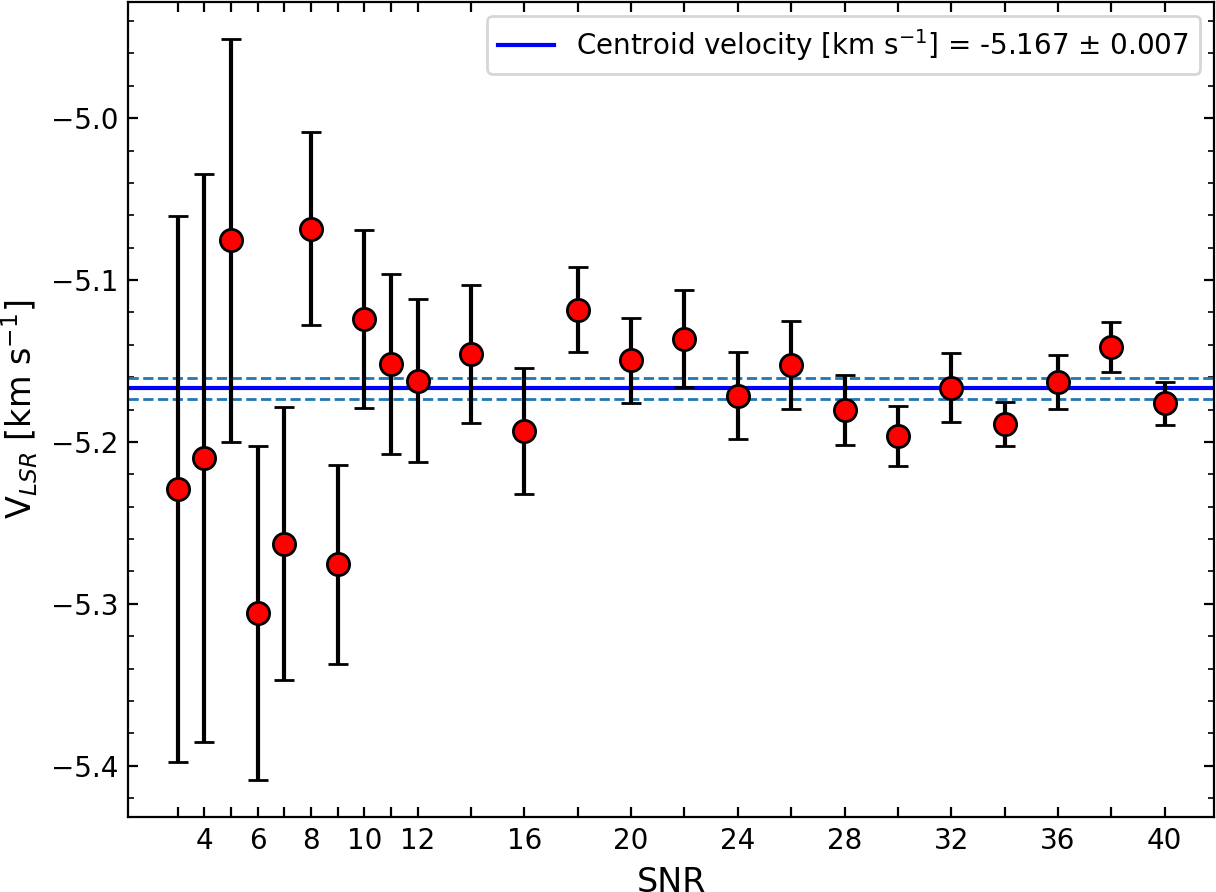}
	\caption{Example of the experiment fitting 1-velocity-component. 
	The red points represent the centroid velocity value of the 
	output models. The blue solid line shows the centroid velocity of the 
	input model, while the blue dashed lines represent the range of the 
	associated error. The error values increase as the S/N decreases. The 
	discrepancy between the centroid velocity values of the output models 
	and the input model increases as the S/N decreases, as we expected.}
\label{fig:experiment1}
\end{figure}

\begin{figure}[h]
	\centering
	\includegraphics[width= \columnwidth]{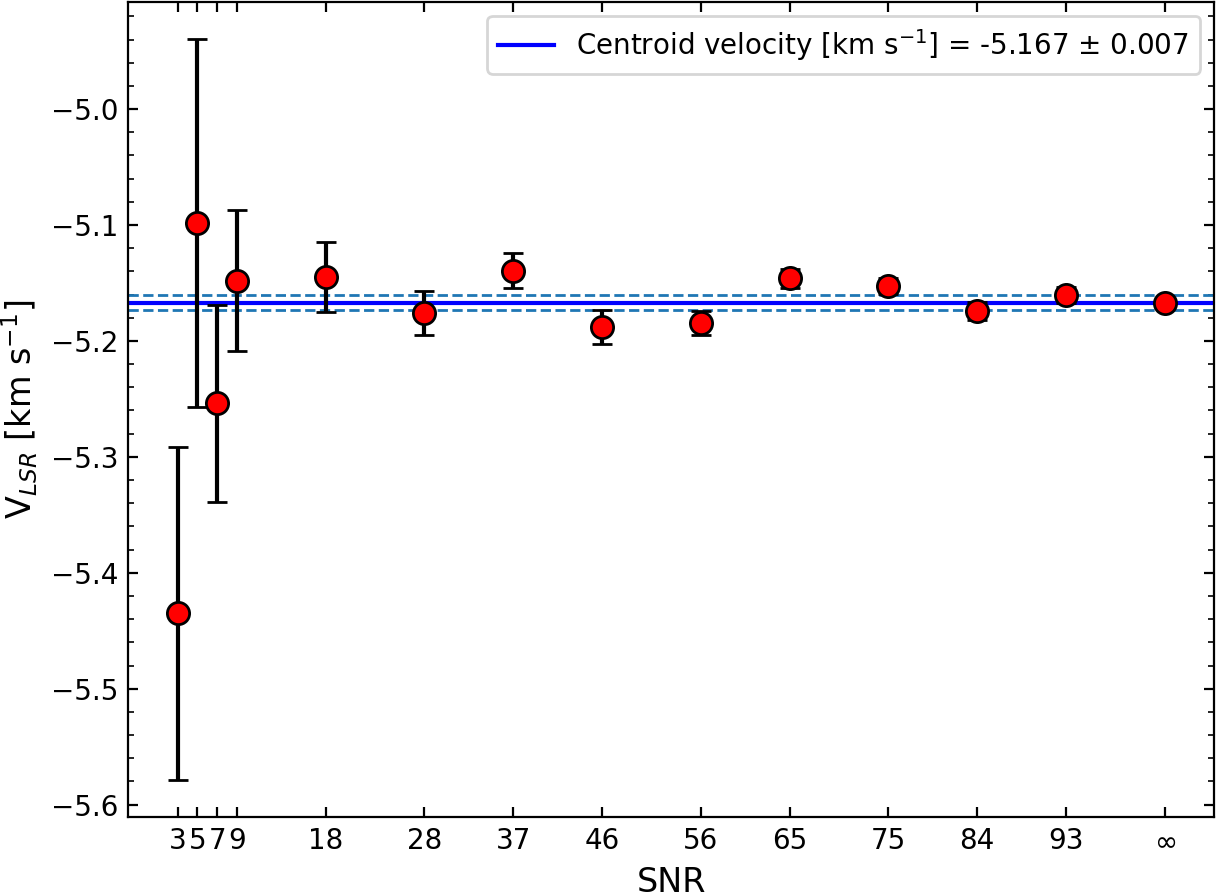}
	\caption{Same caption as Fig.~\ref{fig:experiment1} but for 2-velocity-component
	fit.}
\label{fig:experiment2}
\end{figure}

To evaluate the performance of PySpecKit and determine the S/N threshold 
at which it provides reliable results, we conducted the following test: 1) We select a single spectrum from the data-cube, from regions with S/N > 20, to fit it employing 
PySpecKit (with defined guesses and limits), obtaining a model and its 
respective parameters (input model). 2) We create a grid using this 
model, assigning a different S/N to each spectrum in the grid, 
artificially adding noise. 3) We fit the grid with PySpecKit (using the same guesses and limits as before), 
obtaining a new model for each spectrum (output models). By comparing parameters such as excitation temperature (T$_{\rm ex}$), optical 
depth ($\tau$), centroid velocity (V$_{\rm LSR}$), and line width ($\sigma_{\rm v}$)
returned from the single spectrum (input model) versus 
the parameters returned from the grid (output model), we could assess 
the consistency of PySpecKit's performance. This process was repeated 
for different spectra from the data-cube, fitting 1- and 
2-velocity-components, conducting more than 200 different tests. 
The results of this experiment are consistent 
for both 1- and 2-velocity-component fits, showing that  PySpecKit 
returns reliable values for S/N > 9. In contrast, for spectra with S/N $<$ 9, 
the difference between the parameters from the input model and output models increases significantly, 
reaching values 3 times or more the standard deviation. Additionally, 
the uncertainties for each parameter rise to over the 30\%, as expected. 
The Fig.~\ref{fig:experiment1} and Fig.~\ref{fig:experiment2} show the centroid velocity
for the input and output models at different S/N levels, whose parameter exhibits the
lowest uncertainties.

\section{PySpecKit parameters}\label{pyspeckit_parameters}

\begin{figure}[h]
\centering
\includegraphics[width = \columnwidth]{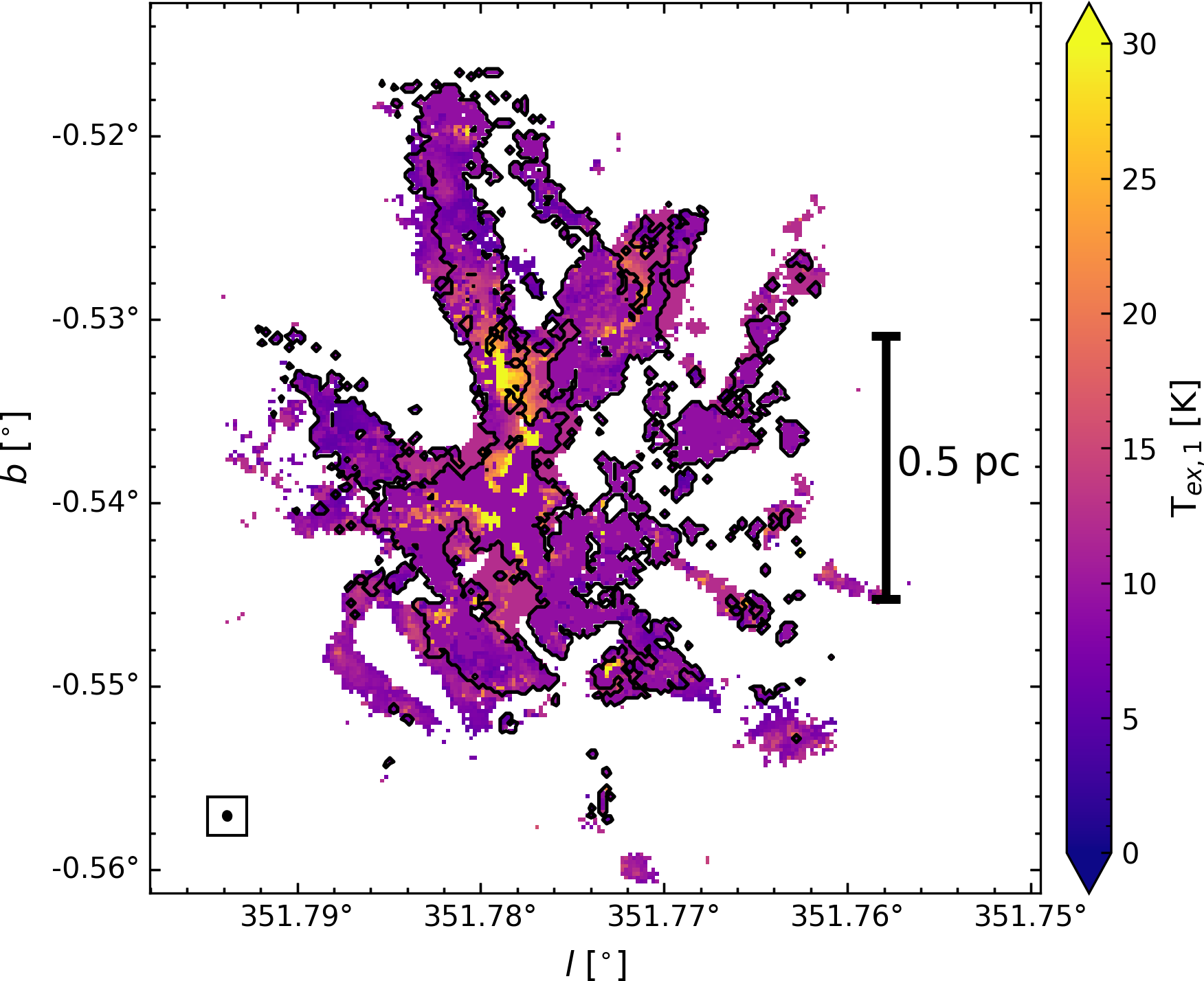}
\includegraphics[width = \columnwidth]{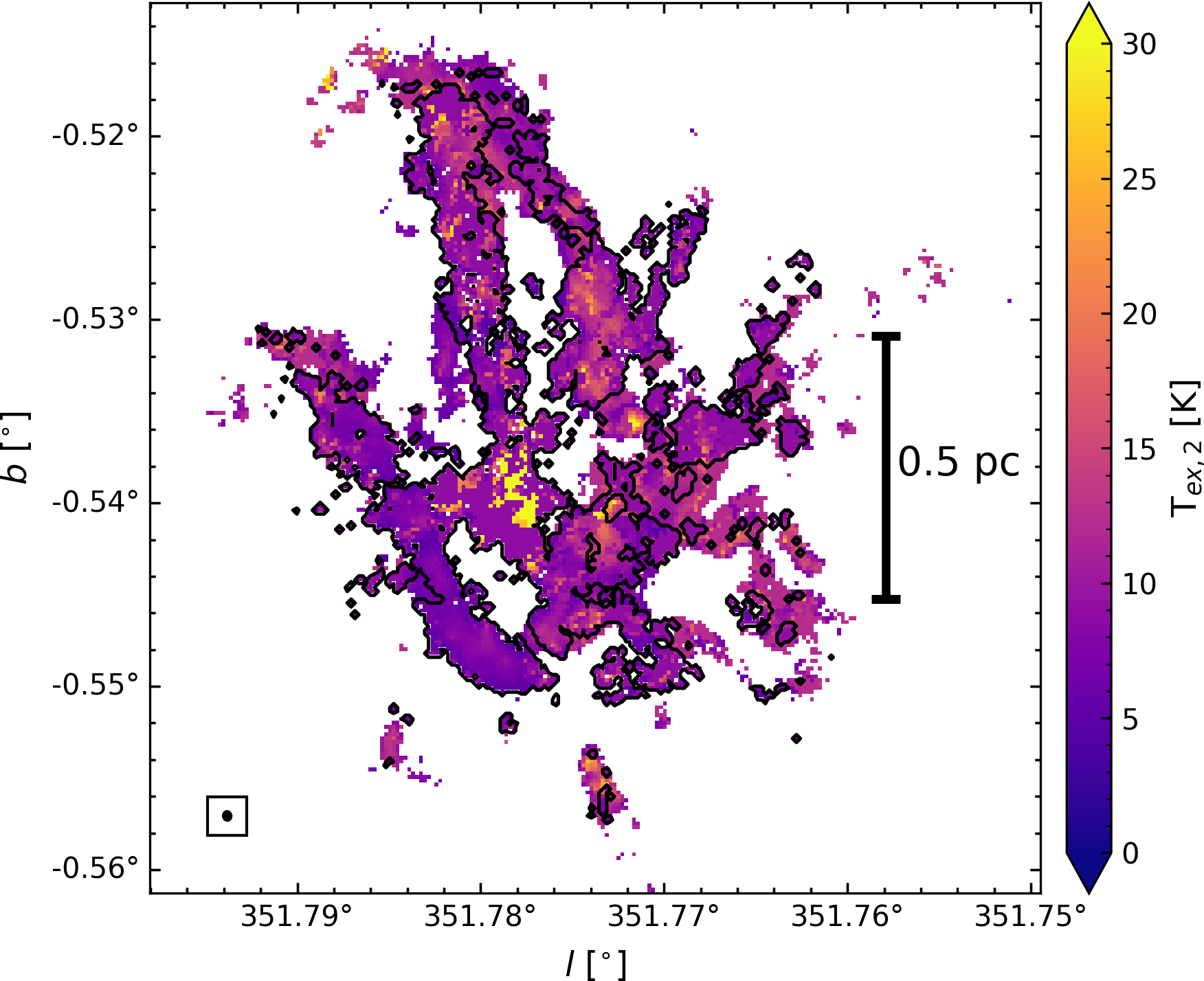}
\caption{Top: excitation temperature map of the Blue-velocity component. 
Bottom: excitation temperature map of the Red-velocity component. The
spectra inside the black contour are fitted with 2-velocity-components.
The ellipse in the bottom-left corner represents the beam size of the
\nhp data.}
\label{fig:temp_map}
\end{figure}

\begin{figure}[h]
\centering
\includegraphics[width = \columnwidth]{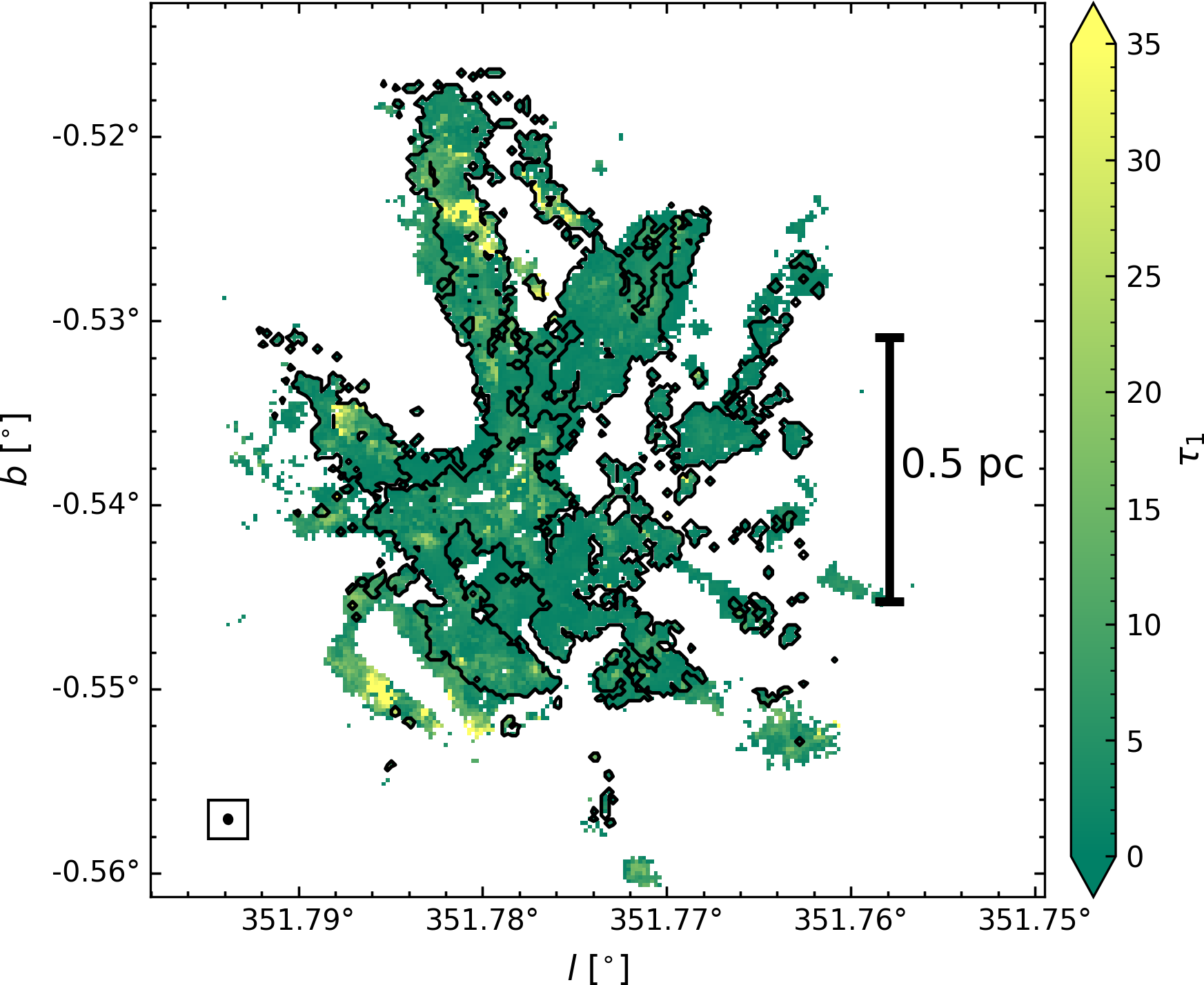}
\includegraphics[width = \columnwidth]{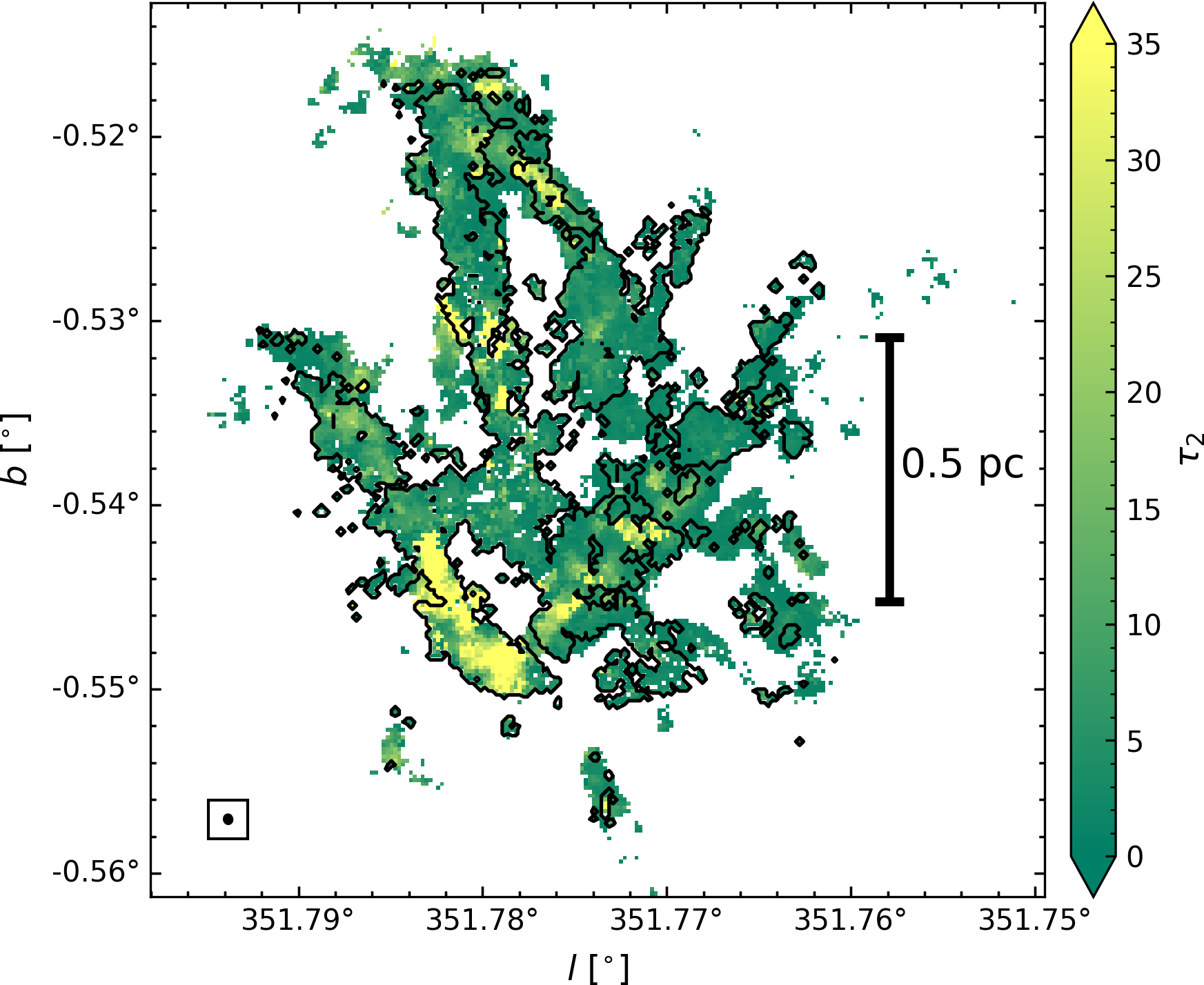}
\caption{Top: optical depth map of the Blue-velocity component. 
Bottom: optical depth map of the Red-velocity component. The
spectra inside the black contour are fitted with 2-velocity-components.
The ellipse in the bottom-left corner represents the beam size of the
\nhp data.}
\label{fig:tau_map}
\end{figure}

The models returned by PySpecKit produce parameter maps of the 
excitation temperature (T$_{\rm ex}$), optical depth ($\tau$),
centroid velocity (V$_{\rm LSR}$), and line width 
($\sigma_{\rm v}$) for each velocity component fitted. 
Figures~\ref{fig:temp_map} and \ref{fig:tau_map} show the
T$_{\rm ex}$ and $\tau$ maps for the Blue- and Red-velocity
component defined in \S~\ref{sec:bluest-and-reddest-velocity-component},
while the Figures~\ref{fig:momentmaps0} and \ref{fig:velocity_map}
display the centroid velocity and line width maps for both
same velocity components (see \S~\ref{sec:bluest-and-reddest-velocity-component}), respectively.

\section{Good and bad fitting examples}\label{sec:goodandbadfit}

\begin{figure}[h]
\centering
	\includegraphics[width = \columnwidth]{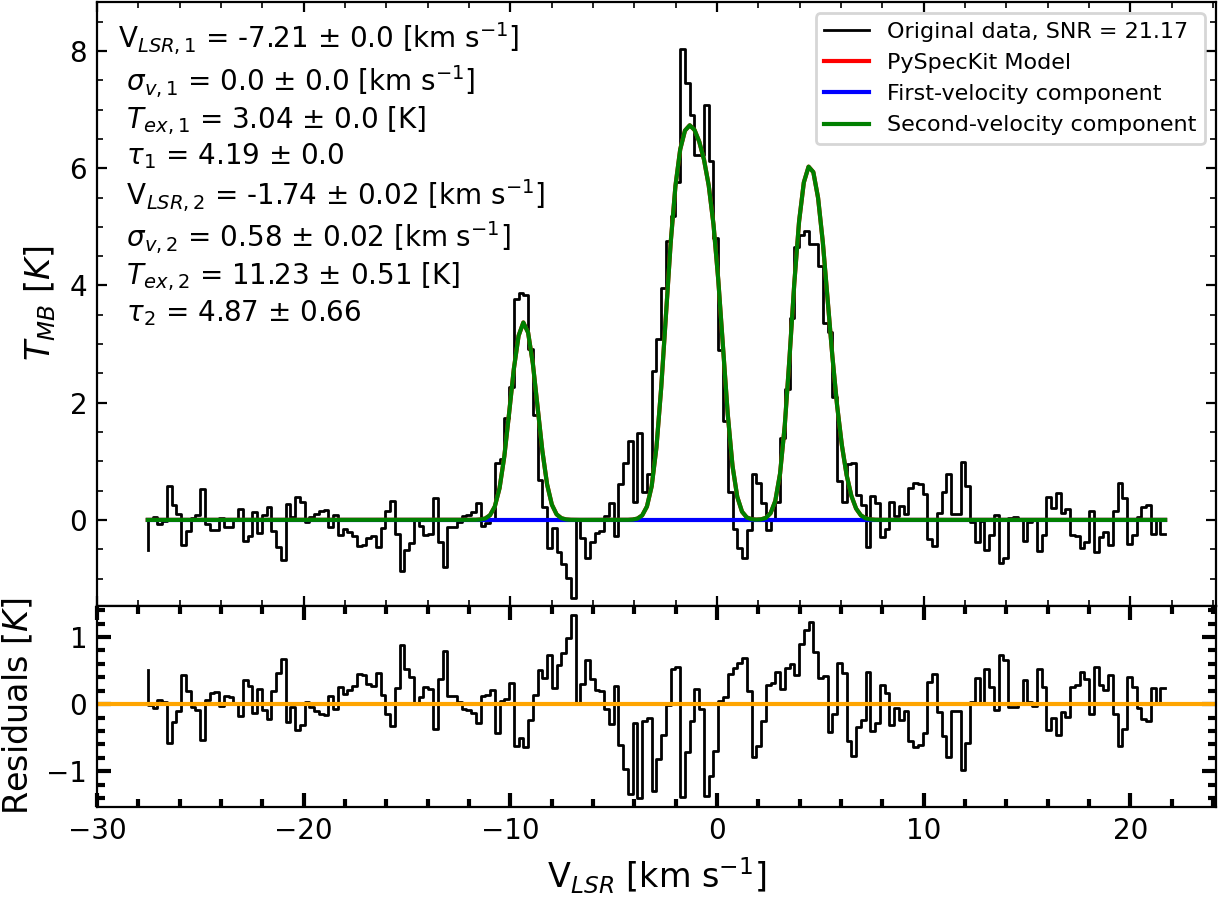}
	\caption{Example of 2-velocity-component fits. The \nhp data is represented
	by the histogram with their corresponding fits (colored curves), and the
	residuals are shown on the bottom panel. The orange line refers to the nil 
	value. The modeled first-velocity-component has a $\sigma_{v,1} = 0$ $\kms$ 
	and no uncertainties in its parameters, indicating that it is not possible
	to fit 2-velocity-component in the spectrum.}  
\label{fig:specdef1comp_1}
\end{figure}

\begin{figure}[h]
\centering
	\includegraphics[width = \columnwidth]{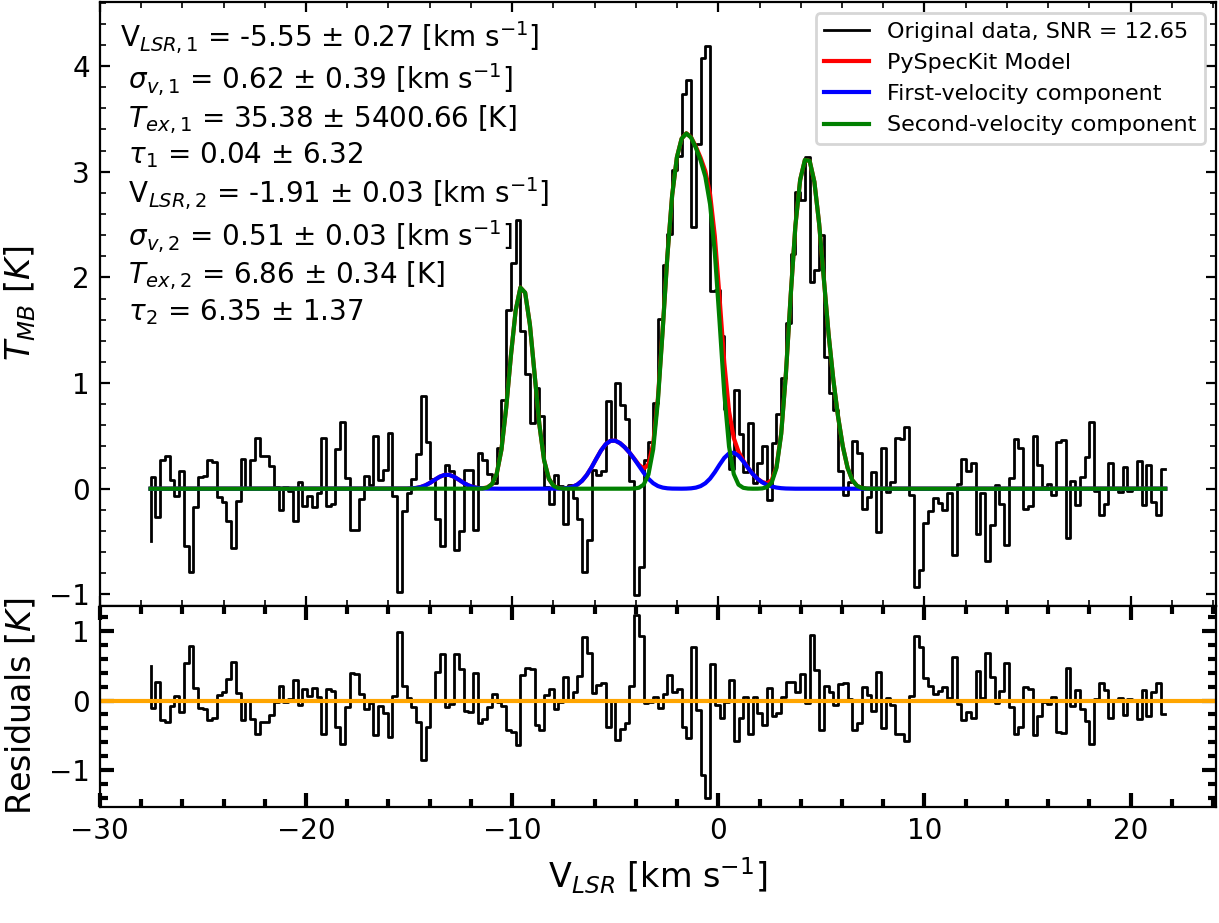}
	\caption{Example of 2-velocity-component fits. The \nhp data is represented
	by the histogram with their corresponding fits (colored curves), and the
	residuals are shown on the bottom panel. The orange line refers to the nil 
	value. The intensity peak of the first-velocity-component is similar as 
	what we measure for noise, indicating that it is not possible
	to fit 2-velocity-component in the spectrum.}
\label{fig:specdef1comp_2}
\end{figure}

The 2-velocity-component fit of the \nhp emission in G351.77 protocluster 
provides a more reliable fit for spectra that present multiple velocity 
components. However, this forces PySpecKit to fit ``false velocity 
components'' in spectra that reveal only one clear velocity component 
(see Fig.~\ref{fig:specdef1comp_1} and Fig.~\ref{fig:specdef1comp_2}). 
The two methods applied over the model fitted with 1- and 
2-velocity-components (see \S~\ref{sec:best-fit-and-final-model}) 
enable us to define the best fit for each spectrum (see 
Fig.~\ref{fig:specwellfited1comp} and Fig.~\ref{fig:specwellfited2comp}).

\section{Infalling and rotating sphere model}\label{sec:model}

\begin{figure}[h]
\centering
\includegraphics[width = \columnwidth]{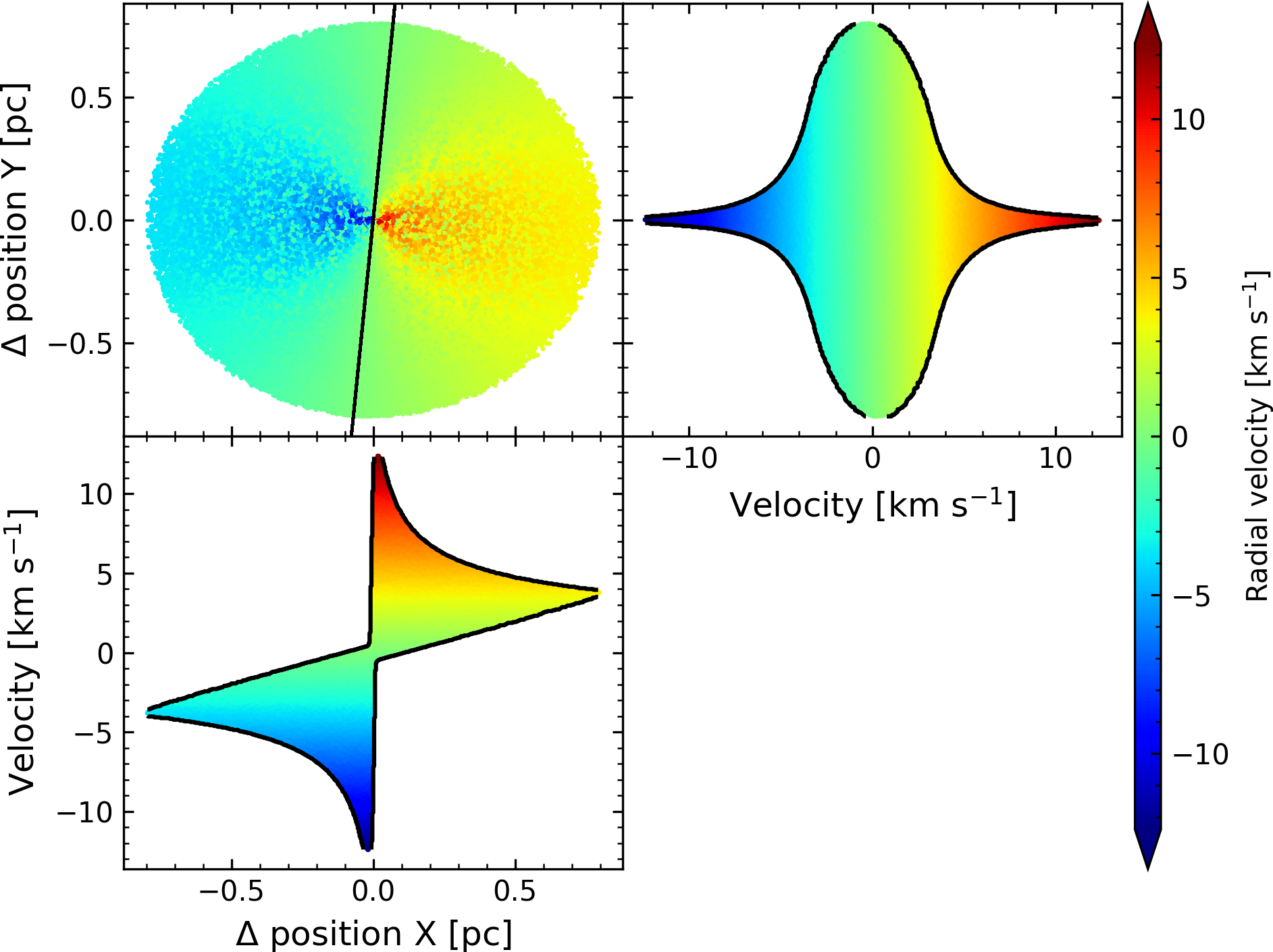}
\caption{Top left: Position-position of a rotation-only sphere of 2700~\msun. 
The black solid line represents the rotating axis whose inclination is of 
85$^{\circ}$ respect to the x axis. Top right and bottom left: PV diagrams. 
The colorbar represent the radial velocities derived from the Eq.~\ref{eq:model}, 
while the black contours show the PV features.}
\label{fig:model_rot}
\end{figure}

\begin{figure}[h]
\centering
\includegraphics[width = \columnwidth]{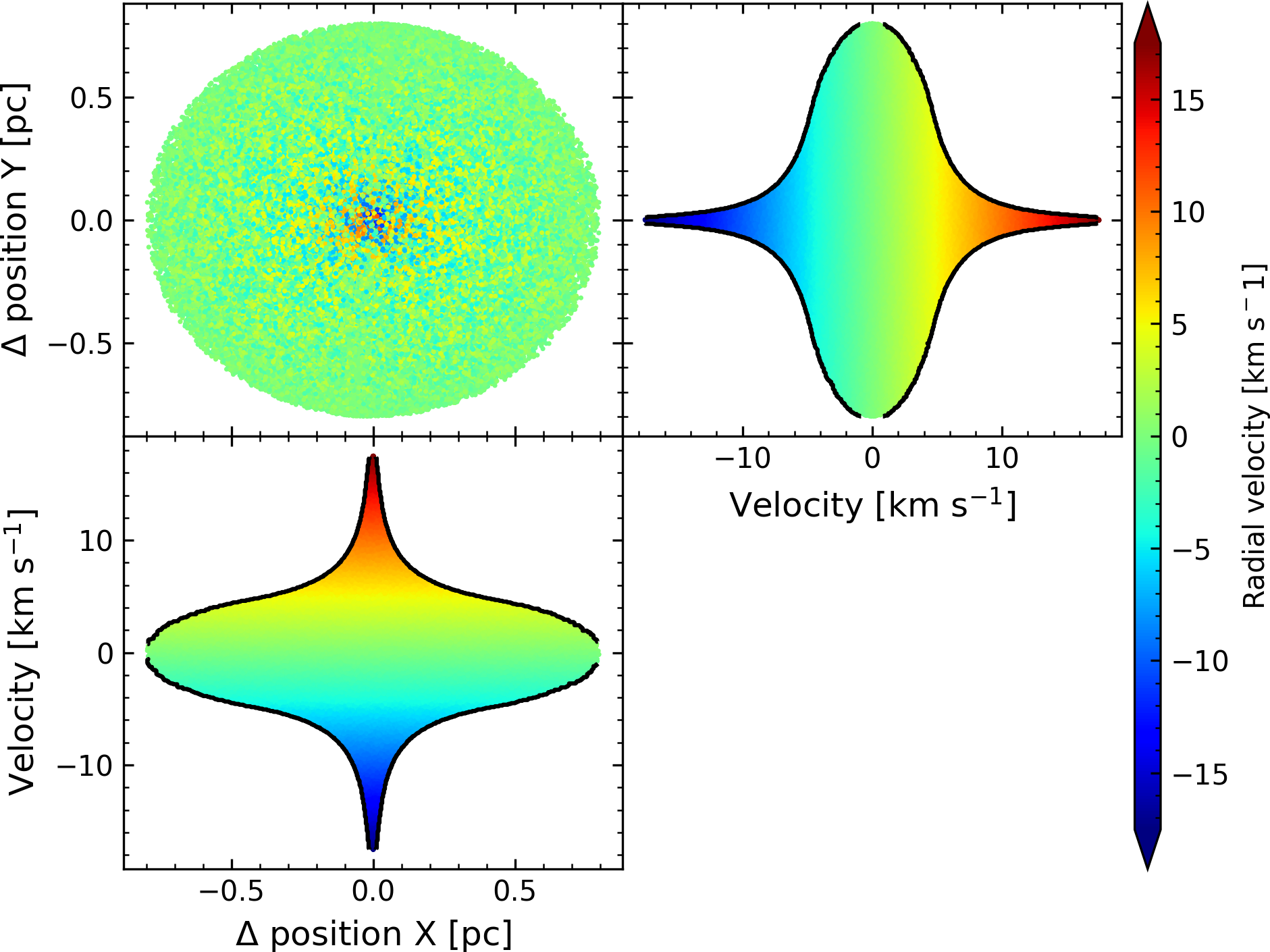}
\caption{Top left: Position-position of an infalling-only sphere of 2700~\msun. 
Top right and bottom left: PV diagrams. The colorbar represent the radial 
velocities derived from the Eq.~\ref{eq:model}, while the black contours show 
the PV features.}
\label{fig:model_in}
\end{figure}

\begin{figure}[h]
\centering
\includegraphics[width = \columnwidth]{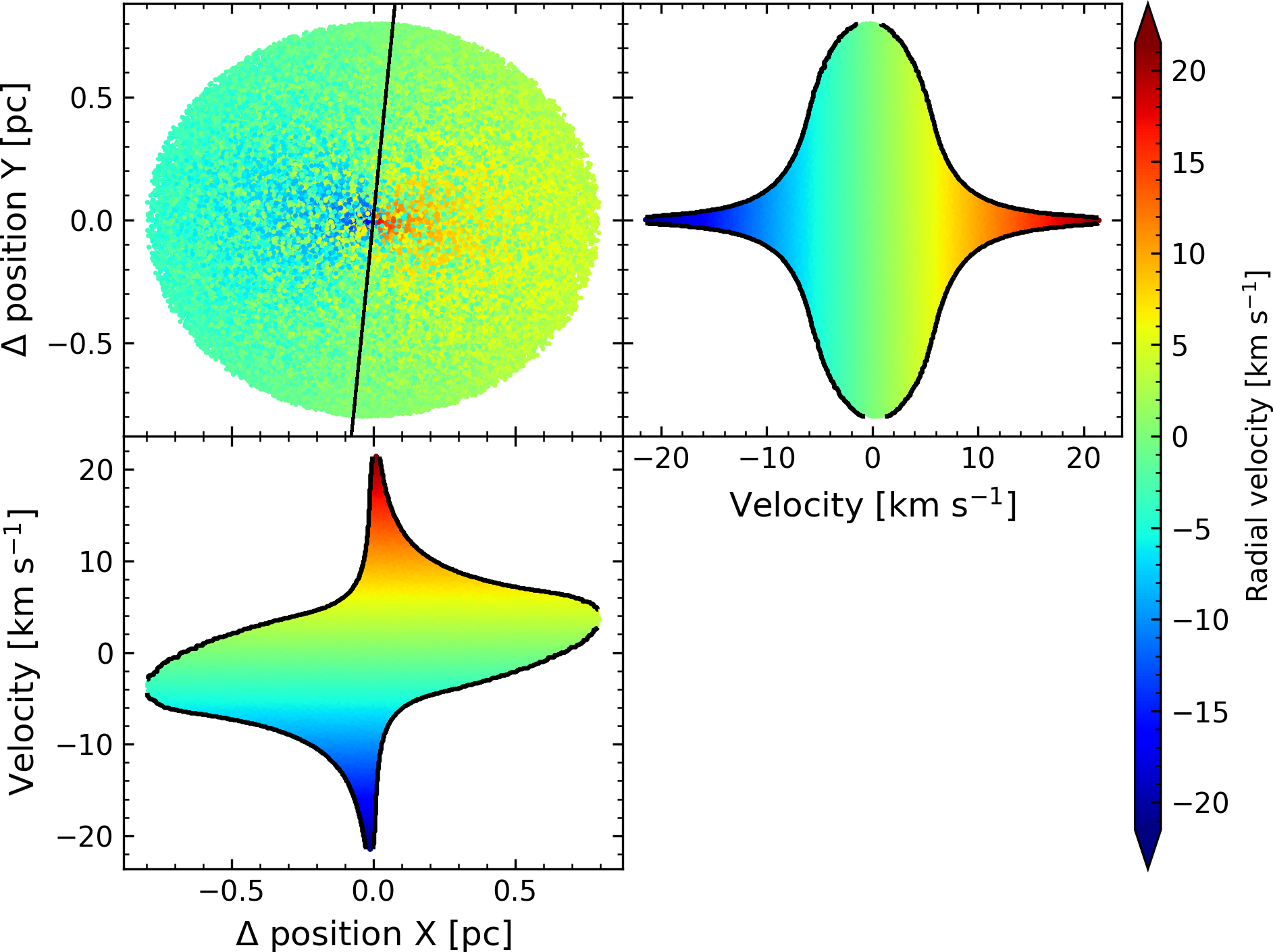}
\caption{Top left: Position-position of a rotating and infalling sphere of 
2700~\msun. The black solid line represent the rotating axis whose inclination 
is of 85$^{\circ}$ respect to the x axis. Top right and bottom left: PV diagrams. 
The colorbar represent the radial velocities derived from the Eq.~\ref{eq:model}, 
while the black contours show the PV features.}
\label{fig:model_in+rot}
\end{figure}

\begin{figure}[h]
\centering
\includegraphics[width = \columnwidth]{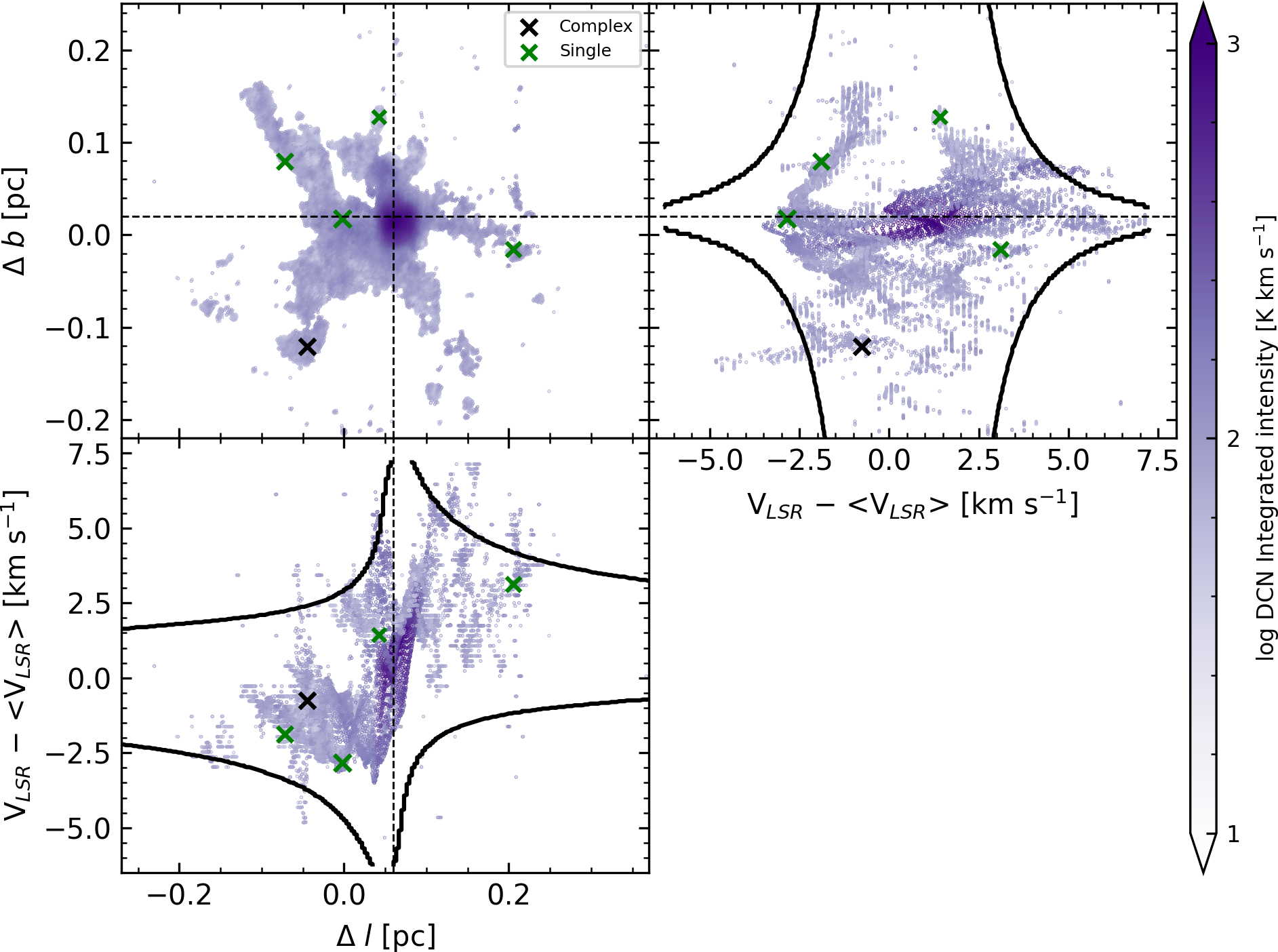}
\caption{Same caption as Fig.~\ref{fig:dcn_rot} with the black contours represent the 
  PV features produced by a rotating and infalling modeled sphere in the 
  PV diagram.}
\label{fig:dcn_in+rot}
\end{figure}

\begin{figure}[h]
\centering
\includegraphics[width = \columnwidth]{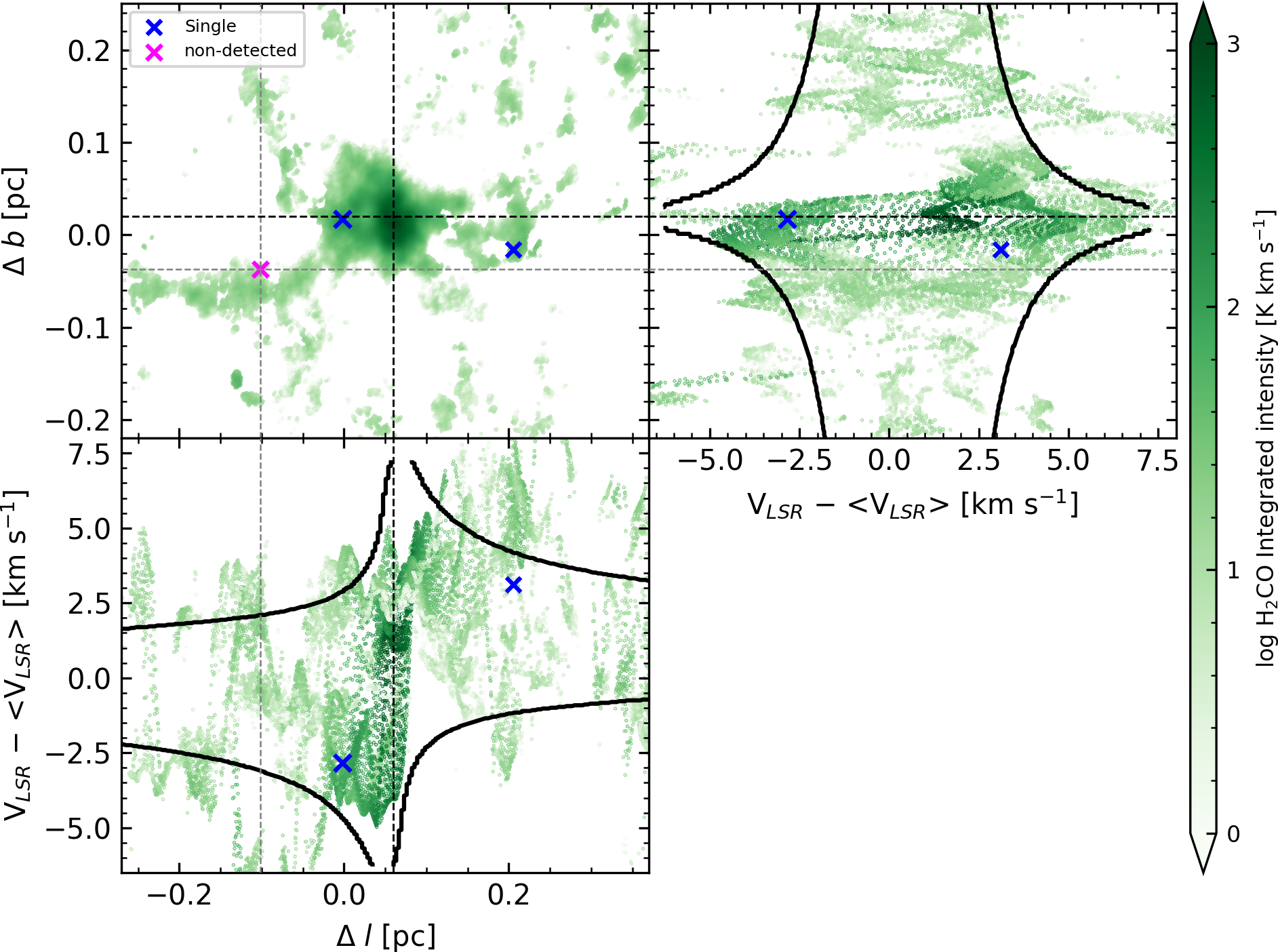}
\caption{Integrated intensity and position-velocity (PV) diagrams of 
  the \hco emission. Top left: Spatial distribution of \hco emission in G351.77. 
  The blue and magenta $\times$ markers indicate the position of
  the dense cores \citep{Fabien}, where each color 
  represents the \citet{Nichol} DCN spectral classification: single, 
  and non-detected, respectively. Dashed gray curves indicate the
  positions of the cores that do not have identifiable velocities
  (non-detected). The intersection of the dashed black curves represent 
  the center of the rotating and infalling modeled sphere 
  (see \S~\ref{sec:model}). Top right and bottom left: PV diagrams along 
  the two perpendicular directions. The black contours represent the 
  PV features produced by a rotating and infalling modeled sphere in the 
  PV diagram.}
\label{fig:h2co_in+rot}
\end{figure}

\begin{figure}[h]
\centering
\includegraphics[width = \columnwidth]{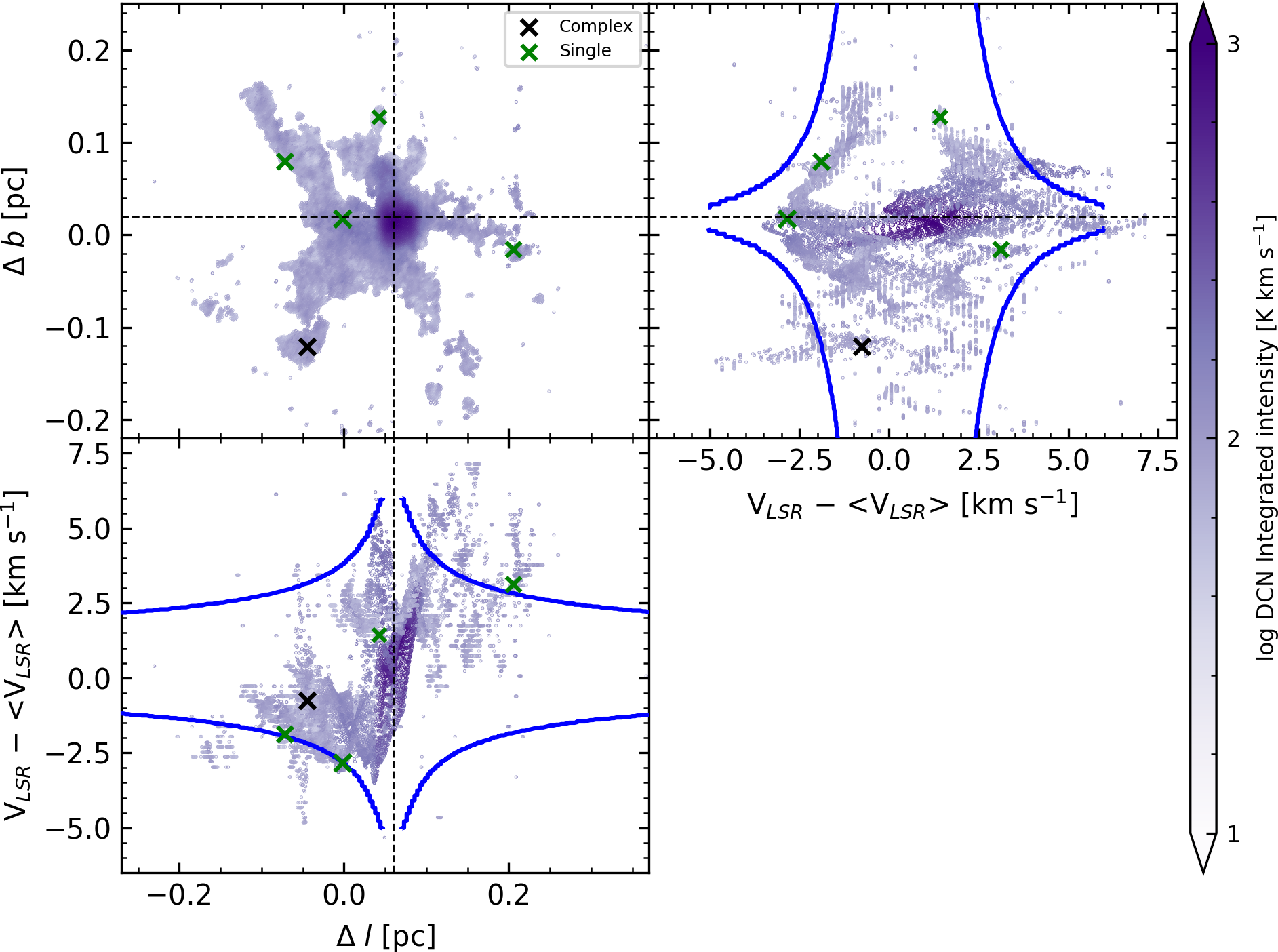}
\caption{Same caption as Fig.~\ref{fig:dcn_rot} with the blue contours represent the 
  PV features produced by a infalling modeled sphere in the 
  PV diagram.}
\label{fig:dcn_in}
\end{figure}

\begin{figure}[h]
\centering
\includegraphics[width = \columnwidth]{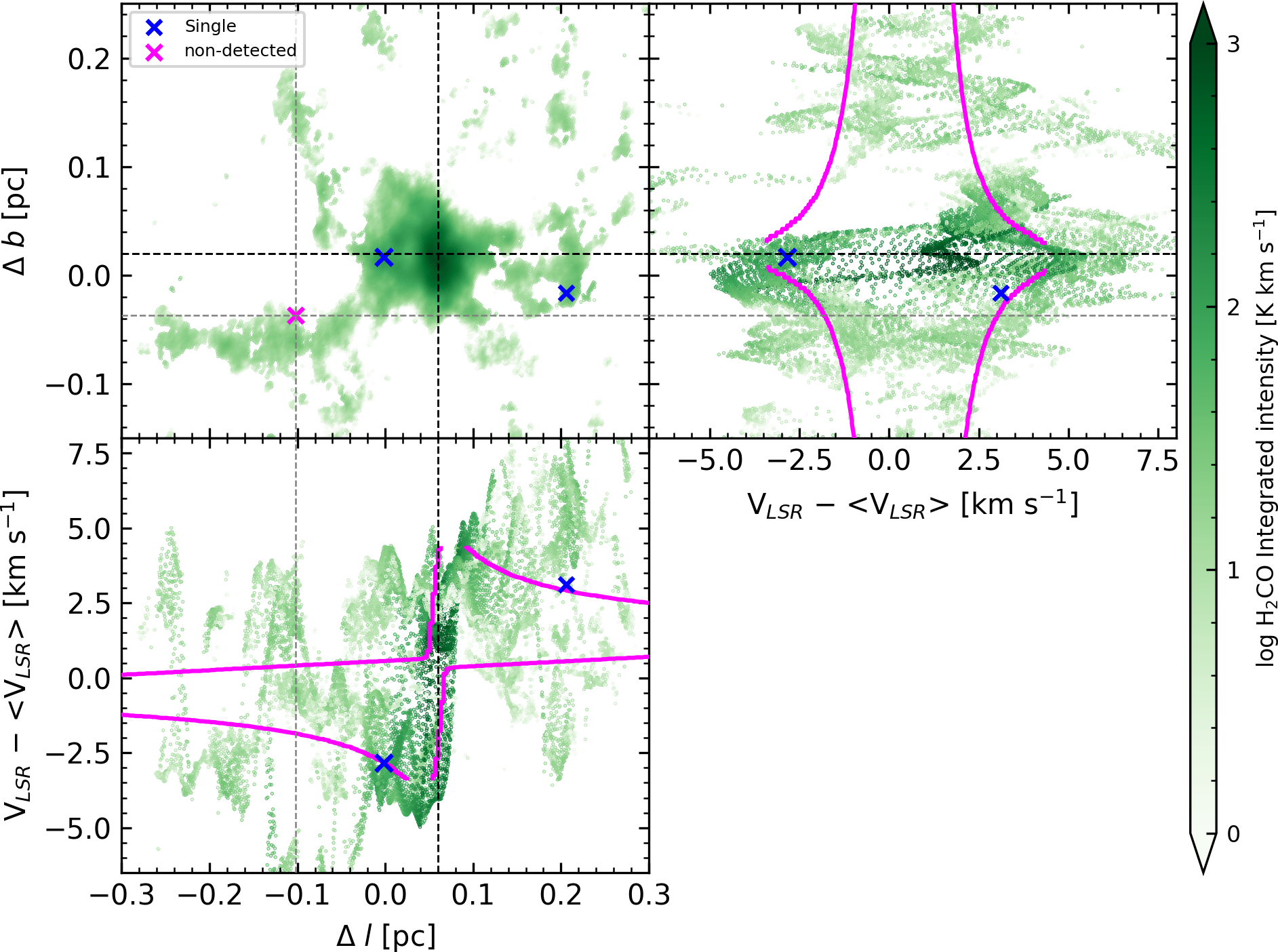}
\caption{Same caption as Fig.~\ref{fig:h2co_in+rot} with the magenta contours 
  represent the 
  PV features produced by a rotating modeled sphere in the 
  PV diagram.}
\label{fig:h2co_rot}
\end{figure}

\begin{figure}[h]
\centering
\includegraphics[width = \columnwidth]{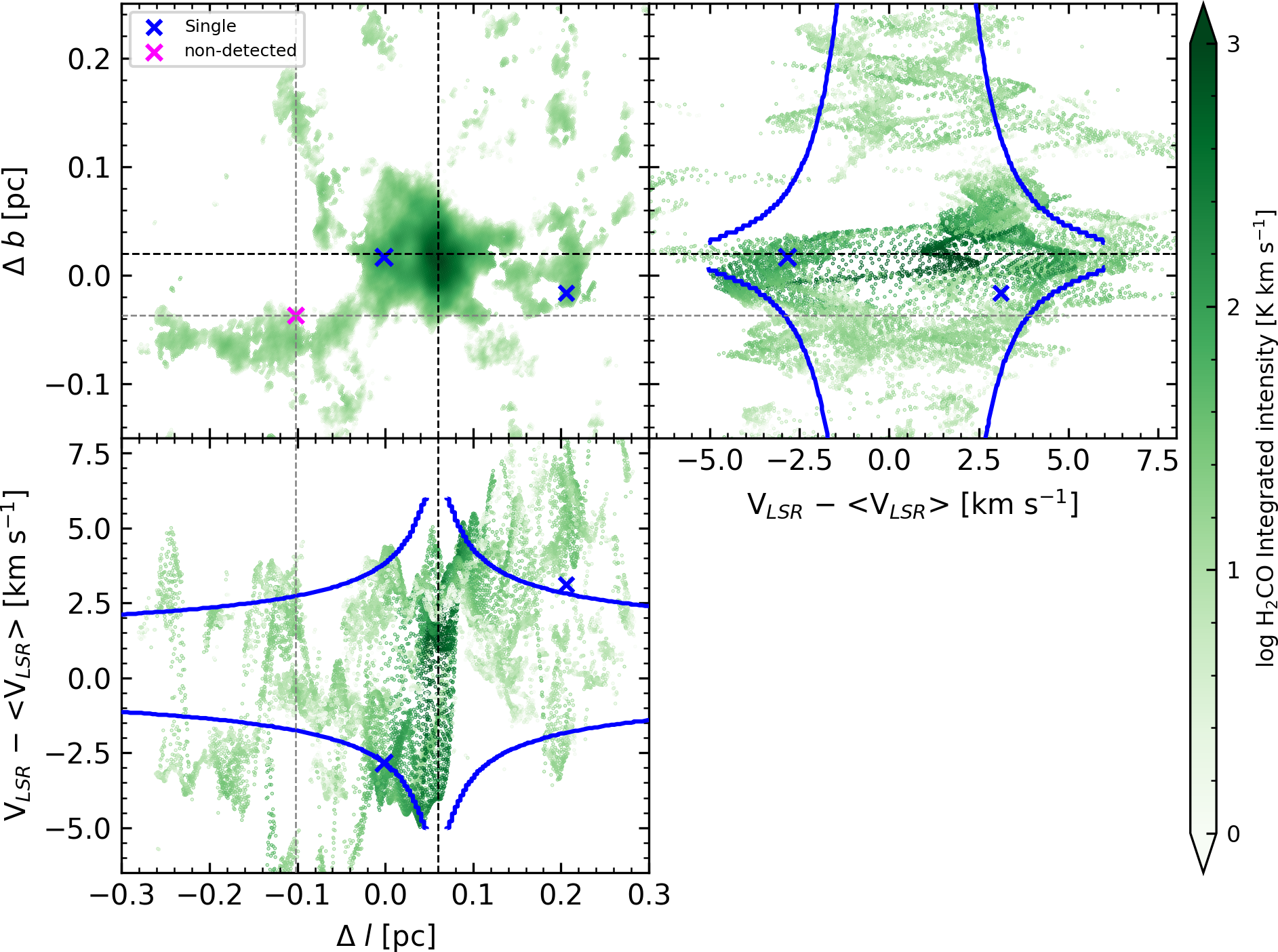}
\caption{Same caption as Fig.~\ref{fig:h2co_in+rot} with the blue contours represent the 
  PV features produced by a infalling modeled sphere in the 
  PV diagram.}
\label{fig:h2co_in}
\end{figure}

The Fig.~\ref{fig:model_rot} displays the modeled sphere under a Keplerian
rotation (see Eq.~\ref{eq:model}) and its PV features produced in the PV diagrams. The Fig.~\ref{fig:model_in}
represent the infalling modeled sphere (see Eq.~\ref{eq:model}) 
and its PV features produced in the PV diagrams. The Fig.~\ref{fig:model_in+rot}
represent the modeled sphere under both Keplerian rotation and infall at the same
time and its PV features observed in the PV diagrams.  
The figures Fig.~\ref{fig:h2co_in+rot}, Fig.~\ref{fig:dcn_in+rot}, Fig.~\ref{fig:h2co_rot}, Fig.~\ref{fig:dcn_in}, and Fig.~\ref{fig:h2co_in} display the H$_2$CO and DCN spectral lines
and its PV features observed in the PV diagrams, along with the PV features produced by the modeled
sphere. 

\end{document}